\documentclass[%
prc, reprint,
superscriptaddress,
altaffillsymbol,
 amsmath,amssymb,
 aps,
]{revtex4-2}

\usepackage{graphicx}% Include figure files
\usepackage{dcolumn}% Align table columns on decimal point
\usepackage{bm}% bold math
\usepackage{multirow}
\usepackage{booktabs}
\usepackage{siunitx}
\usepackage{comment}
\begin{document}

%\preprint{APS/123-QED}

\title{Exploring $\beta$ decay and $\beta$-delayed neutron emission in exotic $^{46,47}$Cl isotopes}

\author{Vandana Tripathi}
\altaffiliation{Corresponding author: vtripath@fsu.edu}
\affiliation{Department of Physics, Florida State University, Tallahassee, Florida 32306, USA}

\author{B.~Longfellow}
\altaffiliation{Corresponding author: longfellow1@llnl.gov}
\affiliation{National Superconducting Cyclotron Laboratory, Michigan State University, East Lansing, Michigan 48824, USA}
\affiliation{Department of Physics and Astronomy, Michigan State University, East Lansing, Michigan 48824, USA}
\affiliation{Lawrence Livermore National Laboratory, Livermore, California 94550, USA}

\author{A.~Volya} 
\affiliation{Department of Physics, Florida State University, Tallahassee, Florida 32306, USA}

\author{E.~Rubino}
\affiliation{Department of Physics, Florida State University, Tallahassee, Florida 32306, USA}

\author{C.~Benetti} 
\affiliation{Department of Physics, Florida State University, Tallahassee, Florida 32306, USA}

\author{J.~F.~Perello}
\affiliation{Department of Physics, Florida State University, Tallahassee, Florida 32306, USA}

\author{S.~L.~Tabor}
\affiliation{Department of Physics, Florida State University, Tallahassee, Florida 32306, USA}

\author{S.~N.~Liddick}
\affiliation{National Superconducting Cyclotron Laboratory, Michigan State University, East Lansing, Michigan 48824, USA}
\affiliation{Facility for Rare Isotope Beams, Michigan State University, East Lansing, Michigan 48824, USA}
\affiliation{Department of Chemistry, Michigan State University, East Lansing, Michigan 48824, USA}

\author{P.~C.~Bender}
\affiliation{Department of Physics, University of Massachusetts Lowell, Lowell, Massachusetts 01854, USA}

\author{M.~P.~Carpenter}
\affiliation{Physics Division, Argonne National Laboratory, Argonne, Illinois 60439, USA}

\author{J.~J.~Carroll}
\affiliation{U.S. Army Combat Capabilities Development Command Army Research Laboratory, Adelphi, Maryland 20783 USA}

\author{A.~Chester}
\affiliation{National Superconducting Cyclotron Laboratory, Michigan State University, East Lansing, Michigan 48824, USA}
\affiliation{Facility for Rare Isotope Beams, Michigan State University, East Lansing, Michigan 48824, USA}

\author{C.~J.~Chiara}
\affiliation{U.S. Army Combat Capabilities Development Command Army Research Laboratory, Adelphi, Maryland 20783 USA}

\author{K.~Childers}
\affiliation{National Superconducting Cyclotron Laboratory, Michigan State University, East Lansing, Michigan 48824, USA}
\affiliation{Department of Chemistry, Michigan State University, East Lansing, Michigan 48824, USA}

\author{B.~R.~Clark}
\affiliation{Department of Physics and Astronomy, Mississippi State University, Mississippi State, Mississippi 39762, USA}

\author{B.~P.~Crider}
\affiliation{Department of Physics and Astronomy, Mississippi State University, Mississippi State, Mississippi 39762, USA}

\author{J.~T.~Harke}
\affiliation{Lawrence Livermore National Laboratory, Livermore, California 94550, USA}

\author{R.~Jain}
\affiliation{National Superconducting Cyclotron Laboratory, Michigan State University, East Lansing, Michigan 48824, USA}
\affiliation{Department of Physics and Astronomy, Michigan State University, East Lansing, Michigan 48824, USA}
\affiliation{Lawrence Livermore National Laboratory, Livermore, California 94550, USA}

\author{S.~Luitel}
\affiliation{Department of Physics and Astronomy, Mississippi State University, Mississippi State, Mississippi 39762, USA}

\author{M.~J.~Mogannam}
\affiliation{National Superconducting Cyclotron Laboratory, Michigan State University, East Lansing, Michigan 48824, USA}
\affiliation{Facility for Rare Isotope Beams, Michigan State University, East Lansing, Michigan 48824, USA}
\affiliation{Department of Chemistry, Michigan State University, East Lansing, Michigan 48824, USA}

\author{T.~H.~Ogunbeku}
\affiliation{Lawrence Livermore National Laboratory, Livermore, California 94550, USA}
\affiliation{Department of Physics and Astronomy, Mississippi State University, Mississippi State, Mississippi 39762, USA}

\author{A.~L.~Richard}
\affiliation{National Superconducting Cyclotron Laboratory, Michigan State University, East Lansing, Michigan 48824, USA}
\affiliation{Department of Physics and Astronomy, Ohio University, Athens, Ohio 45701, USA}
%\affiliation{Lawrence Livermore National Laboratory, Livermore, California 94550, USA}

\author{S.~Saha}
\affiliation{Department of Physics, University of Massachusetts Lowell, Lowell, Massachusetts 01854, USA}

\author{O.~A.~Shehu}
\affiliation{Department of Physics and Astronomy, Mississippi State University, Mississippi State, Mississippi 39762, USA}

\author{R.~Unz}
\affiliation{Department of Physics and Astronomy, Mississippi State University, Mississippi State, Mississippi 39762, USA}

\author{Y.~Xiao}
\affiliation{Department of Physics and Astronomy, Mississippi State University, Mississippi State, Mississippi 39762, USA}

\author{Yiyi Zhu}
\affiliation{Department of Physics, University of Massachusetts Lowell, Lowell, Massachusetts 01854, USA}


\date{\today}% It is always \today, today,

\begin{abstract}
In this paper, $\beta^-$ and $\beta$-delayed neutron decays of $^{46,47}$Cl 
are reported from an experiment carried out at the National Superconducting Cyclotron Laboratory
using the Beta Counting System. The half-lives of both $^{46}$Cl and $^{47}$Cl 
were extracted. Based on the delayed $\gamma$-ray transitions observed, the
level structure of $N = 28$ $^{46}$Ar was determined. Completely different sets of 
excited states above the first $2^+$ state in $^{46}$Ar were populated in the 
$^{46}$Cl $\beta0n$ and $^{47}$Cl $\beta1n$ decay channels. 
Two new $\gamma$-ray 
transitions in $^{47}$Ar were identified from the very weak $^{47}$Cl $\beta0n$ decay. 
Furthermore, $^{46}$Cl $\beta1n$ and $^{47}$Cl $\beta2n$ were also observed to yield different
population patterns for levels in $^{45}$Ar, including states of different parities.
The experimental results allow us to address some of the open questions related to
the delayed neutron emission process. For isotopes with large neutron excess and 
high $Q_{\beta}$ values, delayed neutron emission remains an important decay mode 
and can be utilized as a powerful spectroscopic tool.
Experimental results were compared with shell-model calculations using the FSU and 
$V_{MU}$ effective interactions.

\end{abstract}

\maketitle

%\tableofcontents

\newpage

\section{\label{sec:intro}Introduction}

Experimental investigations away from the valley of stability have highlighted unique and 
interesting properties of exotic nuclei, including neutron halos, neutron skins, 
and the disappearance of the standard magic numbers. 
For extremely neutron-rich nuclei, $\beta$-delayed neutron emission becomes an 
important pathway in the decay process. With increasing neutron number, the $Q$ value
for $\beta^-$ decay increases and, simultaneously, the neutron-separation energy ($S_n$) in the daughter 
nucleus decreases. This leads to an increase in the population of neutron-unbound states 
in the allowed $\beta^-$ decay process. 
Additionally, for nuclei with a large excess of neutrons, the parent and daughter nuclei often
have different ground-state parities due to the valence
protons and neutrons occupying opposite-parity shells. 
As a result, 1-particle, 1-hole ($1p1h$) states, which typically lie at high energies, are preferentially populated in allowed
$\beta^-$ decay. This, in part, drives the shifting of
the Gamow-Teller strength $B(GT)$ above the neutron-emission
threshold. These states typically decay by emitting a neutron, leading to a daughter
nucleus with $A-1$, though, in principle, decay by $\gamma$-ray emission cannot be completely ruled out.
For extremely neutron-rich nuclei, some Gamow-Teller strength can be located above the two-neutron 
separation energy, which will result in the emission of more than one neutron following $\beta^-$ decay.

The $\beta^-$-delayed neutron emission process is often modeled in two steps, where the 
neutron-rich precursor nucleus undergoes $\beta^-$ decay to a highly-excited state in the 
daughter nucleus which then subsequently emits a neutron. It is assumed that neutron emission occurs from an equilibrated system
completely independent of the formation process. Therefore, it is a statistical process and
depends only on the spin, parity, and excitation energy of the level in the daughter nucleus. 
With very limited experimental information for confirmation, this assumption may be
too simplistic, especially for lighter nuclei, such as those in the $sd$ shell where the 
density of states above $S_n$ is relatively low, putting the statistical decay assumption in doubt \cite{jutta}.
For example, there is some experimental evidence that points toward neutron emission proceeding 
from an intermediate doorway 
state with a large probability for neutron decay \cite{Xu_doorway}. 
The quantum dynamics of sequential decay proceeding via unbound neutron resonances is 
of significant theoretical interest because the intermediate state partially retains 
the memory of its initial creation. This evolving state, which comprises both 
resonant and background components, is non-stationary and therefore subject to 
internal mixing as well as explicit decay \cite{Peshkin2014,Wang2023}.

Thus, $\beta$-delayed neutron emission remains an important and unresolved 
challenge for both experiment and theory. Delayed neutron emission plays an 
important role in the determination of elemental abundances in the r-process and 
is also crucial in 
nuclear reactor technology. In addition, $\beta$-delayed neutron emission from exotic
nuclei is a powerful spectroscopic tool for the study of the $A-1$ 
daughter nucleus. In this work, $\beta^-$ decays of $^{46,47}$Cl were 
utilized to 
populate excited states in 
$^{45,46,47}$Ar, close to the neutron magic number $N=28$, by means of both 
direct $\beta^-$ decay and $\beta$-delayed neutron emission.

Studies on the breakdown of the conventional magic number $N=28$ below doubly-magic 
$^{48}$Ca ($Z=20$, $N=28$) have provided considerable insight into the evolution of 
nuclear structure toward the neutron dripline \cite{newmagicnumbers,shell_evolution}.
At the center of the $N=28$ Island of Inversion \cite{N28_shellgap}, spectroscopic 
measurements have shown that $^{42}$Si ($Z=14$, $N=28$) has a low first 
$2^+$ energy \cite{Bastin07,Takeuchi2012,Gade2019}.  
Similarly, $^{44}$S ($Z=16$, $N=28$) has shown characteristics of a diminished $N=28$ 
gap, such as large collectivity \cite{Glasmacher_44S,longfellow_ce} and multiple shape 
coexistence \cite{Force2010,Santiago2011,Parker2017}.

Properties of the transitional nucleus $^{46}$Ar ($Z=18$, $N=28$) on the other hand,
have proven to be more difficult to describe theoretically. Successful shell-model interactions 
in this region of the chart of nuclides overpredict the 
B$(E2;0^+_1\rightarrow2^+_1)$ strength by a factor of 
two \cite{Scheit1996,Gade2003,46Ar} and this issue persists for $^{47}$Ar 
\cite{Winkler2012}, $^{45}$Cl \cite{Ibbotson1999}, and $^{43,44}$S 
\cite{43S_longfellow,longfellow_ce}. The data on excited states in $^{46}$Ar beyond 
the first $2^+$ are limited and come from $^9$Be($^{48}$Ca,$^{46}$Ar$+\gamma$)X 
\cite{Dombradi2003}, inverse-kinematics proton scattering \cite{Riley2005}, and 
$^{44}$Ar$(t,p)$ \cite{Nowak2016}. Although each study reported several $\gamma$-ray 
de-exictations, there is no consistency between the respective level schemes, beyond the 
$2^+_1\rightarrow0^+_1$ transition and, tentatively, the $4^+_1\rightarrow2^+_1$ transition. 
More information on the levels in $^{46}$Ar will provide critical 
benchmarks for shell model Hamiltonians aiming to describe the underlying mechanisms driving 
nuclear shell evolution from $^{48}$Ca into the $N=28$ Island of Inversion.

In the present work, the level structure of $^{46}$Ar was obtained 
from both the $\beta0n$ decay of $^{46}$Cl and the $\beta1n$ decay of $^{47}$Cl.
From the $\beta^-$ decay of $^{46}$Cl, indications of the population of
a previously-unreported negative-parity state in $^{46}$Ar was observed. New states were 
also observed in the $\beta1n$ decay of $^{47}$Cl, although no conjectures on their parities 
could be made. 
The $\beta0n$ decay of $^{47}$Cl to $^{47}$Ar was found to be very 
weak, with $\approx$98\% of the total decay strength instead going toward delayed neutron emission. 
Despite the very small decay branch, we have tentatively identified two 
new $\gamma$-ray transitions in $^{47}$Ar. 
In addition, we observed transitions in $^{45}$Ar from  $^{47}$Cl $\beta2n$ decay. 
These transitions were compared with those seen from $^{46}$Cl $\beta1n$ decay 
and differences are highlighted.

Shell model calculations were performed using the FSU \cite{FSU} and 
$V_{MU}$ \cite{utsuno_new} effective interactions with the valence space including the $sd$ and $fp$ shells. One particle was allowed to move from the $sd$ shell to the $fp$ shell 
to create opposite-parity states.
For calculations using the $V_{MU}$ interaction, both allowed Gamow-Teller
(GT) and First Forbidden (FF) transitions were included when calculating $\beta$-decay
half-lives and delayed neutron emission probabilities.  Inclusion of FF transitions 
in the calculations had the effect of reducing the calculated half-lives and 
neutron emission probabilities as they add a pathway to the decay.

\section{\label{sec:exp}Experimental Setup}

The experiment was carried out at the Coupled Cyclotron Facility at the National
Superconducting Cyclotron Laboratory (NSCL) \cite{brad}. A 140-MeV/u $^{48}$Ca primary
beam was fragmented on a thick Be target at the mid-acceptance position of the A1900 
fragment separator \cite{A1900}. The resulting secondary beam was purified using an 
achromatic Al wedge degrader at the intermediate dispersive image of the A1900 to select
exotic isotopes of P, S, and Cl with neutron number around $N=28$ in two separate magnetic 
rigidity settings. Data from this experiment on the decays of $^{42,43,44}$P and 
$^{44,46}$S \cite{tripathi_new}, $^{45}$Cl \cite{Soumik2023}, and $^{43,45}$S 
\cite{tripathi_submitted} have been published and provide further details of the setup.
Here, we report on the decays of $^{46,47}$Cl, which were produced in the higher 
rigidity setting with a $2\%$ momentum acceptance.

% Figure 1: PID

\begin{figure}
	\includegraphics[width=\columnwidth]{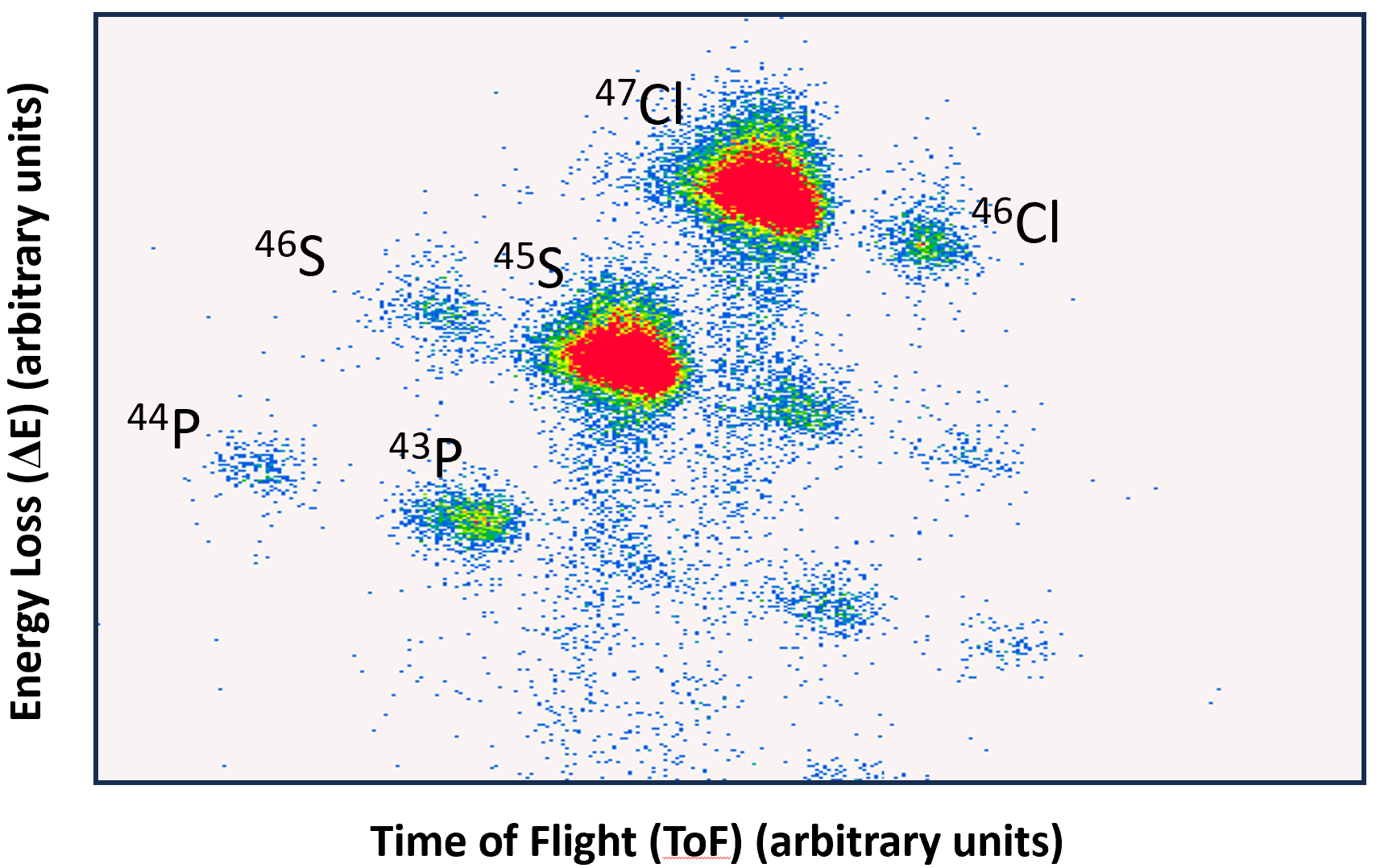}
	\caption{\label{fig:PID}
	Particle identification plot using energy loss ($\Delta$E) in the upstream Si PIN detector 
 and timing ($ToF$) of the upstream Si PIN detector with respect to the plastic scintillator 
 at the A1900 intermediate dispersive image.
	}
\end{figure}

As shown in Fig.~\ref{fig:PID}, particle identification was performed event-by-event 
using energy loss in two Si PIN detectors and their timing information relative to the 
plastic scintillator at the A1900 intermediate dispersive image. Downstream of the Si 
PIN detectors, the secondary beam particles, including the ions of interest, $^{46,47}$Cl,
were implanted in the 40 strips x 40 strips pixelated Double-Sided Silicon Strip 
Detector (DSSD) (active area of about 40 mm x 40 mm) belonging to the Beta Counting System (BCS) \cite{BCS}. An Al 
degrader was placed upstream before the DSSD to ensure that implants were stopped near the middle of the 986-$\mu$m 
thick DSSD. Furthermore, the straggling through the degrader ensured that a large number 
of the DSSD pixels were illuminated. The implantation rate was about 150/s spread over the greater than 1000 pixels of the DSSD. This rate was sufficient to allow the decay 
of the implanted radioactive ion (half-lives typically 100s of ms or less)
before a second implantation in the same or neighboring pixel, enabling clean correlations between the 
chosen implant and its decay products. As the DSSD was $\sim$1 mm thick, the $\beta$ 
detection efficiency was only 50-60\%. A Single-Sided Silicon Strip Detector (SSSD) was installed downstream to 
veto light particles punching through the DSSD. 

The BCS was surrounded by 16 HPGe Clover 
detectors to record $\beta$-delayed $\gamma$ rays. Energy and efficiency calibration was 
performed using standard $\gamma$-ray sources up to 3.5 MeV. The efficiency was about 
5\% at 1 MeV. The setup did not include neutron detection, and, hence, information 
about delayed neutron emission was obtained from delayed $\gamma$-ray transitions instead.
The data were collected using the NSCL digital data acquisition system 
which provides timestamps for each channel, allowing coincidences and correlations to be 
built in the offline analysis \cite{prokop}.

% Figure 2:decay curves

\begin{figure}
	\includegraphics[width=\columnwidth]{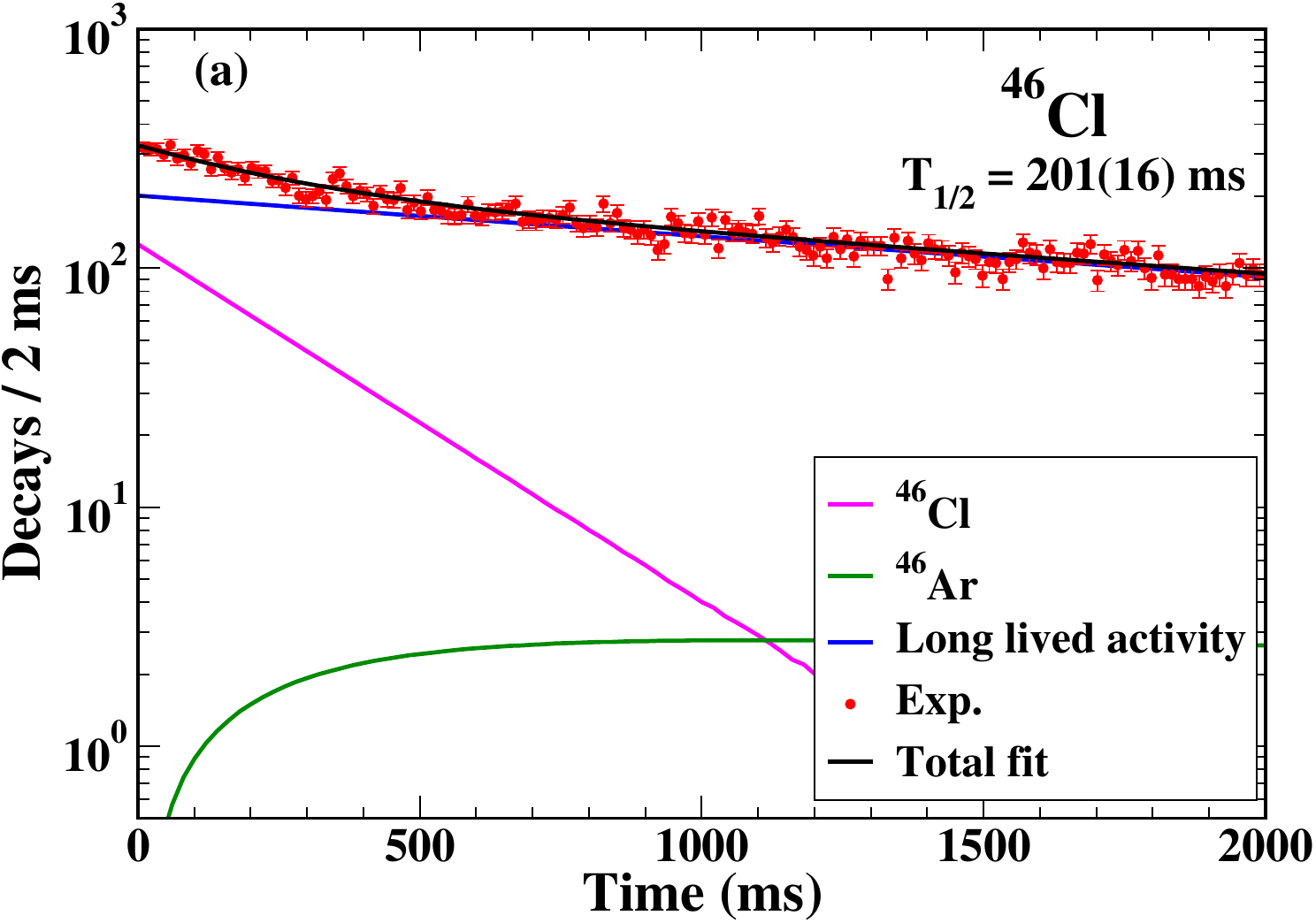}
	\includegraphics[width=\columnwidth]{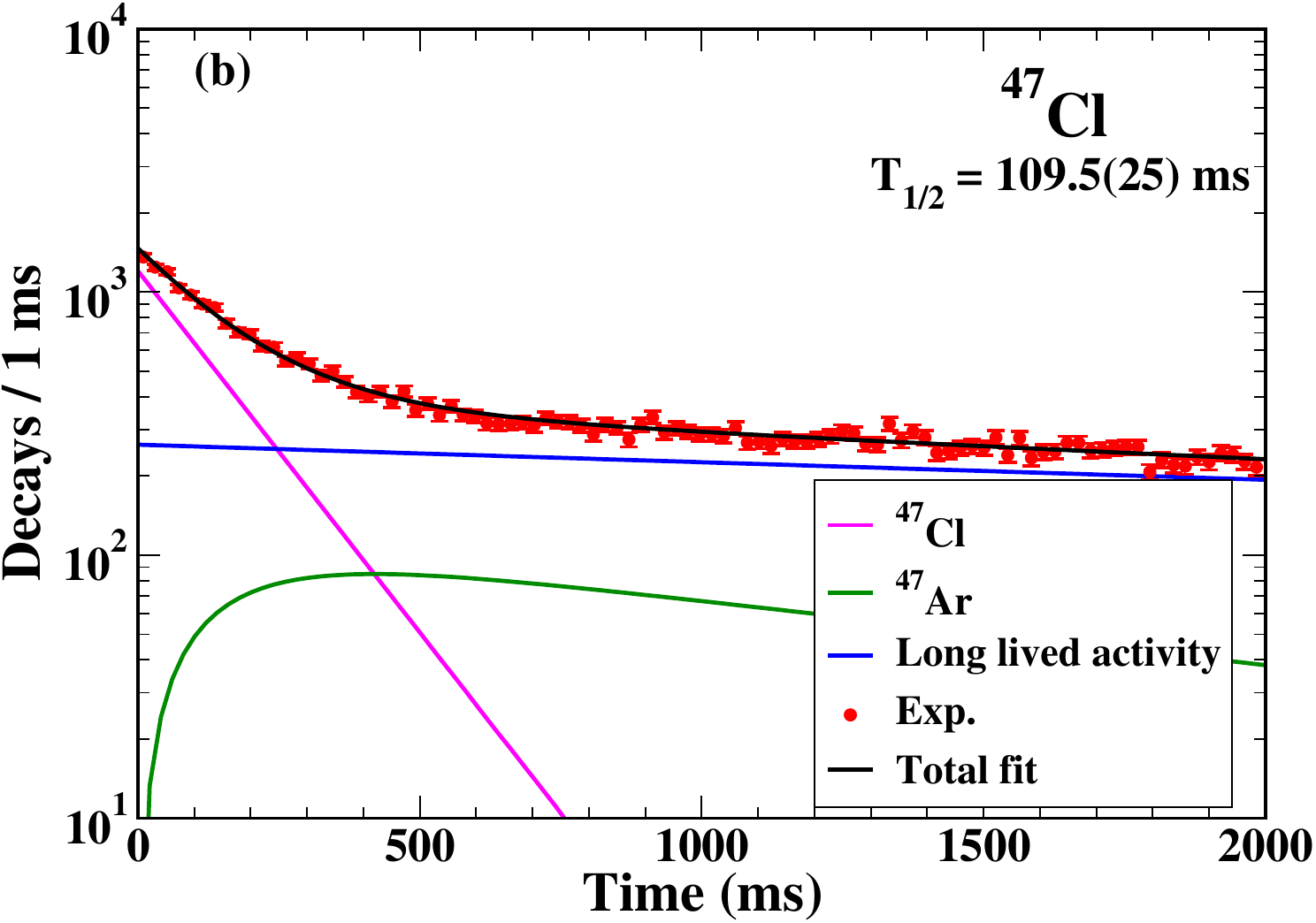}
    \includegraphics[width=\columnwidth]{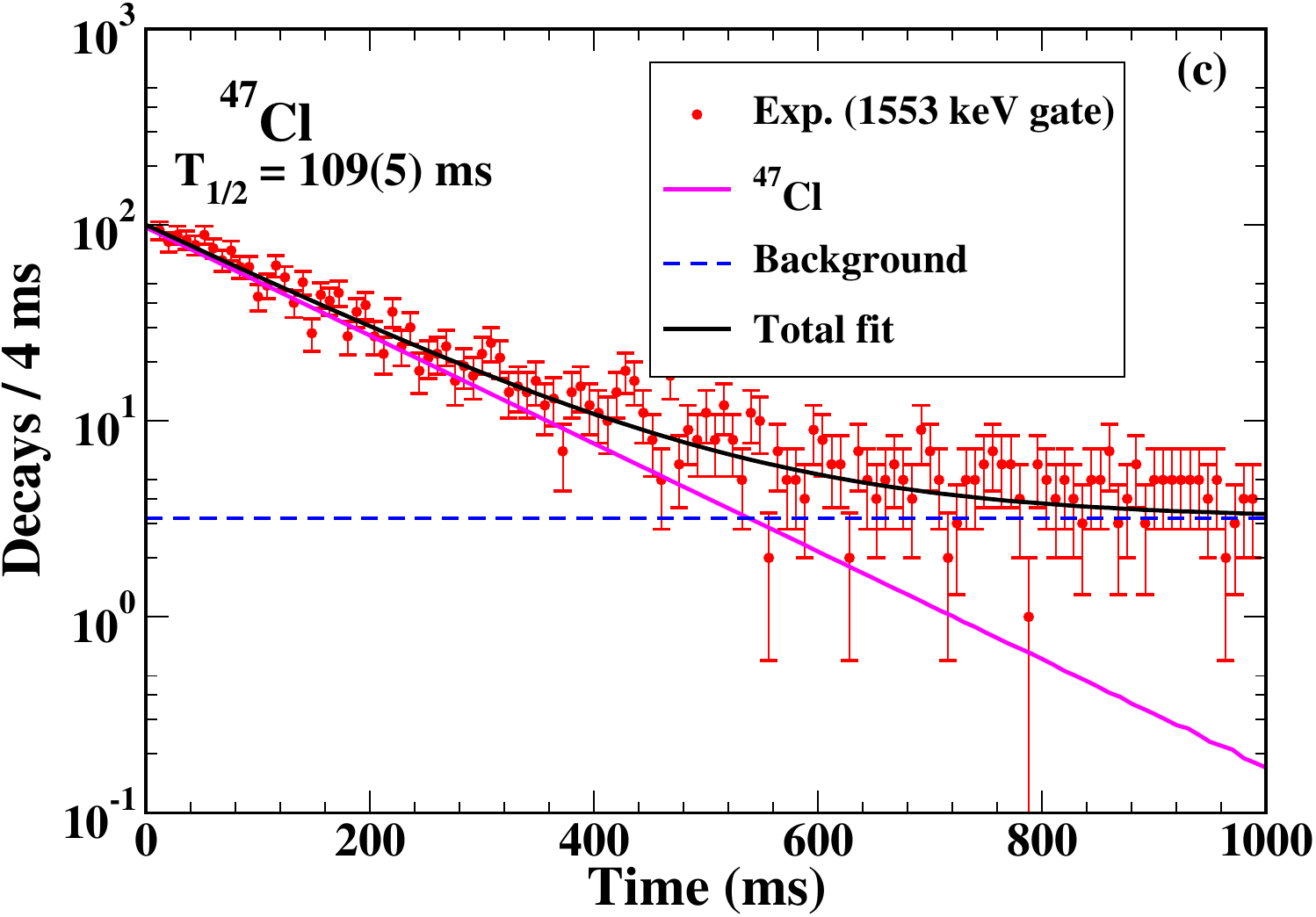}
	\caption{\label{fig:46_47Cl_decaycurve}
		(a)  Decay curve derived for $^{46}$Cl from 
   $\beta$-correlated implants within a grid of nine pixels using a 2 s time correlation 
   window along with the fit used to extract the half-life. The components of the fit are 
   (i) exponential decay of the parent nucleus,  $^{46}$Cl, (ii) exponential growth and 
   decay of the daughter nuclei,  $^{46}$Ar ($\beta0n$) and (iii) exponential background 
   to account for the long-lived activities not considered explicitly.  Known half-lives 
   were used for the daughter nucleus \cite{nndc}.
        (b) Same as (a) for $^{47}$Cl 
	(c)  Decay curve for $^{47}$Cl gated on the 1553-keV $\gamma$-ray transition, 
  which is from the first excited state in the $\beta1n$ daughter $^{46}$Ar and the most 
  intense $\gamma$ transition in the $\beta$-delayed $\gamma$-ray spectrum correlated 
  to $^{47}$Cl implants.
	}
\end{figure}
 
\section{\label{sec:results}Experimental Results}

\subsection{\label{subsec:thalf}Half-life measurements}

% Table1

\begin{table}
\caption{\label{tab:t1/2} Measured and calculated $T_{1/2}$ values for $^{46,47}$Cl. 
Shell model calculations (GT only and GT+FF transitions) were performed for $^{46,47}$Cl using the $V_{MU}$ interaction \cite{utsuno_new}. 
For the parent nucleus, several calculated levels at very low excitation energy $E^*$ with 
different spin-parity were considered for the $\beta$ decay and the corresponding $T_{1/2}$ 
values are listed.
The half-lives in boldface are those closest to the experimental values and indicate 
probable candidates for the spin-parity of the parent ground state.}
  \begin{tabular}{lc|cccc}
    \toprule
    \multirow{2}{*}{Isotope} &
      \multicolumn{3}{c}{} &
      \multicolumn{2}{c}{$T_{1/2}$ (ms)} \\
      & ~{Exp.}~~~&~~{$J^\pi$} &~~~ {$E^*$(keV)} & ~~{GT+FF} ~~& ~~{GT}  \\
      \midrule
    $^{46}$Cl & 201(16) ms & {$0^-$} & 0 & {~271.4} & 345.9 \\
              & & {$1^-$} & 310 & {\textbf{221.9}} & 255.7 \\
              & & {$2^-$} & 160 & {\textbf{240.8}} & 287.6 \\
              & & {$3^-$} & 370 & {~333.8} & 400.7 \\
              \hline
    $^{47}$Cl & 109.5(25) ms& {$1/2^+$} & 0 & {~82.9} & 103.5 \\
              & & {$3/2^+$} & 30 & {\textbf{109.5}} & 129.3 \\   
    \bottomrule
  \end{tabular}
\end{table}

The decay curves for $^{46,47}$Cl are shown in Figs.~\ref{fig:46_47Cl_decaycurve}(a) and \ref{fig:46_47Cl_decaycurve}(b) 
respectively. 
These curves were generated by histogramming the time difference between the selected 
implanted ion ($^{46}$Cl or $^{47}$Cl) and 
its correlated decay event. The $\beta$ particle was required to be detected 
within the same pixel as the implant or in one of the neighboring eight pixels of the 
implant within a time correlation window of 2 s. 
Each decay curve was fitted with an expression which included the  
exponential decay of the parent nucleus and the growth and decay of the daughter nuclei ($\beta$0n)
following the Bateman equations. In both cases, 
literature values of half-lives of daughter nuclei were used \cite{nndc}. 
As the $\beta$n daughters namely, $^{46}$Ar ($t_{1/2} \approx 8.4 s$) and $^{45}$Ar ($t_{1/2} \approx 21.5 s$), are very long lived they 
were not explicitly included. Instead, an additional slowly falling exponential functional with free parameters was added to account for long-lived activities and constant background for each isotope. 
The resulting half-life for $^{46}$Cl is 201(16) ms which agrees with the previous 
measurement of 223(37) ms \cite{Sorlin1993,sorlin_Pn} within uncertainties and is 
slightly lower than the value of 232(2) ms from Ref.~\cite{grevy} but within 2$\sigma$. 
The half-life measured for $^{47}$Cl in this work is 109.5(25) ms, which is slightly higher 
than the literature value of 101(6) ms \cite{grevy}, but within their mutual uncertainties. Additionally, the $^{47}$Cl half-life was 
determined by gating on the 1553-keV $\gamma$-ray transition in the $\beta1n$ daughter,
$^{46}$Ar (Fig.~\ref{fig:46_47Cl_decaycurve}(c)).
An exponential (parent decay) and a constant (background) were fitted to the curve, 
yielding a half-life of 109(5) ms, in excellent 
agreement with the ungated fit.

% Figure 3: beta delayed gamma spectrum

\begin{figure}
	\includegraphics[width=\columnwidth]{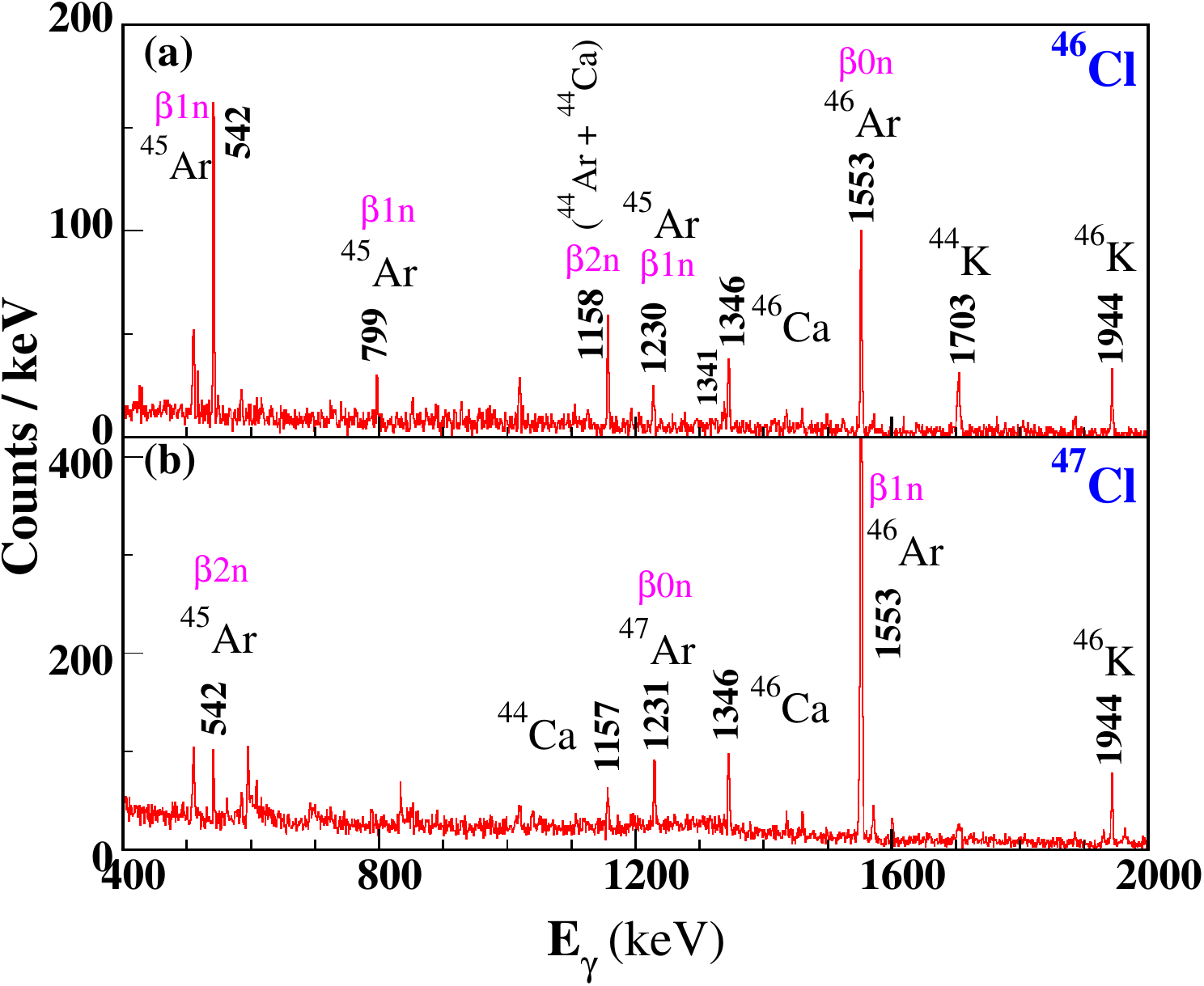}
	\caption{\label{fig:46Cl_47Cl_gamma}
	(a) $\beta$-delayed $\gamma$-ray spectrum for $^{46}$Cl using a time 
    correlation window of 250 ms. 
 (b) $\beta$-delayed $\gamma$-ray spectrum for $^{47}$Cl decay using a 100 ms time 
 correlation window. Intense transitions from $\beta0n$, $\beta1n$, 
 and $\beta2n$ are indicated for both cases. The y-scale in (b) was clipped to 
 highlight the weak transitions.
 }
\end{figure}

The calculated half-lives for $^{46}$Cl and $^{47}$Cl using the $V_{MU}$ interaction 
\cite{utsuno_new} for different assumptions for the spin-parity of the parent ground 
state are listed in Table~\ref{tab:t1/2}. 
The half-lives are calculated both with GT transitions alone and
with GT+FF transitions and experimental $Q_{\beta^-}$ values were used. 
For $^{46}$Cl, the calculated GT+FF values are closest to the 
experimental values of 223(37) ms \cite{Sorlin1993,sorlin_Pn}, 
and 232(2) ms \cite{grevy}, and 201(16) ms from this work if the ground state 
is either $1^-$ or $2^-$. 
For $^{47}$Cl, the calculated value of 109.54 ms assuming a $3/2^+$ ground state and 
using GT+FF agrees remarkably well with the measured half-life in the present work of 109.5(25) ms.
A $3/2^+$ ground state for $^{47}$Cl is also 
consistent with the systematics for the Cl isotopes \cite{tripathi_submitted}.

\subsection{\label{subsec:delayedgamma}$\beta$ delayed $\gamma$-ray transitions}

$\beta$-delayed $\gamma$-ray spectra correlated with $^{46,47}$Cl implants are shown in 
Figs.~\ref{fig:46Cl_47Cl_gamma}(a) and (b). 
Time correlation windows of 250 ms and 100 ms were used for 
$^{46}$Cl and $^{47}$Cl, respectively. These time intervals correspond to roughly 
one half-life for the two cases and highlight decays from the parent nucleus. 
For $^{46}$Cl decay (Fig.~\ref{fig:46Cl_47Cl_gamma}(a)), 
the spectrum is dominated by the 1553(2)-keV and 542(1)-keV transitions which are the 
ground-state decays of the first excited states in the $\beta0n$ daughter ($^{46}$Ar) and 
the $\beta1n$ daughter ($^{45}$Ar), respectively. For $^{47}$Cl 
decay (Fig.~\ref{fig:46Cl_47Cl_gamma}(b)), the spectrum is 
again dominated by the 1553(2)-keV transition in $^{46}$Ar ($\beta1n$ daughter), 
signaling that the $\beta$-delayed neutron emission channel is stronger than the 
$\beta^-$ decay to neutron-bound states. The 1231-keV transition is from the decay of the first excited state in $^{47}$Ar, the $\beta0n$ daughter.

%\subsubsection{$^{47}$Cl decay}

\begin{figure}
	\includegraphics[width=\columnwidth]{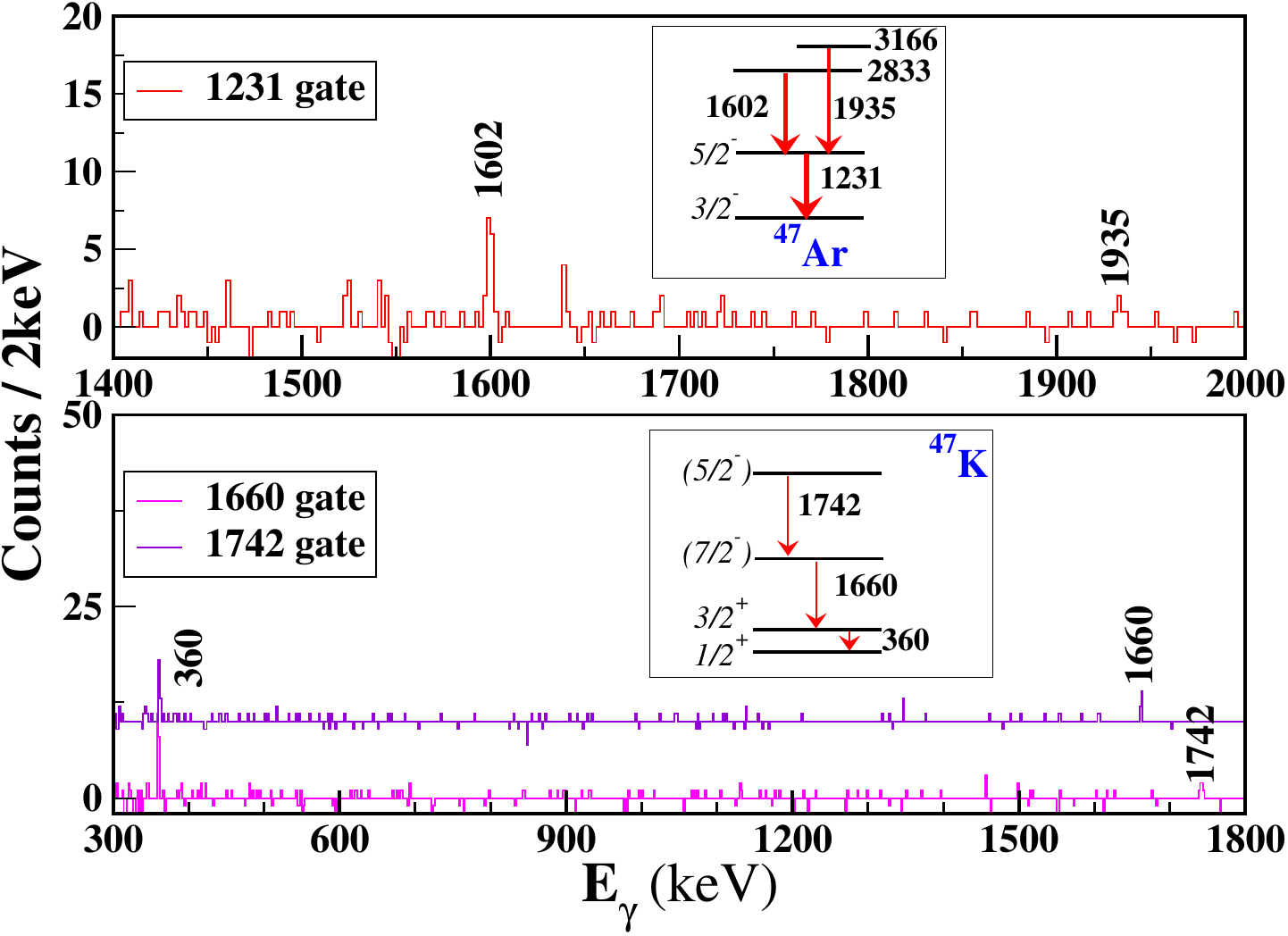}
	\caption{\label{fig:47Cl_coincidence}
	(a) $\gamma$-ray transitions in $^{47}$Ar observed in coincidence with the 
    1231-keV $5/2^-_1 \rightarrow 3/2^-_{gs}$ transition. 
    (b) The coincidences observed in the granddaughter $^{47}$K which 
 confirm that $^{47}$Ar was produced in the $\beta0n$ decay of $^{47}$Cl. The spectrum in 
 coincidence with the 1742-keV transition is offset from zero for clarity. 
	}
\end{figure}

$\gamma$-$\gamma$ coincidences observed in $^{47}$Ar following $^{47}$Cl $\beta^-$ 
decay are shown in Fig.~\ref{fig:47Cl_coincidence}(a). 
A gate on the known 1231(1)-keV transition ($5/2^-_1 \rightarrow 3/2^-_1$) in $^{47}$Ar 
highlights tentative coincidences with the 1602(2)-keV and 1935(2)-keV transitions.  
These transitions have not been reported previously, and correspond to
new levels at 2833(2) keV and 3166(2) keV, respectively. 
The ground state of $^{47}$Cl is conjectured to be $J^\pi$ = $3/2^+$ from the 
present half-life measurement (see Table~\ref{tab:t1/2}), 
and, thus, these states are candidates for positive-parity states populated by 
allowed Gamow-Teller (GT) transitions.  
Fig.~\ref{fig:47Cl_coincidence}(b) shows known $\gamma$-$\gamma$ 
coincidences in the granddaughter nucleus $^{47}$K \cite{Weissman2004} which 
follows the decay of $^{47}$Ar confirming its production in the relatively weak 
$\beta0n$ branch of the decay of $^{47}$Cl.

%\subsubsection{$^{46}$Cl decay}

% Figure 4

\begin{figure}
	\includegraphics[width=\columnwidth]{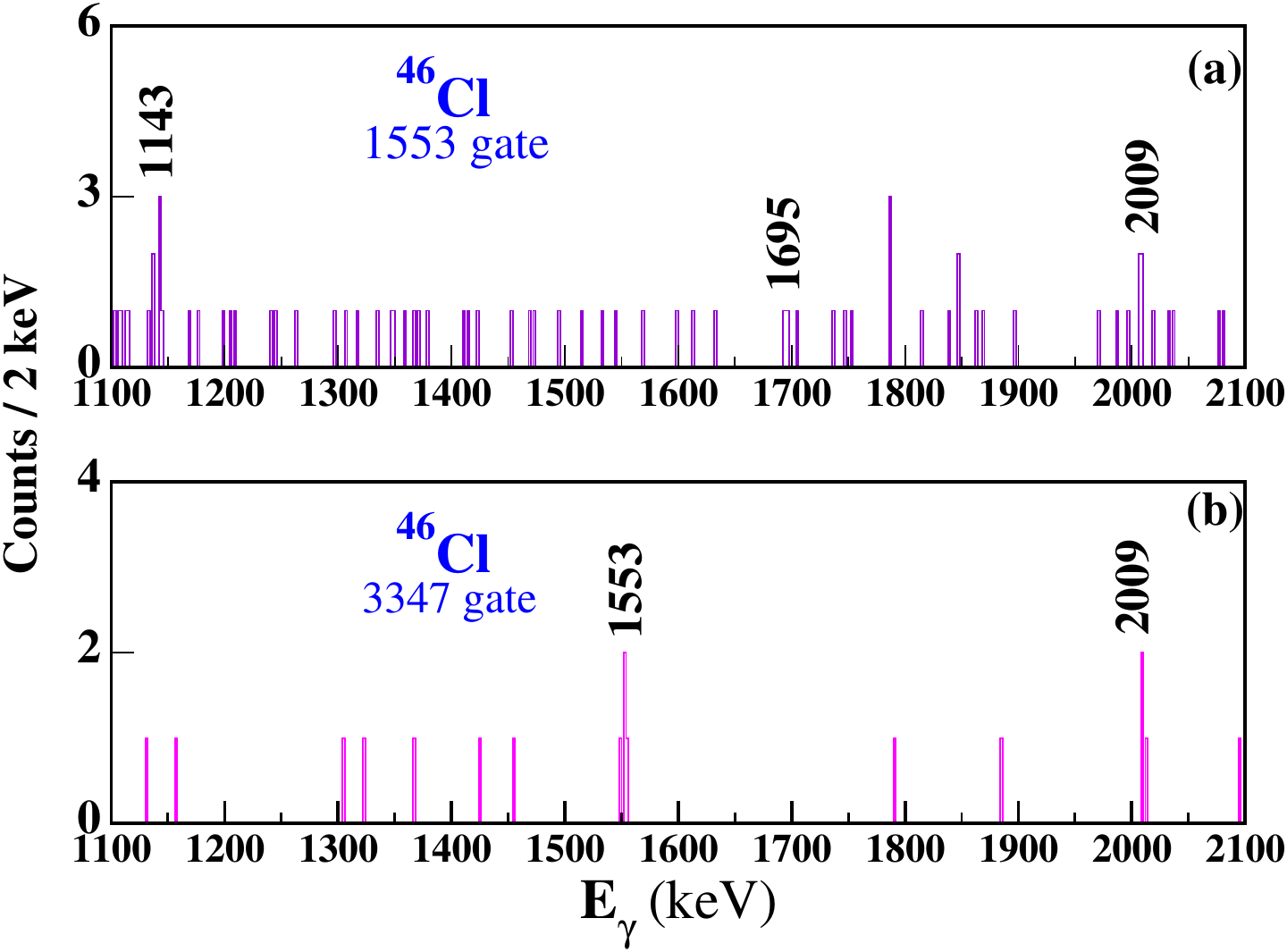}
	\caption{\label{fig:46Cl_1553coincidence}
	(a) Coincident $\gamma$-ray transitions with the 1553-keV transition in $^{46}$Ar as observed 
 in the $\beta0n$ decay of $^{46}$Cl. 
 (b) $\gamma$ rays coincident with the 3347-keV transition in $^{46}$Ar from $^{46}$Cl $\beta^-$ decay.
	}
\end{figure}

\begin{figure}
	\includegraphics[width=\columnwidth]{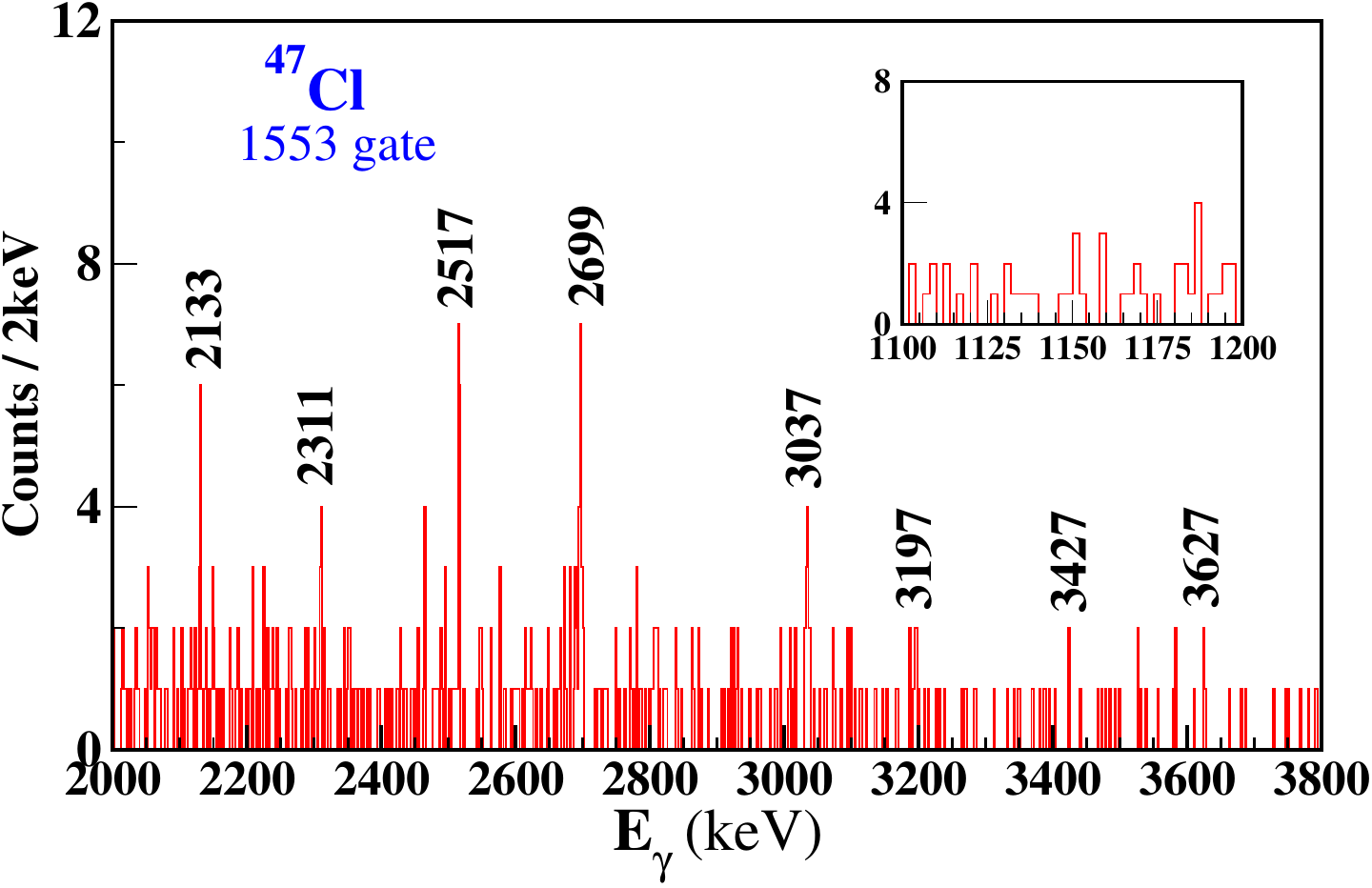}
	\caption{\label{fig:47Cl_1553coincidence}
	Coincident $\gamma$-ray transitions with the 1553-keV transition in $^{46}$Ar as observed 
 from the $\beta1n$ decay of $^{47}$Cl. The inset shows that the 1143 keV transition is 
 not seen here in contrast to Fig.~\ref{fig:46Cl_1553coincidence}(a).
	}
\end{figure}

The $2^+$ $\rightarrow$ $0^+$ transition in $^{46}$Ar at 1553 keV 
is the most intense transition observed in both the $^{46}$Cl and $^{47}$Cl decays 
(Figs.~\ref{fig:46Cl_47Cl_gamma}(a) and (b)). The coincidences observed with the 1553-keV transition 
in the two cases are shown in Figs.~\ref{fig:46Cl_1553coincidence} and \ref{fig:47Cl_1553coincidence}.
From $^{46}$Cl $\beta^-$ decay, a 1143(3)-keV transition was observed 
in coincidence with the 1553(2)-keV transition, as seen in Fig.~\ref{fig:46Cl_1553coincidence}(a).
This could correspond to the 1140(20)-keV peak from Ref.~\cite{Dombradi2003}. 
$^{46}$Cl $\beta^-$ decay was also reported in a conference proceedings 
where 2006-keV and 3350-keV $\gamma$ rays were observed in
 coincidence with each other and with the 1553-keV transition \cite{Mrazek2004}. 
 We were able to confirm these coincidence relationships 
 (Fig.~\ref{fig:46Cl_1553coincidence}(b)) and further suggest that the 2009(3)-keV 
 transition feeds into the 3347(3)-keV transition, although the reverse cannot be ruled 
 out, as both transitions are equally intense within the estimated uncertainties. The absolute intensities of the 1553, 1141, 1695, 2009, and 3347-keV were estimated to be 70(13)\%, 2.6(1)\%, 3.6(1.5)\%, 7(2)\%, and 7.5(2.5)\% respectively.
 The statistics for $^{46}$Cl $\beta0n$ decay were rather low in this experiment and no further 
 $\gamma$-ray transitions could be identified. 
 The absolute $\beta$ feeding to the observed levels in 
 $^{46}$Ar from $^{46}$Cl $\beta^-$ decay  and the corresponding log$ft$ values 
 are given in Table~\ref{tab:46Cl_betaintensity}.

 % Figure 5
 % Figure 6

% Table 2

\begin{table}
\caption{\label{tab:46Cl_betaintensity} Absolute $\beta$ feeding intensity $I$
(per 100 decays of $^{46}$Cl)
to levels in $^{46}$Ar from $^{46}$Cl $\beta0n$ decay. 
The number of $^{46}$Cl implants were extracted from the fit to the decay curve. 
The log$ft$ values were calculated using the measured $t_{1/2}$ of 201(16) ms
and a $Q_{\beta^-}$ of 16040(100) keV \cite{AME2020} with the LOGFT CALCULATOR from NNDC \cite{nndc}. For the FF transitions 
both possibilities (unique and non-unique) are listed. For the decay branch to the 1553-keV level, first-forbidden unique (1FU) is not possible considering 
the likely spin of the $^{46}$Cl ground state of $1^-$ or $2^-$. The first-forbidden 
non-unique (1FNU) log$ft$ values have been calculated as allowed $\beta$ transitions as recommended in Ref.~\cite{forbidden_TURKAT}.
}
  \begin{tabular}{ccccc}
    \toprule
    $E_{level}$  & {$I$}({$\Delta$I}) & & {log$_{10}ft$} &   \\
     keV & \% & GT & 1FU & 1FNU \\
      \midrule
     1553(2) & 56(28)\footnote{should be considered an upper limit due to 
     possible unobserved $\gamma$-ray transitions}  & - & - & 5.59($^{+34}_{-21}$) \\
     2696(2) & 2.6(7) & - & 9.35($^{+17}_{-14}$) & 6.75($^{+17}_{-14}$) \\
     3248(3) & 3.7(10) & - & 9.10($^{+17}_{-14}$) &  6.51($^{+17}_{-14}$) \\
     4900(3) & $\approx$~0 & - & - & - \\
     6909(4) & 7.5(25) & 5.51($^{+21}_{-16}$) & - & -\\
    \bottomrule
  \end{tabular}
\end{table}

From $^{47}$Cl $\beta1n$ decay, several transitions were observed in 
coincidence with the 1553-keV peak in $^{46}$Ar, as seen in 
Fig.~\ref{fig:47Cl_1553coincidence}. 
Due to low statistics, some of these transitions were difficult to observe in an ungated spectrum.
The 2133(2), 2311(2), 2517(2), and 2699(2)-keV transitions in the present work may 
correspond to the 2141(3), 2318(3), 2518(2), 2707(2)-keV transitions cited
in Ref.~\cite{Nowak2016}, although they are somewhat lower in energy. 
The 2307(13) and 2692(16)-keV $\gamma$ rays reported in the 
inverse kinematics proton scattering of $^{46}$Ar in Ref.~\cite{Riley2005} are similar 
in energy to the 2311 and 2699-keV transitions seen 
here. The 3037(2), 3197(3), 3427(3), and 3627(3)-keV transitions are 
reported for the first time in this work. For these transitions with low intensities, effort was made to check them in various time correlation windows and gates (when possible) to avoid possible 
misidentification, though some ambiguity cannot be ruled out.
In Ref.~\cite{Riley2005} a 3430(26)-keV $\gamma$ ray was observed in 
coincidence with the $2^+_1 \rightarrow 0^+_1$ transition and could correspond to 
the 3427-keV transition seen here. However, other decays  
observed in Ref.~\cite{Riley2005} from that state were not observed in this work, 
leading us to suggest that they could be distinct.
The relative intensities of the $\gamma$ rays observed in $^{46}$Ar 
which were possible to observe in a singles spectrum are provided in 
Table~\ref{tab:intensity46Ar} for both $^{46}$Cl and $^{47}$Cl decay.

%Figure 7
\begin{figure}
	\includegraphics[width=\columnwidth]{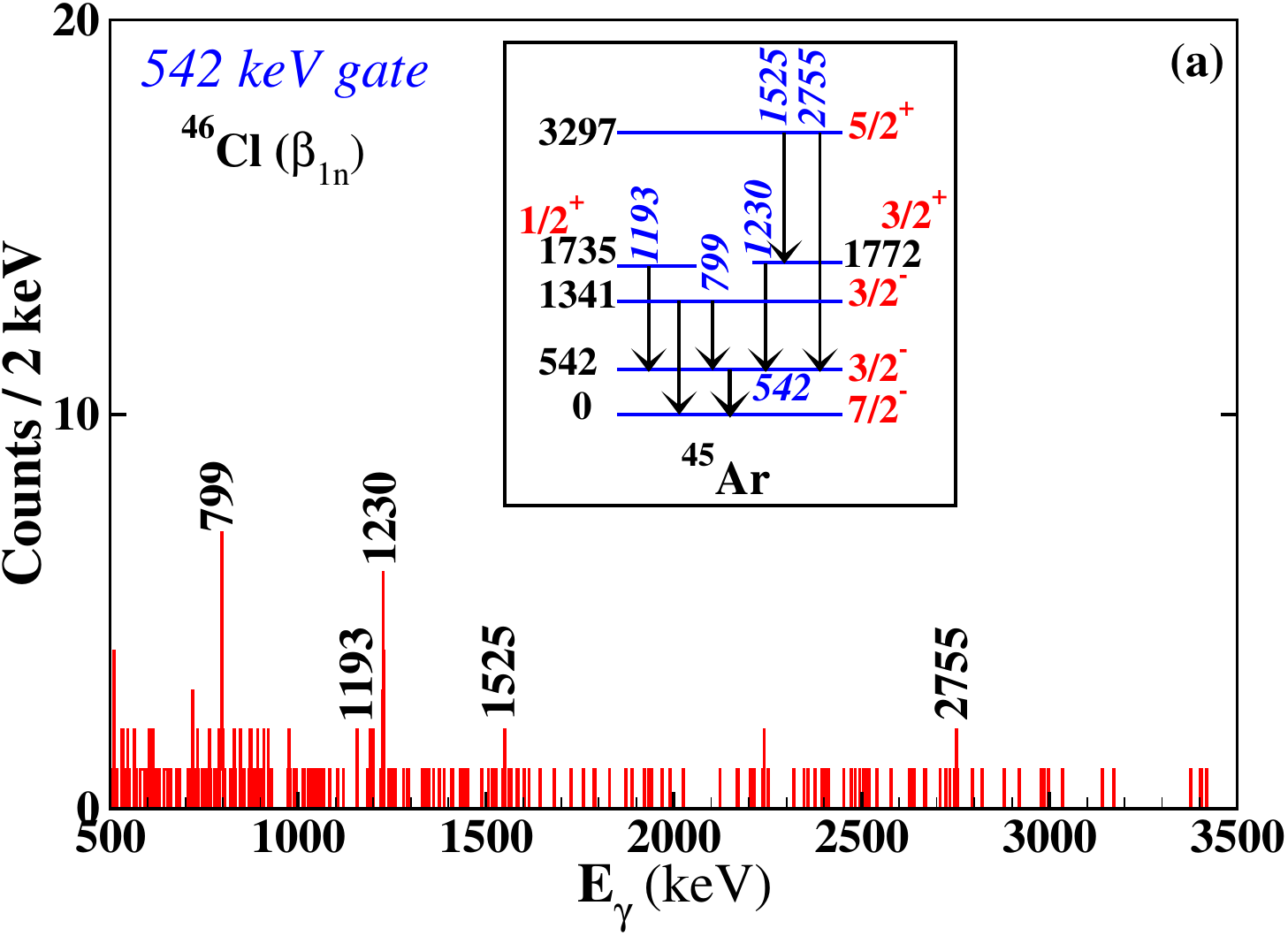}
	\includegraphics[width=\columnwidth]{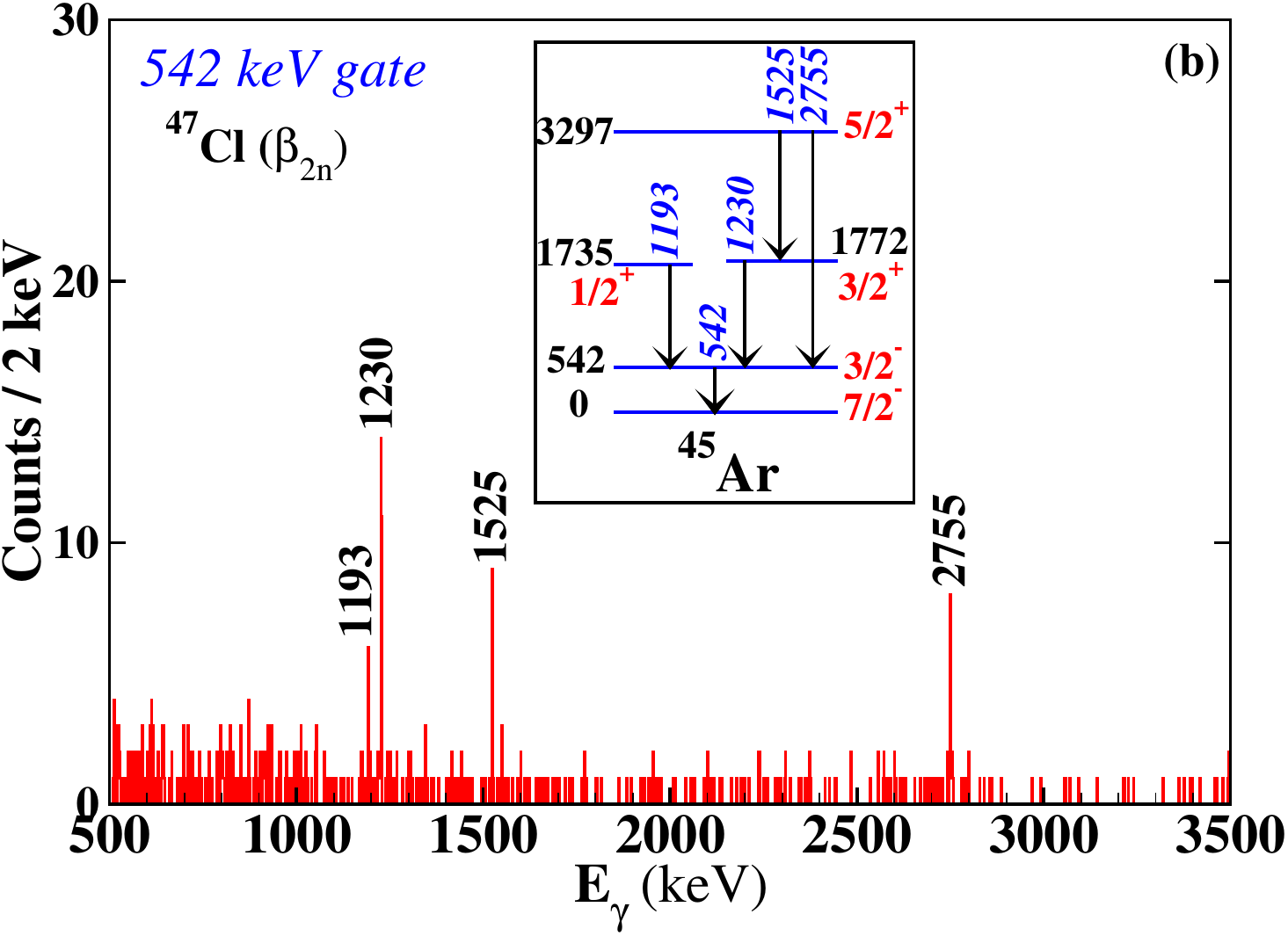}
	\caption{\label{fig:46_47Cl_beta-xn}
		%($^{46}$Cl: likely $J^\pi$ =$1^-,2^-$) 
 (a) $\gamma$-ray transitions observed in coincidence with the 
  542-keV transition in $^{45}$Ar from the $\beta1n$ channel of $^{46}$Cl decay. 
  %($^{47}$Cl: likely $J^\pi$ = $3/2^+$) 
 (b) $\gamma$-ray transitions observed in coincidence with the 
  542-keV transition in $^{45}$Ar from the $\beta2n$ decay of $^{47}$Cl. 
   Two-neutron emission from the unbound states in $^{47}$Ar feeds only positive-parity 
   states in $^{45}$Ar while a negative-parity state is strongly populated in $^{46}$Cl 
   $\beta1n$ decay. The $^{46}$Cl ground state likely has $J^\pi$ =
   $1^-$ or $2^-$, see section \ref{subsec:thalf} while for $^{47}$Cl, $J^\pi$ = $3/2^+$ 
   is favored.
	}
\end{figure}

%Table 3

\begin{table}
\caption{\label{tab:intensity46Ar} Relative intensities of $\gamma$ rays observed in 
$^{46}$Ar from both $^{46}$Cl ($\beta0n$) and $^{47}$Cl ($\beta1n$) decay. The intensity of the $\gamma$ transitions above 3.2 MeV is about 4 times smaller than the 2699 keV transition and could not be ascertained.}
  \begin{tabular}{cc|cc}
    \toprule
    \multicolumn{2}{c}{$^{46}$Cl $\beta0n$ decay} &
   \multicolumn{2}{c}{$^{46}$Cl $\beta1n$ decay} \\
    {$E_\gamma$ (keV)} &~~~~ {$I_\gamma $}~~~~ & ~~{$E_\gamma$ (keV)} ~~& ~~{$I_\gamma $}  \\
      \midrule
     1553(2) & 100(15) & 1553(2) & 100(12) \\
    1143(3)  & 3.7(10) & 2133(2) & 3.0(5)  \\   
     1695(3) & 5.2(15) & 2311(2) & 2.5(5)  \\
     2009(3) & 10(2)   & 2517(2) & 5.3(10) \\
     3347(3) & 11(3)   & 2699(2) & 10(2)   \\
           - & -       & 3037(2) & 6.5(15) \\
           - & -       & 3197(3) & 7(3) \\
     \bottomrule
       \end{tabular}
\end{table}

%Table 4 

\begin{table}
\caption{\label{tab:intensity45Ar} Relative intensities of $\gamma$ rays observed in 
$^{45}$Ar from both $^{46}$Cl ($\beta1n$) and $^{47}$Cl ($\beta2n$) decay. }
  \begin{tabular}{cc|cc}
  \toprule
    \multicolumn{2}{c}{$^{46}$Cl $\beta1n$ decay} &
   \multicolumn{2}{c}{$^{47}$Cl $\beta2n$ decay} \\
    {$E_\gamma$ (keV)} &~~~~ {$I_\gamma $}~~~~ & ~~{$E_\gamma$ (keV)} ~~& ~~{$I_\gamma $}  \\
      \midrule
     542(1)  & 100(12) &  542(1)   & 100(12)  \\
     799(1)  & 30(4)   &  1193(1)  & 14(3) \\   
     1193(1) & 13(2)   &  1230(1)  & \footnote{intensity could not be extracted due to mixing with the 1231 keV transition from $^{47}$Ar.} \\
     1230(1) & 35(5)   &  1525(1)  & 18(3) \\
     1341(1) & 10(3)   &  2755(2)  & 40(6) \\
     1525(1) & 20(4)   &  - & - \\
     2755(2) & 6.5(15) &  - & - \\
    \bottomrule
  \end{tabular}
\end{table}

\begin{comment}
%Table3

\begin{table}
\caption{\label{tab:intensity46Ar} Relative intensities of $\gamma$ rays observed 
in coincidence with the 
1553-keV transition in $^{46}$Ar for both $^{47}$Cl and $^{46}$Cl decay.}
  \begin{tabular}{cc|cc}
    \toprule
    \multicolumn{2}{c}{$^{47}$Cl $\beta1n$ decay} &
   \multicolumn{2}{c}{$^{46}$Cl $\beta0n$ decay} \\
    {$E_\gamma$ (keV)} &~~~~ {$I_\gamma $}~~~~ & ~~{$E_\gamma$ (keV)} ~~& ~~{$I_\gamma $}  \\
      \midrule
     2133(2) & 52(16) & 1143(3) & 38(19) \\
     2311(2) & 40(14) & 1695(3) & 39(23) \\
     2517(2) & 99(23) & 2009(3) & 61(30) \\
     2699(2) & 100 & 3347(3) & 100 \\
     3037(2) & 86(23) & & \\
     3197(3) & 35(16) & & \\
     3427(3) & 30(15) & & \\
     3627(3) & 24(14) & & \\
     \bottomrule
       \end{tabular}
\end{table}

%Table 4 

\begin{table}
\caption{\label{tab:intensity45Ar} Relative intensities of $\gamma$ rays 
observed in coincidence with the 542-keV transition in $^{45}$Ar from both $^{47}$Cl 
and $^{46}$Cl decay.}
  \begin{tabular}{cc|cc}
  \toprule
    \multicolumn{2}{c}{$^{47}$Cl $\beta2n$ decay} &
   \multicolumn{2}{c}{$^{46}$Cl $\beta1n$ decay} \\
    {$E_\gamma$ (keV)} &~~~~ {$I_\gamma $}~~~~ & ~~{$E_\gamma$ (keV)} ~~& ~~{$I_\gamma $}  \\
      \midrule
     1193(1) & 24(9) & 1193(1) & 22(13) \\
     1230(1) & 91(18) & 1230(1) & 100 \\
     1525(1) & 63(16) & 1525(1) & 27(16) \\
     2755(2) & 100 & 2755(2) & 63(31) \\
             &     & 799(1) & 68(20) \\
    \bottomrule
  \end{tabular}
\end{table}
\end{comment}

The $Q_{\beta^-}$ value for $^{47}$Cl is 15790(200) keV, while $Q_{\beta2n}$ is 4.0(3) MeV \cite{AME2020}. 
This allowed us to observe $\gamma$-ray transitions in the 
$\beta2n$ daughter, $^{45}$Ar, which were also independently populated via the 
$^{46}$Cl $\beta1n$ channel ($Q_{\beta1n}$ = 7.96(10) MeV \cite{AME2020}). For both cases, 
clean coincidences with the 542(1)-keV $\gamma$ ray 
(de-excitation of the first excited state to the ground state in $^{45}$Ar) were observed.
The $^{46}$Cl $\beta1n$ and $^{47}$Cl $\beta2n$ delayed $\gamma$-ray spectra 
gated on the 542(1)-keV transition are compared in Figs.~\ref{fig:46_47Cl_beta-xn}(a) and  (b),
and the corresponding partial level schemes are also displayed. 
For $^{45}$Ar from $^{47}$Cl $\beta2n$ decay, only decays from positive-parity 
states above the 542-keV level were observed~\cite{Soumik2023,Mrazek2004}. 
For $^{45}$Ar populated from $^{46}$Cl $\beta1n$ decay, 
the $\gamma$ rays from the same set of positive-parity states are seen 
along with a strong 799(1)-keV transition. Based on the $^{45}$Cl $\beta0n$ decay scheme
reported in Refs.~\cite{Soumik2023,Mrazek2004},
this $\gamma$ ray is assigned to the de-excitation of the $3/2^-$ state at 1341(1) keV. The direct decay from the 1341-keV state is 
also observed as seen in Fig.~\ref{fig:46Cl_47Cl_gamma}(a).
The relative intensities of the $\gamma$-ray transitions observed in 
$^{45}$Ar are 
provided in Table~\ref{tab:intensity45Ar} for both $^{46}$Cl and $^{47}$Cl decay.

\section{\label{sec:discussion}Discussion}
\subsection{\label{subsec:discus46Ar}$^{46}$Ar}

%Figure 8

\begin{figure*}
	\includegraphics[width=16cm]{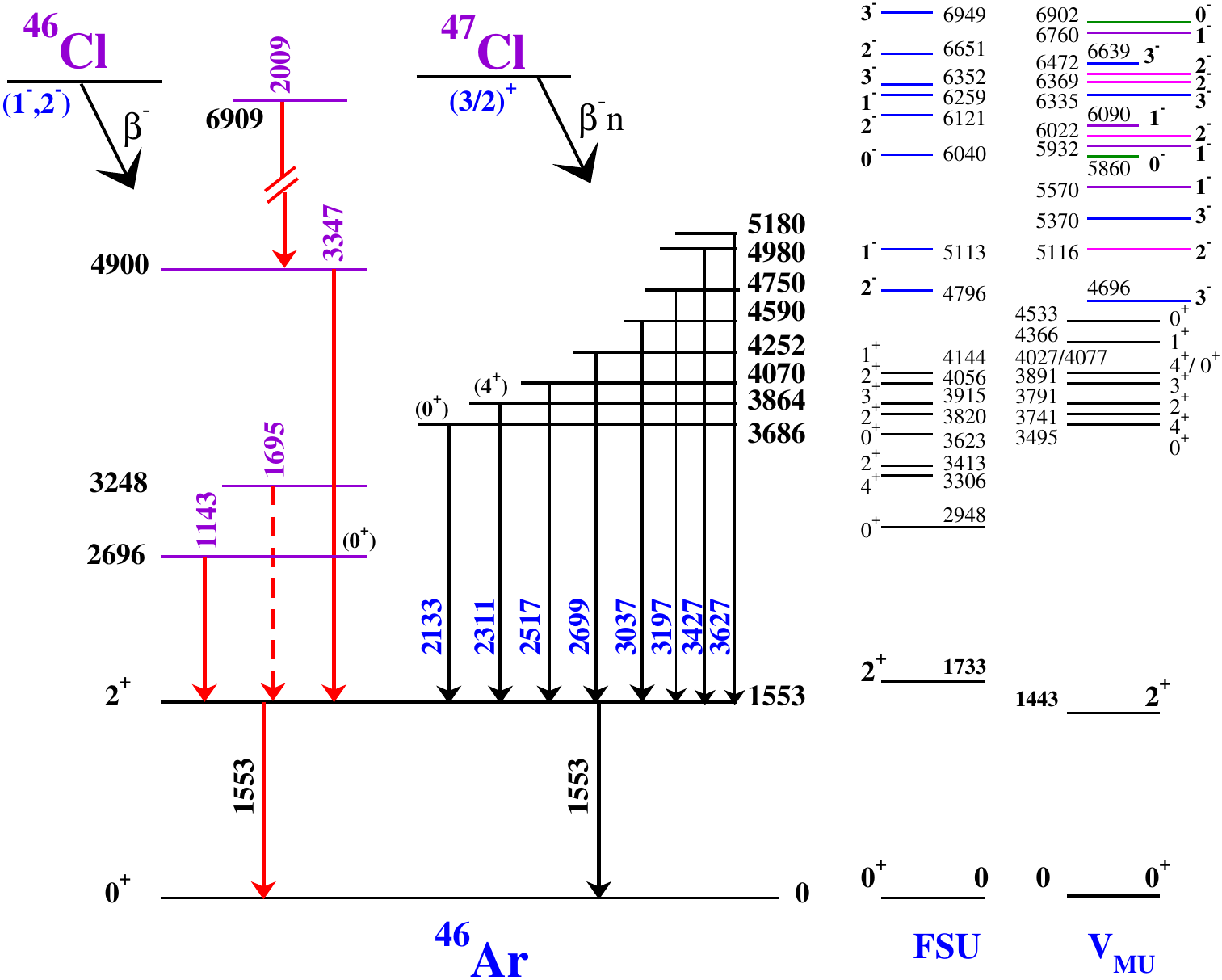}
	\caption{\label{fig:46Ar_levelscheme}
	A Partial level scheme of $^{46}$Ar from this work following the $\beta0n$-decay of 
    $^{46}$Cl is shown on the left.
 The level scheme deduced from $\beta1n$ decay of $^{47}$Cl is shown on the right. 
 Alongside are predictions from
 shell model calculations using the FSU \cite{FSU} and $V_{MU}$ interactions \cite{utsuno_new}.
 The positive-parity states are $0p0h$ states while the negative-parity 
 states are $1p1h$ states with excitations across the $N=20$ shell gap. 
 The first ten positive-parity states are shown for both, while negative parity states 
 ($0^-$, $1^-$, $2^-$, and $3^-$ only) are shown up to 7.0 MeV.
 The neutron separation energy $S_n$ is 8071.8(24) MeV for $^{46}$Ar \cite{AME2020}.
 }
\end{figure*}

%Table 5

\begin{table*}
\caption{\label{tab:logft46Ar_Vmu} Calculated log{\it ft} values 
for GT and FF transitions using the $V_{MU}$ interaction for selected levels in $^{46}$Ar.  
Different spin-parity possibilities are considered for the parent $^{46}$Cl. 
Only levels below 7 MeV excitation energy and with a log{\it ft} value 
lower than 6.0 are listed for GT transitions.}
  \begin{tabular}{ccc|ccc|ccc|ccc}
    \toprule
    \multicolumn{3}{c}{GT:$^{46}$Cl ($0^-$)} &
    \multicolumn{3}{c}{GT:$^{46}$Cl ($1^-$)} &
   \multicolumn{3}{c}{GT:$^{46}$Cl ($2^-$)} &
   \multicolumn{3}{c}{GT:$^{46}$Cl ($3^-$)} \\
    {$J^\pi$}~& ~{$E_x$ (keV)} ~&~ {$log_{10}ft$} ~&~{$J^\pi$} ~& ~{$E_x$ (keV)} ~& ~{$log_{10}ft$~} ~&~ {$J^\pi$}~ &~{$E_x$ (keV)}~ & ~{$log_{10}ft$}~ &~ {$J^\pi$}~& ~{$E_x$ (keV)} ~&~ {$log_{10}ft$}  \\
      \midrule
     $1^-$ & 5570 & 5.37 & $0^-$ & 5860 & 5.94 & $1^-$ & 5570 & 5.64 & $3^-$ & 5370 & 5.70\\
     $1^-$ & 6759 & 5.39 & $0^-$ & 6902 & 5.63 & $1^-$ & 5932 & 5.98 & - & -  & - \\
     -& -&- & $1^-$ & 5570 & 5.27 & $2^-$ & 5116 & 5.86 & -& -&- \\
     -& -& -& $1^-$ & 5932 & 5.81 & $2^-$ & 6022 & 5.60 & -& -&- \\
     -& -& -& $2^-$ & 6369 & 5.61 & $2^-$ & 6369 & 5.48 & -& -&- \\
     \midrule
     - &- & - & $2^+_1$(FF) & 1443 & 7.71 & $2^+_1$(FF) & 1443 & 6.82 & $2^+_1$(FF) & 1443 & 7.32\\
     $0^+_2$(FF) & 3496 & 6.89 & $0^+_2$(FF) & 3496 & 8.62 & - &-  & - & - & - & -\\
     \bottomrule
       \end{tabular}
\end{table*}

The level scheme of $^{46}$Ar, which is the $\beta0n$ daughter of $^{46}$Cl and the $\beta1n$ 
daughter of $^{47}$Cl was established using $\gamma$-$\gamma$ coincidences 
(Fig.~\ref{fig:46Cl_1553coincidence} and Fig.~\ref{fig:47Cl_1553coincidence}) and energy and intensity balance. 
The partial level schemes with states in $^{46}$Ar populated in 
$^{46}$Cl $\beta0n$ decay (left) and $^{47}$Cl $\beta1n$ decay (right) are shown 
in Fig.~\ref{fig:46Ar_levelscheme}. 
Interestingly, though not a complete surprise, the levels beyond the first $2^+$ observed from 
$^{46}$Cl $\beta0n$ decay and $^{47}$Cl $\beta1n$ decay are completely different. 
The low spin states predicted from shell model calculations using the FSU interaction \cite{FSU} 
and the $V_{MU}$ interaction \cite{utsuno_new} are displayed alongside for comparison. 

Both interactions predict $1p1h$ negative-parity levels in $^{46}$Ar occurring above an 
excitation energy of 4.5 MeV. The level at 6909(4) keV which is directly fed in the 
$\beta^-$ decay of $^{46}$Cl is a very likely candidate for a negative-parity state. 
The $^{46}$Cl ground state likely has $J^\pi$ = $1^-$ or $2^-$, as was discussed in section 
\ref{sec:results} A. The intensity observed (Table~\ref{tab:46Cl_betaintensity}) 
for this level corresponds to a log$ft$ value of 5.51 which is consistent with an allowed GT transition. 
Log$ft$ values from shell model calculations using the $V_{MU}$ interaction 
are listed in Table~\ref{tab:logft46Ar_Vmu} assuming different $J^\pi$ values for the $^{46}$Cl 
ground state. The decay from a $1^-$ ground state to a $0^-$ level at 6902 keV has a 
log$ft$ value of 5.63 and to a $2^-$ state at 6369 keV has a 
log$ft$ value of 5.61, making both possible candidates. The $2^-$ state at 
6369 keV can also be populated from the decay of a $2^-$ ground state in $^{46}$Cl and 
is predicted to have a log$ft$ of 5.48. Since the 4900-keV level is fed by this 
probable negative-parity state and decays to the $2^+_1$ level at 1553 keV, it will likely 
have a spin of 1 or 2 with either parity possible.

In Ref.~\cite{Dombradi2003} the $2^+_1$ energy was reported as 1570(5) keV, while the energy 
of the $2^+_1$ is quoted as 1558(9) keV in Ref.~\cite{Riley2005}, 1554(1) keV in Ref.~\cite{Nowak2016}, 
and 1553(2) keV in this work. The $4^+_1$ state is tentatively placed at 3892(9) keV in 
Ref.~\cite{Dombradi2003}, at 3866(16) keV in Ref.~\cite{Riley2005}, and at 
3872(3) keV in Ref. \cite{Nowak2016}. The transition analogous in energy in the current 
work (2311 keV) would place the $4^+_1$ state at 3864 
keV. The shell model calculations place the $4^+_1$ at 3306 keV (FSU) 
or 3741 keV ($V_{MU}$) in reasonable agreement. 

The first excited $0^+$ level was 
tentatively placed at 2710(21) keV through comparison with level 
energies from shell model calculations in Ref.~\cite{Dombradi2003}. On the other 
hand, based on angular distribution analysis in the study of  
$^{44}$Ar$(t,p)$~\cite{Nowak2016}, the level at 3695(3) keV was proposed 
as the $0^+_2$ state instead, while the state from Ref.~\cite{Dombradi2003} 
was not seen. In the current study, levels with similar energies to both the prior candidates,
2696 keV and 3686 keV, are observed in $^{46}$Cl $\beta0n$ decay and $^{47}$Cl $\beta1n$ decay, 
respectively. Shell model calculations generate several yrare $0^+$ states with both 
interactions agreeing on $0^+$ being the third excited state, although 
differing in energy. The $0^+_2$ (2948 keV) and $0^+_3$ (3623 keV) predictions from the 
FSU interaction agree well with the two experimental states in 
question. In our previous paper on $\beta^-$ of $^{45}$ Cl~\cite{Soumik2023}, we had  
identified the $0^+_2$ in $^{44}$Ar produced again in the $\beta1n$ channel at 2978 keV, 
while the FSU interaction predicted the state at 2717 keV, a difference of 261 keV. 
Similarly assigning the 2696 keV as $0^+_2$ in $^{46}$Ar compared to the calculated 
2948 keV would be an energy difference of 252 keV making it a very plausible scenario.
Of course, more experimentation is needed to definitively clarify the 
position of the excited $0^+$ states, which will impact our understanding of 
shell structure in this region and can be correlated to nuclear size and 
deformation.  

Based on the experimental information we have, it is not feasible
to comment on the spin or parity of other states populated in the $^{47}$Cl $\beta1n$ decay. 
With decays to the $2^+_1$ state, their spins can be constrained to be within 0 and
4 if they have positive parity and likely within 1 and 3 if they have negative parity. 
Based on relative intensities (see Table~\ref{tab:intensity46Ar}), it appears that 
the levels at 4070 keV, 4252 keV, and 4590 keV are favored over the others in 
the neutron emission process, 
though a definitive determination of the spin-parity cannot 
be made. In Ref.~\cite{Nowak2016}, these states were populated in a 
$(t,p)$ reaction on $^{44}$Ar and were thought to have positive parity, 
although no spin assignments were made. A positive parity is consistent
with their population in the $\beta1n$ process, as will be discussed in 
Section~\ref{subsec:delayedneutron}.

\subsection{\label{subsec:discus47Ar}$^{47}$Ar}

\begin{figure*}
	\includegraphics[width=14cm]{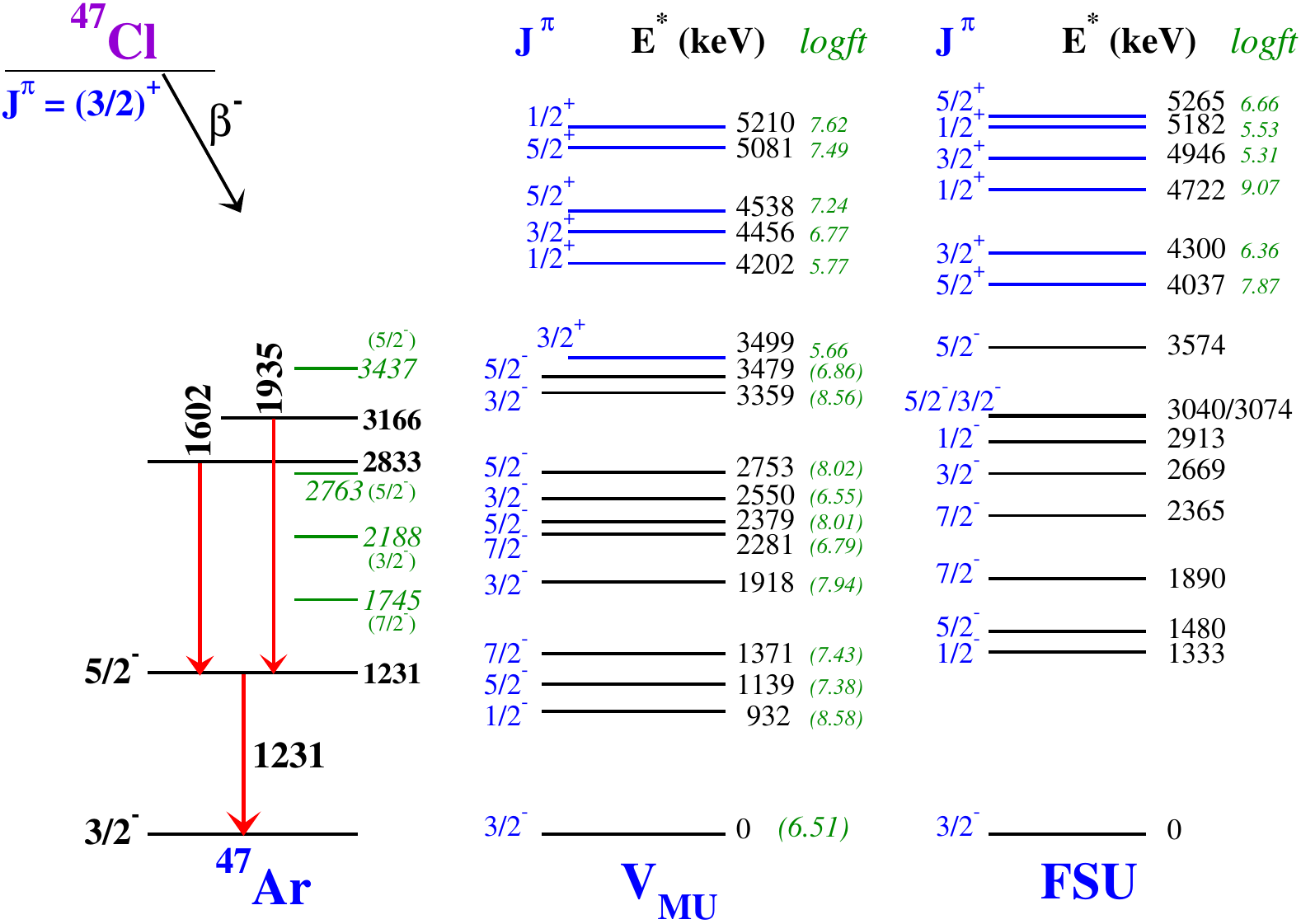}
	\caption{\label{fig:47Ar_levelscheme}
    A Partial level scheme of $^{47}$Ar from this work following $\beta0n$-decay of 
    $^{47}$Cl. Alongside are predictions from
 shell model calculations using the $V_{MU}$ interaction \cite{utsuno_new} and 
 the FSU interaction \cite{FSU}.
 The negative-parity states are $0p0h$ states while the positive-parity 
 states are $1p1h$ states with excitations across the $N=20$ shell gap. 
 The first 10 negative-parity excited states are shown. For the positive-parity levels, 
 two states of 
 each of the spins $1/2^+$, $3/2^+$, and $5/2^+$ from the calculations are shown.
 The log$ft$ values for the positive-parity states correspond to allowed GT transitions, 
 while for negative-parity states they correspond to FF transitions.
 The neutron separation energy $S_n$ is 3667(3) keV for $^{47}$Ar \cite{AME2020}. 
 The experimental states shown in green and indicated by italics are from 
 Ref.~\cite{Gade2016} for comparison with the calculations.
	}
\end{figure*}

The $\gamma$-$\gamma$ coincidences observed in $^{47}$Ar following $^{47}$Cl $\beta0n$ decay were
shown in Fig.~\ref{fig:47Cl_coincidence}(a). Based on these coincidences, a
very tentative level scheme for $^{47}$Ar is proposed and shown in
Fig.~\ref{fig:47Ar_levelscheme}. Selected states predicted by shell-model 
calculations using the FSU interaction \cite{FSU} and
the $V_{MU}$ interaction \cite{utsuno_new} are displayed alongside for comparison. 
The neutron separation energy in $^{47}$Ar is 3667(3) keV and, therefore, Gamow-Teller 
decays
from the likely $3/2^+$ ground state of 
$^{47}$Cl would feed levels predicted to be close to or above the neutron separation energy. 
For example, the log$ft$ value to the shell-model $3/2^+_1$ state at 3499 keV ($V_{MU}$)
is 5.66 and the log$ft$ value to the $5/2^+$ state at 4202 keV is 5.77. 
These predicted positive-parity states are higher in energy than the 
experimentally observed states. Alternatively, the observed levels may have been 
populated by First Forbidden (FF) decay. 
The log$ft$ values to the shell-model states with $J^\pi$ of 
$7/2^-_2$ (2281 keV), $3/2^-_3$ (2550 keV), $5/2^-_4$ (3479 keV)
calculated using $V_{MU}$ are 6.79, 6.55, and 6.86 respectively, making all of 
these predicted states possible candidates for the observed levels. 

A $7/2^-$ state was reported at 1745 keV in a prior in-beam $\gamma$-ray spectroscopy 
experiment following one-proton knockout \cite{Gade2016}. The same experiment 
proposed a
$3/2^-_2$ state at 2188 keV and $5/2^-$ states at 1231 keV, 2761 keV, and 3438 keV 
\cite{Gade2016}. These states are indicated 
by green lines in Fig.~\ref{fig:47Ar_levelscheme} and as can be 
seen they align closely in energy to levels with the same $J^{\pi}$ predicted by $V_{MU}$ 
and FSU interactions. The previously-observed $7/2^-$ level at 1745 keV nicely 
corroborates with the shell-model $7/2^-_1$ state at 1371 keV from the $V_{MU}$ interaction 
and 1890 keV state as predicted using the FSU interaction. The $V_{MU}$ interaction
provides candidate for the 2188 keV and  2763 keV states
in the $3/2^-_2$ and $5/2^-_2$ states.
This would leave open the possibility of the present 2833-keV and 3166-keV levels tentatively 
being the $7/2^-_2$ (2281 KeV) and $3/2^-_3$ (2550 keV) states though 
somewhat lower in energy. The 
equivalent states calculated using the FSU interaction are at 
2365 keV ($7/2^-_2$) and 3074 keV ($3/2^-_3$).

% Figure 9
%level scheme 47Ar

\subsection{\label{subsec:delayedneutron}$\beta$-delayed neutron emission}

In prior studies, the reported experimental probabilities for $\beta$-delayed neutron 
emission showed dramatic changes as a function of neutron number: 24(4)\% for $^{45}$Cl, 
60(9)\% for $^{46}$Cl, and then, very surprisingly, less than 3\% for $^{47}$Cl 
\cite{Sorlin1993}. From the present work, following the activity in the 
granddaughter nuclei from the three branches, $\beta0n$, $\beta1n$, and $\beta2n$, 
we have estimated the $P_n$ values to be 75(15)\% for $^{46}$Cl and 98(12)\% for 
$^{47}$Cl. This smooth and increasing trend is what one would expect as neutron number 
increases. The 
measured $P_n$ values are compared with those from shell model calculations ($V_{MU}$)
with and without the inclusion of FF transitions in Table~\ref{tab:beta-Pn}. Although the $V_{MU}$ calculations reproduce the trend of increasing $P_n$ with neutron number, it is interesting to note that the $P_n$ values are underestimated in both cases. 
For $^{47}$Cl, the $P_n$ value becomes 
much smaller than the experimental value with the inclusion of FF transitions, 
suggesting that the FF contribution is overestimated in the calculation. 
Revisiting Fig.~\ref{fig:46Cl_47Cl_gamma}(b) emphasizes that FF 
transitions are weak for $^{47}$Cl decay, as the spectrum is dominated by 
$\gamma$ rays from 
the $\beta1n$ daughter. In contrast, the measured $T_{1/2}$ values were 
better reproduced when FF transitions were included. 
This conflict needs to be explored 
further with higher-precision measurements of the neutron emission probabilities and 
the investigation of even more neutron-rich nuclei.

% Figure 10
\begin{figure}
	\includegraphics[width=\columnwidth]{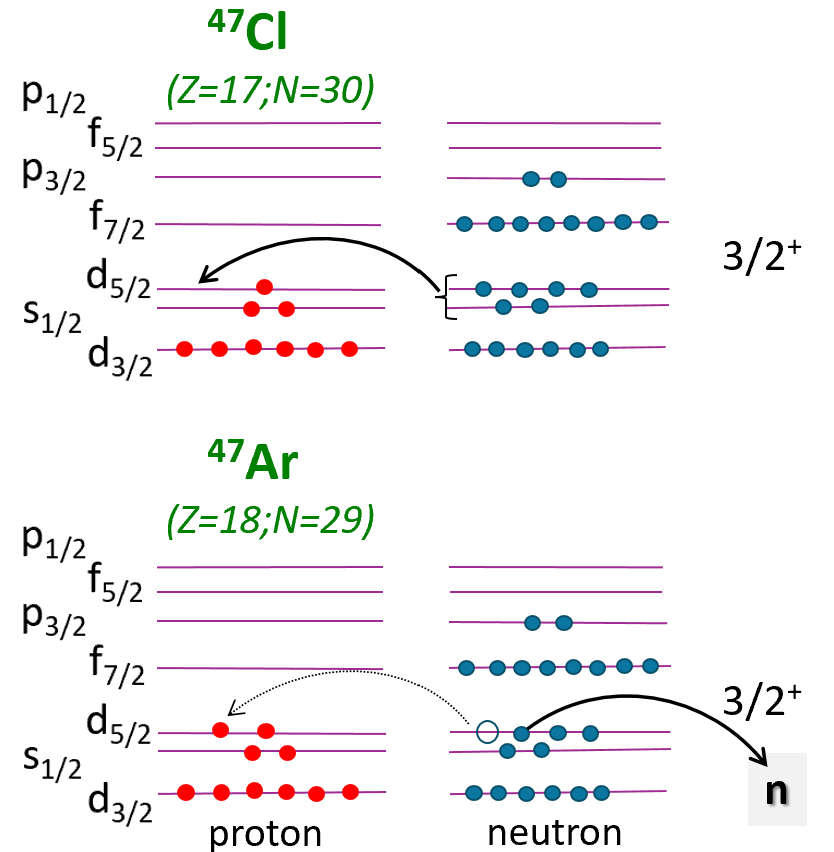}
	\caption{\label{fig:beta_n_47Cl}
	$\beta$-delayed neutron emission: Illustration of the path from the $^{47}$Cl 
    ground state ($3/2^+$) to a $1p1h$ $3/2^+$ state in 
    $^{47}$Ar by a conversion of a neutron to a proton (GT transformation) (top), and 
    then the emission of a neutron from the $sd$ shell to populate a $2p2h$ positive-parity 
    state (for example the $2^+_1$) in the daughter nucleus $^{46}$Ar (bottom).
	}
\end{figure}

 %Figure 11
\begin{figure}
	\includegraphics[width=\columnwidth]{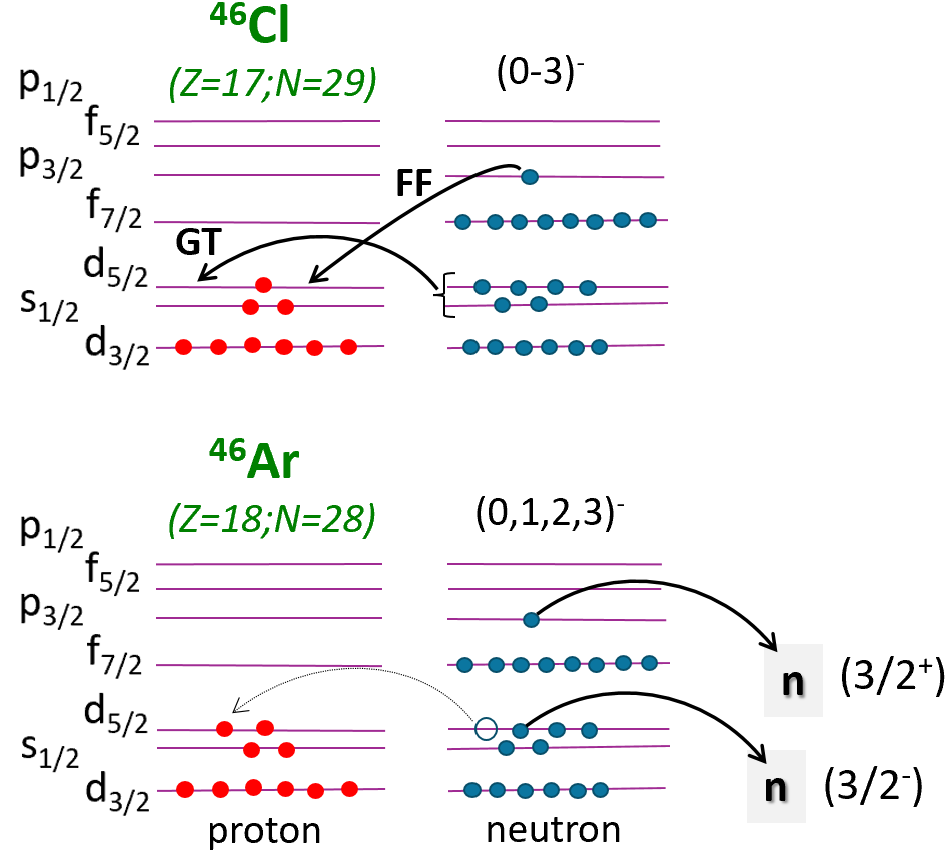}
	\caption{\label{fig:beta_n_46Cl}
	Illustration of the $\beta$-delayed neutron emission process in $^{46}$Cl. 
    Top: GT allowed decay would convert, for example, a $d_{3/2}$ neutron to a $d_{3/2}$ 
    proton or a $d_{3/2}$ proton could be formed from FF transition of a $p_{3/2}$ neutron. 
    Bottom: Emission of the $p_{3/2}$ neutron would populate a $3/2^+$ state in $^{45}$Ar 
    due to the remaining unpaired $d_{3/2}$ neutron. Emission of a $d_{3/2}$ neutron would 
    populate a $3/2^-$ state in $^{45}$Ar due to the remaining $p_{3/2}$ neutron.
	}
\end{figure}

%Table6

\begin{table}
	\caption{\label{tab:beta-Pn}. Measured and calculated delayed neutron 
    emission probabilities in percent.
	$Q_{x\beta}$ values for $^{46}$Cl and $^{47}$Cl are also noted \cite{AME2020}.
    } 
		\begin{tabular}{c|ccccc}
			\hline
			 Isotope & $Q_{\beta_{xn}}$(MeV) & $J^\pi$ & $P_n$(Exp.) & $P_n$($V_{MU}$)&$P_n$($V_{MU}$) \\
               &  &  & \% & GT+FF,$Q_{exp}$ &GT,$Q_{exp}$ \\
			\hline
			 & $16.04(10)_{0n}$  & $0^-$ &  75(15) & 34.90 & 43.34 \\
              $^{46}$Cl            & $7.960(100)_{1n}$ & $1^-$ &           & 23.83 & 26.01 \\
                          & $2.80(14)_{2n}$   & $2^-$ &           & 23.17 & 27.57 \\   
                          &            & $3^-$ &           & 32.28 & 38.28 \\ 
                          \hline
			 & $15.79(20)_{0n}$  & $1/2^+$ &  98(12) & 86.13 & 96.22 \\
               $^{47}$Cl           & $12.12(20)_{1n}$  & $3/2^+$ &        & 77.56 & 87.04 \\
                          & $4.0(3)_{2n}$     &         &       &       &       \\   
			
			\hline		
		\end{tabular}
\end{table}

As noted before, a different set of excited states in $^{46}$Ar 
were populated in the $\beta1n$ decay of $^{47}$Cl than those populated in the 
direct $\beta0n$ decay of $^{46}$Cl. Comparison with prior studies and 
predictions from shell model calculations seem to favor positive parities 
for the states seen in $\beta1n$ decay. In Fig.~\ref{fig:beta_n_47Cl}, a simple 
illustration of how $\beta$-delayed 
neutron emission can proceed based on the occupancies suggested by 
shell model calculations performed using the FSU interaction is shown.

The ground-state configuration of $^{47}$Cl ($J^\pi = 3/2^+$) is 
$\pi[d_{5/2}^6s_{1/2}^2d_{3/2}^1]\nu[sd(20)f_{7/2}^8p_{3/2}^2]$ with a 
filled neutron $sd$ shell and a hole in the proton $d_{3/2}$ orbital.
The states with proton excitations into the $fp$ shell are energetically 
too high in $^{46}$Ar to be reached by GT $\beta^-$ decay ({\it i.e} 
conversion of one of the $fp$ neutrons into a $fp$ proton). 
Instead, a GT transformation changes one of the neutrons in either 
the $\nu s_{1/2}$ or 
$\nu d_{3/2}$ orbitals to a proton, filling the $\pi d_{3/2}$ orbital, as shown by 
the arrow in the top cartoon. This leaves a hole in the $\nu d_{3/2}$ or $\nu s_{1/2}$ 
orbital resulting in a positive parity $1p1h$ state of $^{47}$Ar. 
The ground state configuration of $^{47}$Ar ($J^\pi = 3/2^-$) is 
$\pi[d_{5/2}^6s_{1/2}^2d_{3/2}^2]\nu[sd(20)f_{7/2}^8p_{3/2}^1]$ with a full neutron 
$sd$ shell and an unpaired neutron in the $p_{3/2}$ orbital.
We can assume that the $3/2^+_2$ (log$ft$ = 5.3) and $1/2^+_2$ (log$ft$ = 5.5) 
states in $^{47}$Ar predicted at 4.95 MeV and 5.18 MeV, respectively, are populated 
in this allowed GT decay from the ground state of $^{47}$Cl. 
With $S_n$ equal to 3.667(3) MeV \cite{AME2020}, these states lie above the neutron separation 
threshold in $^{47}$Ar.

In the next step, neutron emission from these $1p1h$ states populates 
low-lying positive-parity states in $^{46}$Ar, as seen experimentally. 
For that, an unpaired neutron is emitted from the $sd$ shell creating a 
$2p2h$ positive parity state in the daughter nucleus $^{46}$Ar, 
as seen in the bottom cartoon in Fig.~\ref{fig:beta_n_47Cl}. 
This emission is energetically 
favored due to nuclear pairing and the low $l$ orbitals involved. 
Breaking a pair would typically cost an additional one or two MeV.
According to shell model calculations using the FSU interaction, 
the $2^+_1$ state in $^{46}$Ar is a $0p0h$ state with a full neutron $sd$ shell 
($\pi[d_{5/2}^6s_{1/2}^2d_{3/2}^2]\nu[sd(20)f_{7/2}^8]$). Hence, the neutron 
emission has to be facilitated by a small admixture of $2p2h$ configuration 
if the state is populated directly. Therefore, delayed neutron emission is 
probing and manifesting mixing between configurations 
($0p0h$ and $2p2h$ in this case).
In a recent publication \cite{IPM_Maheshwari_2024}, IBM calculations including 
configuration mixing (IBM-CM) predicted low-lying states in 
$^{46}$Ar. In that work, it was shown that the yrast $0^+_1$ and $2^+_1$ states
have a low admixture of intruder [n+2] space while the $4^+_1$, $0^+_2$, $2^+_2$, 
and $4^+_2$ states are dominated by $2p2h$ intruder configurations. 
This can explain the experimental observation of their
population in the $\beta$-delayed neutron emission channel.

A schematic view of $\beta$-delayed neutron emission from $^{46}$Cl, which has possible 
ground-state spin-parities of $1^-$ and $2^-$, is shown in Fig.~\ref{fig:beta_n_46Cl}. 
GT allowed decay would transform one of the $sd$ 
neutrons into an $sd$ proton, while transformation of the unpaired neutron in the $p_{3/2}$
orbital to a $d_{3/2}$ proton would be a FF transition. Following GT decay, the emission of 
a $p_{3/2}$ neutron would create a $3/2^+$ state in $^{45}$Ar due to the unpaired $d_{3/2}$ 
neutron. On the other hand, if the $d_{3/2}$ neutron is emitted, a $3/2^-$ state would be 
populated in the daughter nucleus due to the unpaired $p_{3/2}$ neutron. 
This simple picture seems to be consistent with the observed 
population of both positive-parity (e.g.~3297 keV, $5/2^+$) and negative-parity 
(e.g.~1341 keV, $3/2^-$) states, as shown in Fig.~\ref{fig:46_47Cl_beta-xn}.
The occupancy of the $5/2^+$ state from the shell-model calculation shows contributions
 from a $d_{5/2}$ hole along with an $s_{1/2}$ hole \cite{Soumik2023}.

\section{Summary}

The Beta Counting System surrounded by an array of 16 HPGe Clover detectors was used to study the 
$\beta$ and $\beta$-delayed neutron decays of the exotic $^{46,47}$Cl isotopes at the National 
Superconducting Cyclotron Laboratory. The level structures of $^{45,46,47}$Ar were elucidated from the 
various decays, and comparisons were made to shell model calculations when possible. 
Following the activity in the granddaughter nuclei, 
the $P_n$ values for $^{46}$Cl and $^{47}$Cl are estimated to be 75(15)\% and 98(12)\%, respectively. 
The large $P_n$ for $^{47}$Cl is at odds with an extremely low prior measurement 
of $P_n$. However the increasing trend for $P_n$ going from $^{45}$Cl to $^{47}$Cl, 
is expected as the neutron dripline is approached. 
The half-lives of $^{46,47}$Cl were measured 
and compared with calculations using the $V_{MU}$ interaction for a variety of possible ground-state 
spins. When including both GT and FF transitions, good agreement between 
experiment and theory was found assuming spin-parities of $1^-$ or $2^-$ for the $^{46}$Cl ground state 
and a spin-parity of $3/2^+$ for the $^{47}$Cl ground state. These spin-parities values are also 
consistent with the observed decays. Two new $\gamma$-ray transitions in $^{47}$Ar were 
identified following 
the weak $\beta0n$ decay of $^{47}$Cl. It was also observed that the levels populated in
$N=28$ $^{46}$Ar from $^{46}$Cl $\beta0n$ and $^{47}$Cl $\beta1n$ decay are completely 
different beyond the first $2^+$ excited state. Differences were also seen 
in the states populated in $^{45}$Ar from $^{46}$Cl 
$\beta1n$ decay compared to $^{47}$Cl $\beta2n$ decay. 
Attempts were made to explain states populated in delayed neutron emission using a 
simple shell-model picture.

The large $P_n$ values observed for $^{46,47}$Cl indicate a very small decay 
branch to bound states, likely through FF transitions. This observation seems to be 
consistent for $^{47}$Cl decay. However, for $^{46}$Cl, a large branch 
to the low-lying $2^+$ state was observed, somewhat contradictory to 
the large $P_n$ value.
Furthermore, shell model calculations using the $V_{MU}$ interaction
which reproduced the measured half-lives after including FF decays 
severely underestimated the measured $P_n$ values. 
This highlights that further work is necessary, both in experiments and in the
refinement of theoretical approaches, to quantify the impact of FF transitions.
This region of neutron-rich isotopes around the 
$N=28$ Island of Inversion remains a fertile testing ground for state-of-the-art nuclear theory. 
With the start of operations at the new Facility for Rare Isotope Beams (FRIB), nuclei with 
large $\beta$-delayed neutron branches, like $^{46}$Cl and $^{47}$Cl isotopes 
studied in the present work, will feature prominently in experimental campaigns. 
Key experimental devices like 
the FRIB Decay Station Initiator will allow measurements of both neutrons and $\gamma$ rays from 
these exotic systems, allowing for complete decay spectroscopy \cite{Cox2024}. 
With the new radioactive-beam facilities beginning to provide much-needed experimental data, 
a clearer picture of $\beta$-delayed neutron decay and the role 
of FF transitions may soon emerge.

\section*{Acknowledgments}

We thank the NSCL operations team and the A1900 team, especially Tom Ginter for 
the production and optimization of the secondary beam.  
%We acknowledge the contribution of C.J. Chiara and J.J. Carroll to the experiment. 
This work was supported by the US National Science Foundation under 
Grant Nos.~PHY-2012522 (FSU) and PHY-1848177 (CAREER), the   
US Department of Energy, Office of Science, Office of Nuclear Physics
under award Nos.~DE-SC0020451 (FRIB), DE-FG02-94ER40848 (UML), DE-AC52-07NA27344 (LLNL),
DE-AC02-06CH11357 (ANL), and DE-SC0009883 (FSU) and also by the US Department of Energy 
National Nuclear Security Administration Grant No. DOE-DE-NA0003906 and
the Nuclear Science and Security Consortium under Award No.~DE-NA0003180.

\bibliography{apssamp}% Produces the bibliography via BibTeX.

\providecommand{\noopsort}[1]{}\providecommand{\singleletter}[1]{#1}%
\begin{thebibliography}{44}%
\makeatletter
\providecommand \@ifxundefined [1]{%
 \@ifx{#1\undefined}
}%
\providecommand \@ifnum [1]{%
 \ifnum #1\expandafter \@firstoftwo
 \else \expandafter \@secondoftwo
 \fi
}%
\providecommand \@ifx [1]{%
 \ifx #1\expandafter \@firstoftwo
 \else \expandafter \@secondoftwo
 \fi
}%
\providecommand \natexlab [1]{#1}%
\providecommand \enquote  [1]{``#1''}%
\providecommand \bibnamefont  [1]{#1}%
\providecommand \bibfnamefont [1]{#1}%
\providecommand \citenamefont [1]{#1}%
\providecommand \href@noop [0]{\@secondoftwo}%
\providecommand \href [0]{\begingroup \@sanitize@url \@href}%
\providecommand \@href[1]{\@@startlink{#1}\@@href}%
\providecommand \@@href[1]{\endgroup#1\@@endlink}%
\providecommand \@sanitize@url [0]{\catcode `\\12\catcode `\$12\catcode `\&12\catcode `\#12\catcode `\^12\catcode `\_12\catcode `\%12\relax}%
\providecommand \@@startlink[1]{}%
\providecommand \@@endlink[0]{}%
\providecommand \url  [0]{\begingroup\@sanitize@url \@url }%
\providecommand \@url [1]{\endgroup\@href {#1}{\urlprefix }}%
\providecommand \urlprefix  [0]{URL }%
\providecommand \Eprint [0]{\href }%
\providecommand \doibase [0]{https://doi.org/}%
\providecommand \selectlanguage [0]{\@gobble}%
\providecommand \bibinfo  [0]{\@secondoftwo}%
\providecommand \bibfield  [0]{\@secondoftwo}%
\providecommand \translation [1]{[#1]}%
\providecommand \BibitemOpen [0]{}%
\providecommand \bibitemStop [0]{}%
\providecommand \bibitemNoStop [0]{.\EOS\space}%
\providecommand \EOS [0]{\spacefactor3000\relax}%
\providecommand \BibitemShut  [1]{\csname bibitem#1\endcsname}%
\let\auto@bib@innerbib\@empty
%</preamble>
\bibitem [{\citenamefont {{Gorton, Oliver C.}}\ \emph {et~al.}(2023)\citenamefont {{Gorton, Oliver C.}}, \citenamefont {{Johnson, Calvin W.}},\ and\ \citenamefont {{Escher, Jutta E.}}}]{jutta}%
  \BibitemOpen
  \bibfield  {author} {\bibinfo {author} {\bibnamefont {{Gorton, Oliver C.}}}, \bibinfo {author} {\bibnamefont {{Johnson, Calvin W.}}},\ and\ \bibinfo {author} {\bibnamefont {{Escher, Jutta E.}}},\ }\bibfield  {title} {\bibinfo {title} {A problem in the statistical description of beta-delayed neutron emission},\ }\href {https://doi.org/10.1051/epjconf/202328403013} {\bibfield  {journal} {\bibinfo  {journal} {EPJ Web of Conf.}\ }\textbf {\bibinfo {volume} {284}},\ \bibinfo {pages} {03013} (\bibinfo {year} {2023})}\BibitemShut {NoStop}%
\bibitem [{\citenamefont {Xu}\ \emph {et~al.}(2024)\citenamefont {Xu}, \citenamefont {Grzywacz}, \citenamefont {Gottardo}, \citenamefont {Madurga}, \citenamefont {Alonso}, \citenamefont {Andreyev}, \citenamefont {Benzoni}, \citenamefont {Borge}, \citenamefont {Cap}, \citenamefont {Costache}, \citenamefont {De~Witte}, \citenamefont {Dimitrov}, \citenamefont {Escher}, \citenamefont {Fijalkowska}, \citenamefont {Fraile}, \citenamefont {Franchoo}, \citenamefont {Fynbo}, \citenamefont {Gonsalves}, \citenamefont {Gross}, \citenamefont {Harkness-Brennan}, \citenamefont {Heideman}, \citenamefont {Huyse}, \citenamefont {Judson}, \citenamefont {Kawano}, \citenamefont {King}, \citenamefont {Kisyov}, \citenamefont {Kolos}, \citenamefont {Korgul}, \citenamefont {Lazarus}, \citenamefont {Lic\ifmmode~\u{a}\else \u{a}\fi{}}, \citenamefont {Liu}, \citenamefont {Lynch}, \citenamefont {Marginean}, \citenamefont {Marginean}, \citenamefont {Mazzocchi}, \citenamefont {Mengoni}, \citenamefont {Mihai}, \citenamefont {Morales},
  \citenamefont {Page}, \citenamefont {Pakarinen}, \citenamefont {Paulauskas}, \citenamefont {Perea}, \citenamefont {Piersa-Si\l{}kowska}, \citenamefont {Podoly\'ak}, \citenamefont {Sotty}, \citenamefont {Taylor}, \citenamefont {Tengblad}, \citenamefont {Van~Duppen}, \citenamefont {Vedia}, \citenamefont {Verney}, \citenamefont {Warr},\ and\ \citenamefont {Yuan}}]{Xu_doorway}%
  \BibitemOpen
  \bibfield  {author} {\bibinfo {author} {\bibfnamefont {Z.~Y.}\ \bibnamefont {Xu}}, \bibinfo {author} {\bibfnamefont {R.}~\bibnamefont {Grzywacz}}, \bibinfo {author} {\bibfnamefont {A.}~\bibnamefont {Gottardo}}, \bibinfo {author} {\bibfnamefont {M.}~\bibnamefont {Madurga}}, \bibinfo {author} {\bibfnamefont {I.~M.}\ \bibnamefont {Alonso}}, \bibinfo {author} {\bibfnamefont {A.~N.}\ \bibnamefont {Andreyev}}, \bibinfo {author} {\bibfnamefont {G.}~\bibnamefont {Benzoni}}, \bibinfo {author} {\bibfnamefont {M.~J.~G.}\ \bibnamefont {Borge}}, \bibinfo {author} {\bibfnamefont {T.}~\bibnamefont {Cap}}, \bibinfo {author} {\bibfnamefont {C.}~\bibnamefont {Costache}}, \bibinfo {author} {\bibfnamefont {H.}~\bibnamefont {De~Witte}}, \bibinfo {author} {\bibfnamefont {B.~I.}\ \bibnamefont {Dimitrov}}, \bibinfo {author} {\bibfnamefont {J.~E.}\ \bibnamefont {Escher}}, \bibinfo {author} {\bibfnamefont {A.}~\bibnamefont {Fijalkowska}}, \bibinfo {author} {\bibfnamefont {L.~M.}\ \bibnamefont {Fraile}}, \bibinfo {author}
  {\bibfnamefont {S.}~\bibnamefont {Franchoo}}, \bibinfo {author} {\bibfnamefont {H.~O.~U.}\ \bibnamefont {Fynbo}}, \bibinfo {author} {\bibfnamefont {B.~C.}\ \bibnamefont {Gonsalves}}, \bibinfo {author} {\bibfnamefont {C.~J.}\ \bibnamefont {Gross}}, \bibinfo {author} {\bibfnamefont {L.~J.}\ \bibnamefont {Harkness-Brennan}}, \bibinfo {author} {\bibfnamefont {J.}~\bibnamefont {Heideman}}, \bibinfo {author} {\bibfnamefont {M.}~\bibnamefont {Huyse}}, \bibinfo {author} {\bibfnamefont {D.~S.}\ \bibnamefont {Judson}}, \bibinfo {author} {\bibfnamefont {T.}~\bibnamefont {Kawano}}, \bibinfo {author} {\bibfnamefont {T.~T.}\ \bibnamefont {King}}, \bibinfo {author} {\bibfnamefont {S.}~\bibnamefont {Kisyov}}, \bibinfo {author} {\bibfnamefont {K.}~\bibnamefont {Kolos}}, \bibinfo {author} {\bibfnamefont {A.}~\bibnamefont {Korgul}}, \bibinfo {author} {\bibfnamefont {I.}~\bibnamefont {Lazarus}}, \bibinfo {author} {\bibfnamefont {R.}~\bibnamefont {Lic\ifmmode~\u{a}\else \u{a}\fi{}}}, \bibinfo {author} {\bibfnamefont {M.~L.}\
  \bibnamefont {Liu}}, \bibinfo {author} {\bibfnamefont {L.}~\bibnamefont {Lynch}}, \bibinfo {author} {\bibfnamefont {N.}~\bibnamefont {Marginean}}, \bibinfo {author} {\bibfnamefont {R.}~\bibnamefont {Marginean}}, \bibinfo {author} {\bibfnamefont {C.}~\bibnamefont {Mazzocchi}}, \bibinfo {author} {\bibfnamefont {D.}~\bibnamefont {Mengoni}}, \bibinfo {author} {\bibfnamefont {C.}~\bibnamefont {Mihai}}, \bibinfo {author} {\bibfnamefont {A.~I.}\ \bibnamefont {Morales}}, \bibinfo {author} {\bibfnamefont {R.~D.}\ \bibnamefont {Page}}, \bibinfo {author} {\bibfnamefont {J.}~\bibnamefont {Pakarinen}}, \bibinfo {author} {\bibfnamefont {S.~V.}\ \bibnamefont {Paulauskas}}, \bibinfo {author} {\bibfnamefont {A.}~\bibnamefont {Perea}}, \bibinfo {author} {\bibfnamefont {M.}~\bibnamefont {Piersa-Si\l{}kowska}}, \bibinfo {author} {\bibfnamefont {Z.}~\bibnamefont {Podoly\'ak}}, \bibinfo {author} {\bibfnamefont {C.}~\bibnamefont {Sotty}}, \bibinfo {author} {\bibfnamefont {S.}~\bibnamefont {Taylor}}, \bibinfo {author}
  {\bibfnamefont {O.}~\bibnamefont {Tengblad}}, \bibinfo {author} {\bibfnamefont {P.}~\bibnamefont {Van~Duppen}}, \bibinfo {author} {\bibfnamefont {V.}~\bibnamefont {Vedia}}, \bibinfo {author} {\bibfnamefont {D.}~\bibnamefont {Verney}}, \bibinfo {author} {\bibfnamefont {N.}~\bibnamefont {Warr}},\ and\ \bibinfo {author} {\bibfnamefont {C.~X.}\ \bibnamefont {Yuan}},\ }\bibfield  {title} {\bibinfo {title} {Compound-nucleus and doorway-state decays of $\ensuremath{\beta}$-delayed neutron emitters $^{51,52,53}\mathrm{K}$},\ }\href {https://doi.org/10.1103/PhysRevLett.133.042501} {\bibfield  {journal} {\bibinfo  {journal} {Phys. Rev. Lett.}\ }\textbf {\bibinfo {volume} {133}},\ \bibinfo {pages} {042501} (\bibinfo {year} {2024})}\BibitemShut {NoStop}%
\bibitem [{\citenamefont {Peshkin}\ \emph {et~al.}(2014)\citenamefont {Peshkin}, \citenamefont {Volya},\ and\ \citenamefont {Zelevinsky}}]{Peshkin2014}%
  \BibitemOpen
  \bibfield  {author} {\bibinfo {author} {\bibfnamefont {M.}~\bibnamefont {Peshkin}}, \bibinfo {author} {\bibfnamefont {A.}~\bibnamefont {Volya}},\ and\ \bibinfo {author} {\bibfnamefont {V.}~\bibnamefont {Zelevinsky}},\ }\bibfield  {title} {\bibinfo {title} {Non-exponential and oscillatory decays in quantum mechanics},\ }\href {https://doi.org/10.1209/0295-5075/107/40001} {\bibfield  {journal} {\bibinfo  {journal} {Europhysics Letters}\ }\textbf {\bibinfo {volume} {107}},\ \bibinfo {pages} {40001} (\bibinfo {year} {2014})}\BibitemShut {NoStop}%
\bibitem [{\citenamefont {Wang}\ \emph {et~al.}(2023)\citenamefont {Wang}, \citenamefont {Nazarewicz}, \citenamefont {Volya},\ and\ \citenamefont {Ma}}]{Wang2023}%
  \BibitemOpen
  \bibfield  {author} {\bibinfo {author} {\bibfnamefont {S.~M.}\ \bibnamefont {Wang}}, \bibinfo {author} {\bibfnamefont {W.}~\bibnamefont {Nazarewicz}}, \bibinfo {author} {\bibfnamefont {A.}~\bibnamefont {Volya}},\ and\ \bibinfo {author} {\bibfnamefont {Y.~G.}\ \bibnamefont {Ma}},\ }\bibfield  {title} {\bibinfo {title} {Probing the nonexponential decay regime in open quantum systems},\ }\href {https://doi.org/10.1103/PhysRevResearch.5.023183} {\bibfield  {journal} {\bibinfo  {journal} {Phys. Rev. Res.}\ }\textbf {\bibinfo {volume} {5}},\ \bibinfo {pages} {023183} (\bibinfo {year} {2023})}\BibitemShut {NoStop}%
\bibitem [{\citenamefont {Sorlin}\ and\ \citenamefont {Porquet}(2008)}]{newmagicnumbers}%
  \BibitemOpen
  \bibfield  {author} {\bibinfo {author} {\bibfnamefont {O.}~\bibnamefont {Sorlin}}\ and\ \bibinfo {author} {\bibfnamefont {M.-G.}\ \bibnamefont {Porquet}},\ }\bibfield  {title} {\bibinfo {title} {Nuclear magic numbers: New features far from stability},\ }\href {https://doi.org/https://doi.org/10.1016/j.ppnp.2008.05.001} {\bibfield  {journal} {\bibinfo  {journal} {Progress in Particle and Nuclear Physics}\ }\textbf {\bibinfo {volume} {61}},\ \bibinfo {pages} {602} (\bibinfo {year} {2008})}\BibitemShut {NoStop}%
\bibitem [{\citenamefont {Otsuka}\ \emph {et~al.}(2020)\citenamefont {Otsuka}, \citenamefont {Gade}, \citenamefont {Sorlin}, \citenamefont {Suzuki},\ and\ \citenamefont {Utsuno}}]{shell_evolution}%
  \BibitemOpen
  \bibfield  {author} {\bibinfo {author} {\bibfnamefont {T.}~\bibnamefont {Otsuka}}, \bibinfo {author} {\bibfnamefont {A.}~\bibnamefont {Gade}}, \bibinfo {author} {\bibfnamefont {O.}~\bibnamefont {Sorlin}}, \bibinfo {author} {\bibfnamefont {T.}~\bibnamefont {Suzuki}},\ and\ \bibinfo {author} {\bibfnamefont {Y.}~\bibnamefont {Utsuno}},\ }\bibfield  {title} {\bibinfo {title} {Evolution of shell structure in exotic nuclei},\ }\href {https://doi.org/10.1103/RevModPhys.92.015002} {\bibfield  {journal} {\bibinfo  {journal} {Rev. Mod. Phys.}\ }\textbf {\bibinfo {volume} {92}},\ \bibinfo {pages} {015002} (\bibinfo {year} {2020})}\BibitemShut {NoStop}%
\bibitem [{\citenamefont {Caurier}\ \emph {et~al.}(2014)\citenamefont {Caurier}, \citenamefont {Nowacki},\ and\ \citenamefont {Poves}}]{N28_shellgap}%
  \BibitemOpen
  \bibfield  {author} {\bibinfo {author} {\bibfnamefont {E.}~\bibnamefont {Caurier}}, \bibinfo {author} {\bibfnamefont {F.}~\bibnamefont {Nowacki}},\ and\ \bibinfo {author} {\bibfnamefont {A.}~\bibnamefont {Poves}},\ }\bibfield  {title} {\bibinfo {title} {Merging of the islands of inversion at $n=20$ and $n=28$},\ }\href {https://doi.org/10.1103/PhysRevC.90.014302} {\bibfield  {journal} {\bibinfo  {journal} {Phys. Rev. C}\ }\textbf {\bibinfo {volume} {90}},\ \bibinfo {pages} {014302} (\bibinfo {year} {2014})}\BibitemShut {NoStop}%
\bibitem [{\citenamefont {Bastin}\ \emph {et~al.}(2007)\citenamefont {Bastin}, \citenamefont {Gr\'evy}, \citenamefont {Sohler}, \citenamefont {Sorlin}, \citenamefont {Dombr\'adi}, \citenamefont {Achouri}, \citenamefont {Ang\'elique}, \citenamefont {Azaiez}, \citenamefont {Baiborodin}, \citenamefont {Borcea}, \citenamefont {Bourgeois}, \citenamefont {Buta}, \citenamefont {B\"urger}, \citenamefont {Chapman}, \citenamefont {Dalouzy}, \citenamefont {Dlouhy}, \citenamefont {Drouard}, \citenamefont {Elekes}, \citenamefont {Franchoo}, \citenamefont {Iacob}, \citenamefont {Laurent}, \citenamefont {Lazar}, \citenamefont {Liang}, \citenamefont {Li\'enard}, \citenamefont {Mrazek}, \citenamefont {Nalpas}, \citenamefont {Negoita}, \citenamefont {Orr}, \citenamefont {Penionzhkevich}, \citenamefont {Podoly\'ak}, \citenamefont {Pougheon}, \citenamefont {Roussel-Chomaz}, \citenamefont {Saint-Laurent}, \citenamefont {Stanoiu}, \citenamefont {Stefan}, \citenamefont {Nowacki},\ and\ \citenamefont {Poves}}]{Bastin07}%
  \BibitemOpen
  \bibfield  {author} {\bibinfo {author} {\bibfnamefont {B.}~\bibnamefont {Bastin}}, \bibinfo {author} {\bibfnamefont {S.}~\bibnamefont {Gr\'evy}}, \bibinfo {author} {\bibfnamefont {D.}~\bibnamefont {Sohler}}, \bibinfo {author} {\bibfnamefont {O.}~\bibnamefont {Sorlin}}, \bibinfo {author} {\bibfnamefont {Z.}~\bibnamefont {Dombr\'adi}}, \bibinfo {author} {\bibfnamefont {N.~L.}\ \bibnamefont {Achouri}}, \bibinfo {author} {\bibfnamefont {J.~C.}\ \bibnamefont {Ang\'elique}}, \bibinfo {author} {\bibfnamefont {F.}~\bibnamefont {Azaiez}}, \bibinfo {author} {\bibfnamefont {D.}~\bibnamefont {Baiborodin}}, \bibinfo {author} {\bibfnamefont {R.}~\bibnamefont {Borcea}}, \bibinfo {author} {\bibfnamefont {C.}~\bibnamefont {Bourgeois}}, \bibinfo {author} {\bibfnamefont {A.}~\bibnamefont {Buta}}, \bibinfo {author} {\bibfnamefont {A.}~\bibnamefont {B\"urger}}, \bibinfo {author} {\bibfnamefont {R.}~\bibnamefont {Chapman}}, \bibinfo {author} {\bibfnamefont {J.~C.}\ \bibnamefont {Dalouzy}}, \bibinfo {author} {\bibfnamefont
  {Z.}~\bibnamefont {Dlouhy}}, \bibinfo {author} {\bibfnamefont {A.}~\bibnamefont {Drouard}}, \bibinfo {author} {\bibfnamefont {Z.}~\bibnamefont {Elekes}}, \bibinfo {author} {\bibfnamefont {S.}~\bibnamefont {Franchoo}}, \bibinfo {author} {\bibfnamefont {S.}~\bibnamefont {Iacob}}, \bibinfo {author} {\bibfnamefont {B.}~\bibnamefont {Laurent}}, \bibinfo {author} {\bibfnamefont {M.}~\bibnamefont {Lazar}}, \bibinfo {author} {\bibfnamefont {X.}~\bibnamefont {Liang}}, \bibinfo {author} {\bibfnamefont {E.}~\bibnamefont {Li\'enard}}, \bibinfo {author} {\bibfnamefont {J.}~\bibnamefont {Mrazek}}, \bibinfo {author} {\bibfnamefont {L.}~\bibnamefont {Nalpas}}, \bibinfo {author} {\bibfnamefont {F.}~\bibnamefont {Negoita}}, \bibinfo {author} {\bibfnamefont {N.~A.}\ \bibnamefont {Orr}}, \bibinfo {author} {\bibfnamefont {Y.}~\bibnamefont {Penionzhkevich}}, \bibinfo {author} {\bibfnamefont {Z.}~\bibnamefont {Podoly\'ak}}, \bibinfo {author} {\bibfnamefont {F.}~\bibnamefont {Pougheon}}, \bibinfo {author} {\bibfnamefont
  {P.}~\bibnamefont {Roussel-Chomaz}}, \bibinfo {author} {\bibfnamefont {M.~G.}\ \bibnamefont {Saint-Laurent}}, \bibinfo {author} {\bibfnamefont {M.}~\bibnamefont {Stanoiu}}, \bibinfo {author} {\bibfnamefont {I.}~\bibnamefont {Stefan}}, \bibinfo {author} {\bibfnamefont {F.}~\bibnamefont {Nowacki}},\ and\ \bibinfo {author} {\bibfnamefont {A.}~\bibnamefont {Poves}},\ }\bibfield  {title} {\bibinfo {title} {Collapse of the $n=28$ shell closure in $^{42}\mathrm{S}\mathrm{i}$},\ }\href {https://doi.org/10.1103/PhysRevLett.99.022503} {\bibfield  {journal} {\bibinfo  {journal} {Phys. Rev. Lett.}\ }\textbf {\bibinfo {volume} {99}},\ \bibinfo {pages} {022503} (\bibinfo {year} {2007})}\BibitemShut {NoStop}%
\bibitem [{\citenamefont {Takeuchi}\ \emph {et~al.}(2012)\citenamefont {Takeuchi}, \citenamefont {Matsushita}, \citenamefont {Aoi}, \citenamefont {Doornenbal}, \citenamefont {Li}, \citenamefont {Motobayashi}, \citenamefont {Scheit}, \citenamefont {Steppenbeck}, \citenamefont {Wang}, \citenamefont {Baba}, \citenamefont {Bazin}, \citenamefont {C\`aceres}, \citenamefont {Crawford}, \citenamefont {Fallon}, \citenamefont {Gernh\"auser}, \citenamefont {Gibelin}, \citenamefont {Go}, \citenamefont {Gr\'evy}, \citenamefont {Hinke}, \citenamefont {Hoffman}, \citenamefont {Hughes}, \citenamefont {Ideguchi}, \citenamefont {Jenkins}, \citenamefont {Kobayashi}, \citenamefont {Kondo}, \citenamefont {Kr\"ucken}, \citenamefont {Le~Bleis}, \citenamefont {Lee}, \citenamefont {Lee}, \citenamefont {Matta}, \citenamefont {Michimasa}, \citenamefont {Nakamura}, \citenamefont {Ota}, \citenamefont {Petri}, \citenamefont {Sako}, \citenamefont {Sakurai}, \citenamefont {Shimoura}, \citenamefont {Steiger}, \citenamefont {Takahashi},
  \citenamefont {Takechi}, \citenamefont {Togano}, \citenamefont {Winkler},\ and\ \citenamefont {Yoneda}}]{Takeuchi2012}%
  \BibitemOpen
  \bibfield  {author} {\bibinfo {author} {\bibfnamefont {S.}~\bibnamefont {Takeuchi}}, \bibinfo {author} {\bibfnamefont {M.}~\bibnamefont {Matsushita}}, \bibinfo {author} {\bibfnamefont {N.}~\bibnamefont {Aoi}}, \bibinfo {author} {\bibfnamefont {P.}~\bibnamefont {Doornenbal}}, \bibinfo {author} {\bibfnamefont {K.}~\bibnamefont {Li}}, \bibinfo {author} {\bibfnamefont {T.}~\bibnamefont {Motobayashi}}, \bibinfo {author} {\bibfnamefont {H.}~\bibnamefont {Scheit}}, \bibinfo {author} {\bibfnamefont {D.}~\bibnamefont {Steppenbeck}}, \bibinfo {author} {\bibfnamefont {H.}~\bibnamefont {Wang}}, \bibinfo {author} {\bibfnamefont {H.}~\bibnamefont {Baba}}, \bibinfo {author} {\bibfnamefont {D.}~\bibnamefont {Bazin}}, \bibinfo {author} {\bibfnamefont {L.}~\bibnamefont {C\`aceres}}, \bibinfo {author} {\bibfnamefont {H.}~\bibnamefont {Crawford}}, \bibinfo {author} {\bibfnamefont {P.}~\bibnamefont {Fallon}}, \bibinfo {author} {\bibfnamefont {R.}~\bibnamefont {Gernh\"auser}}, \bibinfo {author} {\bibfnamefont {J.}~\bibnamefont
  {Gibelin}}, \bibinfo {author} {\bibfnamefont {S.}~\bibnamefont {Go}}, \bibinfo {author} {\bibfnamefont {S.}~\bibnamefont {Gr\'evy}}, \bibinfo {author} {\bibfnamefont {C.}~\bibnamefont {Hinke}}, \bibinfo {author} {\bibfnamefont {C.~R.}\ \bibnamefont {Hoffman}}, \bibinfo {author} {\bibfnamefont {R.}~\bibnamefont {Hughes}}, \bibinfo {author} {\bibfnamefont {E.}~\bibnamefont {Ideguchi}}, \bibinfo {author} {\bibfnamefont {D.}~\bibnamefont {Jenkins}}, \bibinfo {author} {\bibfnamefont {N.}~\bibnamefont {Kobayashi}}, \bibinfo {author} {\bibfnamefont {Y.}~\bibnamefont {Kondo}}, \bibinfo {author} {\bibfnamefont {R.}~\bibnamefont {Kr\"ucken}}, \bibinfo {author} {\bibfnamefont {T.}~\bibnamefont {Le~Bleis}}, \bibinfo {author} {\bibfnamefont {J.}~\bibnamefont {Lee}}, \bibinfo {author} {\bibfnamefont {G.}~\bibnamefont {Lee}}, \bibinfo {author} {\bibfnamefont {A.}~\bibnamefont {Matta}}, \bibinfo {author} {\bibfnamefont {S.}~\bibnamefont {Michimasa}}, \bibinfo {author} {\bibfnamefont {T.}~\bibnamefont {Nakamura}}, \bibinfo
  {author} {\bibfnamefont {S.}~\bibnamefont {Ota}}, \bibinfo {author} {\bibfnamefont {M.}~\bibnamefont {Petri}}, \bibinfo {author} {\bibfnamefont {T.}~\bibnamefont {Sako}}, \bibinfo {author} {\bibfnamefont {H.}~\bibnamefont {Sakurai}}, \bibinfo {author} {\bibfnamefont {S.}~\bibnamefont {Shimoura}}, \bibinfo {author} {\bibfnamefont {K.}~\bibnamefont {Steiger}}, \bibinfo {author} {\bibfnamefont {K.}~\bibnamefont {Takahashi}}, \bibinfo {author} {\bibfnamefont {M.}~\bibnamefont {Takechi}}, \bibinfo {author} {\bibfnamefont {Y.}~\bibnamefont {Togano}}, \bibinfo {author} {\bibfnamefont {R.}~\bibnamefont {Winkler}},\ and\ \bibinfo {author} {\bibfnamefont {K.}~\bibnamefont {Yoneda}},\ }\bibfield  {title} {\bibinfo {title} {Well developed deformation in $^{42}\mathrm{Si}$},\ }\href {https://doi.org/10.1103/PhysRevLett.109.182501} {\bibfield  {journal} {\bibinfo  {journal} {Phys. Rev. Lett.}\ }\textbf {\bibinfo {volume} {109}},\ \bibinfo {pages} {182501} (\bibinfo {year} {2012})}\BibitemShut {NoStop}%
\bibitem [{\citenamefont {Gade}\ \emph {et~al.}(2019)\citenamefont {Gade}, \citenamefont {Brown}, \citenamefont {Tostevin}, \citenamefont {Bazin}, \citenamefont {Bender}, \citenamefont {Campbell}, \citenamefont {Crawford}, \citenamefont {Elman}, \citenamefont {Kemper}, \citenamefont {Longfellow}, \citenamefont {Lunderberg}, \citenamefont {Rhodes},\ and\ \citenamefont {Weisshaar}}]{Gade2019}%
  \BibitemOpen
  \bibfield  {author} {\bibinfo {author} {\bibfnamefont {A.}~\bibnamefont {Gade}}, \bibinfo {author} {\bibfnamefont {B.~A.}\ \bibnamefont {Brown}}, \bibinfo {author} {\bibfnamefont {J.~A.}\ \bibnamefont {Tostevin}}, \bibinfo {author} {\bibfnamefont {D.}~\bibnamefont {Bazin}}, \bibinfo {author} {\bibfnamefont {P.~C.}\ \bibnamefont {Bender}}, \bibinfo {author} {\bibfnamefont {C.~M.}\ \bibnamefont {Campbell}}, \bibinfo {author} {\bibfnamefont {H.~L.}\ \bibnamefont {Crawford}}, \bibinfo {author} {\bibfnamefont {B.}~\bibnamefont {Elman}}, \bibinfo {author} {\bibfnamefont {K.~W.}\ \bibnamefont {Kemper}}, \bibinfo {author} {\bibfnamefont {B.}~\bibnamefont {Longfellow}}, \bibinfo {author} {\bibfnamefont {E.}~\bibnamefont {Lunderberg}}, \bibinfo {author} {\bibfnamefont {D.}~\bibnamefont {Rhodes}},\ and\ \bibinfo {author} {\bibfnamefont {D.}~\bibnamefont {Weisshaar}},\ }\bibfield  {title} {\bibinfo {title} {Is the structure of $^{42}\mathrm{Si}$ understood?},\ }\href {https://doi.org/10.1103/PhysRevLett.122.222501}
  {\bibfield  {journal} {\bibinfo  {journal} {Phys. Rev. Lett.}\ }\textbf {\bibinfo {volume} {122}},\ \bibinfo {pages} {222501} (\bibinfo {year} {2019})}\BibitemShut {NoStop}%
\bibitem [{\citenamefont {Glasmacher}\ \emph {et~al.}(1997)\citenamefont {Glasmacher}, \citenamefont {Brown}, \citenamefont {Chromik}, \citenamefont {Cottle}, \citenamefont {Fauerbach}, \citenamefont {Ibbotson}, \citenamefont {Kemper}, \citenamefont {Morrissey}, \citenamefont {Scheit}, \citenamefont {Sklenicka},\ and\ \citenamefont {Steiner}}]{Glasmacher_44S}%
  \BibitemOpen
  \bibfield  {author} {\bibinfo {author} {\bibfnamefont {T.}~\bibnamefont {Glasmacher}}, \bibinfo {author} {\bibfnamefont {B.}~\bibnamefont {Brown}}, \bibinfo {author} {\bibfnamefont {M.}~\bibnamefont {Chromik}}, \bibinfo {author} {\bibfnamefont {P.}~\bibnamefont {Cottle}}, \bibinfo {author} {\bibfnamefont {M.}~\bibnamefont {Fauerbach}}, \bibinfo {author} {\bibfnamefont {R.}~\bibnamefont {Ibbotson}}, \bibinfo {author} {\bibfnamefont {K.}~\bibnamefont {Kemper}}, \bibinfo {author} {\bibfnamefont {D.}~\bibnamefont {Morrissey}}, \bibinfo {author} {\bibfnamefont {H.}~\bibnamefont {Scheit}}, \bibinfo {author} {\bibfnamefont {D.}~\bibnamefont {Sklenicka}},\ and\ \bibinfo {author} {\bibfnamefont {M.}~\bibnamefont {Steiner}},\ }\bibfield  {title} {\bibinfo {title} {Collectivity in 44s},\ }\href {https://doi.org/https://doi.org/10.1016/S0370-2693(97)00077-4} {\bibfield  {journal} {\bibinfo  {journal} {Physics Letters B}\ }\textbf {\bibinfo {volume} {395}},\ \bibinfo {pages} {163} (\bibinfo {year} {1997})}\BibitemShut
  {NoStop}%
\bibitem [{\citenamefont {Longfellow}\ \emph {et~al.}(2021)\citenamefont {Longfellow}, \citenamefont {Weisshaar}, \citenamefont {Gade}, \citenamefont {Brown}, \citenamefont {Bazin}, \citenamefont {Brown}, \citenamefont {Elman}, \citenamefont {Pereira}, \citenamefont {Rhodes},\ and\ \citenamefont {Spieker}}]{longfellow_ce}%
  \BibitemOpen
  \bibfield  {author} {\bibinfo {author} {\bibfnamefont {B.}~\bibnamefont {Longfellow}}, \bibinfo {author} {\bibfnamefont {D.}~\bibnamefont {Weisshaar}}, \bibinfo {author} {\bibfnamefont {A.}~\bibnamefont {Gade}}, \bibinfo {author} {\bibfnamefont {B.~A.}\ \bibnamefont {Brown}}, \bibinfo {author} {\bibfnamefont {D.}~\bibnamefont {Bazin}}, \bibinfo {author} {\bibfnamefont {K.~W.}\ \bibnamefont {Brown}}, \bibinfo {author} {\bibfnamefont {B.}~\bibnamefont {Elman}}, \bibinfo {author} {\bibfnamefont {J.}~\bibnamefont {Pereira}}, \bibinfo {author} {\bibfnamefont {D.}~\bibnamefont {Rhodes}},\ and\ \bibinfo {author} {\bibfnamefont {M.}~\bibnamefont {Spieker}},\ }\bibfield  {title} {\bibinfo {title} {Quadrupole collectivity in the neutron-rich sulfur isotopes $^{38,40,42,44}\mathrm{S}$},\ }\href {https://doi.org/10.1103/PhysRevC.103.054309} {\bibfield  {journal} {\bibinfo  {journal} {Phys. Rev. C}\ }\textbf {\bibinfo {volume} {103}},\ \bibinfo {pages} {054309} (\bibinfo {year} {2021})}\BibitemShut {NoStop}%
\bibitem [{\citenamefont {Force}\ \emph {et~al.}(2010)\citenamefont {Force}, \citenamefont {Gr\'evy}, \citenamefont {Gaudefroy}, \citenamefont {Sorlin}, \citenamefont {C\'aceres}, \citenamefont {Rotaru}, \citenamefont {Mrazek}, \citenamefont {Achouri}, \citenamefont {Ang\'elique}, \citenamefont {Azaiez}, \citenamefont {Bastin}, \citenamefont {Borcea}, \citenamefont {Buta}, \citenamefont {Daugas}, \citenamefont {Dlouhy}, \citenamefont {Dombr\'adi}, \citenamefont {De~Oliveira}, \citenamefont {Negoita}, \citenamefont {Penionzhkevich}, \citenamefont {Saint-Laurent}, \citenamefont {Sohler}, \citenamefont {Stanoiu}, \citenamefont {Stefan}, \citenamefont {Stodel},\ and\ \citenamefont {Nowacki}}]{Force2010}%
  \BibitemOpen
  \bibfield  {author} {\bibinfo {author} {\bibfnamefont {C.}~\bibnamefont {Force}}, \bibinfo {author} {\bibfnamefont {S.}~\bibnamefont {Gr\'evy}}, \bibinfo {author} {\bibfnamefont {L.}~\bibnamefont {Gaudefroy}}, \bibinfo {author} {\bibfnamefont {O.}~\bibnamefont {Sorlin}}, \bibinfo {author} {\bibfnamefont {L.}~\bibnamefont {C\'aceres}}, \bibinfo {author} {\bibfnamefont {F.}~\bibnamefont {Rotaru}}, \bibinfo {author} {\bibfnamefont {J.}~\bibnamefont {Mrazek}}, \bibinfo {author} {\bibfnamefont {N.~L.}\ \bibnamefont {Achouri}}, \bibinfo {author} {\bibfnamefont {J.~C.}\ \bibnamefont {Ang\'elique}}, \bibinfo {author} {\bibfnamefont {F.}~\bibnamefont {Azaiez}}, \bibinfo {author} {\bibfnamefont {B.}~\bibnamefont {Bastin}}, \bibinfo {author} {\bibfnamefont {R.}~\bibnamefont {Borcea}}, \bibinfo {author} {\bibfnamefont {A.}~\bibnamefont {Buta}}, \bibinfo {author} {\bibfnamefont {J.~M.}\ \bibnamefont {Daugas}}, \bibinfo {author} {\bibfnamefont {Z.}~\bibnamefont {Dlouhy}}, \bibinfo {author} {\bibfnamefont
  {Z.}~\bibnamefont {Dombr\'adi}}, \bibinfo {author} {\bibfnamefont {F.}~\bibnamefont {De~Oliveira}}, \bibinfo {author} {\bibfnamefont {F.}~\bibnamefont {Negoita}}, \bibinfo {author} {\bibfnamefont {Y.}~\bibnamefont {Penionzhkevich}}, \bibinfo {author} {\bibfnamefont {M.~G.}\ \bibnamefont {Saint-Laurent}}, \bibinfo {author} {\bibfnamefont {D.}~\bibnamefont {Sohler}}, \bibinfo {author} {\bibfnamefont {M.}~\bibnamefont {Stanoiu}}, \bibinfo {author} {\bibfnamefont {I.}~\bibnamefont {Stefan}}, \bibinfo {author} {\bibfnamefont {C.}~\bibnamefont {Stodel}},\ and\ \bibinfo {author} {\bibfnamefont {F.}~\bibnamefont {Nowacki}},\ }\bibfield  {title} {\bibinfo {title} {Prolate-spherical shape coexistence at $n=28$ in $^{44}\mathbf{S}$},\ }\href {https://doi.org/10.1103/PhysRevLett.105.102501} {\bibfield  {journal} {\bibinfo  {journal} {Phys. Rev. Lett.}\ }\textbf {\bibinfo {volume} {105}},\ \bibinfo {pages} {102501} (\bibinfo {year} {2010})}\BibitemShut {NoStop}%
\bibitem [{\citenamefont {Santiago-Gonzalez}\ \emph {et~al.}(2011)\citenamefont {Santiago-Gonzalez}, \citenamefont {Wiedenh\"over}, \citenamefont {Abramkina}, \citenamefont {Avila}, \citenamefont {Baugher}, \citenamefont {Bazin}, \citenamefont {Brown}, \citenamefont {Cottle}, \citenamefont {Gade}, \citenamefont {Glasmacher}, \citenamefont {Kemper}, \citenamefont {McDaniel}, \citenamefont {Rojas}, \citenamefont {Ratkiewicz}, \citenamefont {Meharchand}, \citenamefont {Simpson}, \citenamefont {Tostevin}, \citenamefont {Volya},\ and\ \citenamefont {Weisshaar}}]{Santiago2011}%
  \BibitemOpen
  \bibfield  {author} {\bibinfo {author} {\bibfnamefont {D.}~\bibnamefont {Santiago-Gonzalez}}, \bibinfo {author} {\bibfnamefont {I.}~\bibnamefont {Wiedenh\"over}}, \bibinfo {author} {\bibfnamefont {V.}~\bibnamefont {Abramkina}}, \bibinfo {author} {\bibfnamefont {M.~L.}\ \bibnamefont {Avila}}, \bibinfo {author} {\bibfnamefont {T.}~\bibnamefont {Baugher}}, \bibinfo {author} {\bibfnamefont {D.}~\bibnamefont {Bazin}}, \bibinfo {author} {\bibfnamefont {B.~A.}\ \bibnamefont {Brown}}, \bibinfo {author} {\bibfnamefont {P.~D.}\ \bibnamefont {Cottle}}, \bibinfo {author} {\bibfnamefont {A.}~\bibnamefont {Gade}}, \bibinfo {author} {\bibfnamefont {T.}~\bibnamefont {Glasmacher}}, \bibinfo {author} {\bibfnamefont {K.~W.}\ \bibnamefont {Kemper}}, \bibinfo {author} {\bibfnamefont {S.}~\bibnamefont {McDaniel}}, \bibinfo {author} {\bibfnamefont {A.}~\bibnamefont {Rojas}}, \bibinfo {author} {\bibfnamefont {A.}~\bibnamefont {Ratkiewicz}}, \bibinfo {author} {\bibfnamefont {R.}~\bibnamefont {Meharchand}}, \bibinfo {author}
  {\bibfnamefont {E.~C.}\ \bibnamefont {Simpson}}, \bibinfo {author} {\bibfnamefont {J.~A.}\ \bibnamefont {Tostevin}}, \bibinfo {author} {\bibfnamefont {A.}~\bibnamefont {Volya}},\ and\ \bibinfo {author} {\bibfnamefont {D.}~\bibnamefont {Weisshaar}},\ }\bibfield  {title} {\bibinfo {title} {Triple configuration coexistence in $^{44}\mathrm{S}$},\ }\href {https://doi.org/10.1103/PhysRevC.83.061305} {\bibfield  {journal} {\bibinfo  {journal} {Phys. Rev. C}\ }\textbf {\bibinfo {volume} {83}},\ \bibinfo {pages} {061305} (\bibinfo {year} {2011})}\BibitemShut {NoStop}%
\bibitem [{\citenamefont {Parker}\ \emph {et~al.}(2017)\citenamefont {Parker}, \citenamefont {Wiedenh\"over}, \citenamefont {Cottle}, \citenamefont {Baker}, \citenamefont {McPherson}, \citenamefont {Riley}, \citenamefont {Santiago-Gonzalez}, \citenamefont {Volya}, \citenamefont {Bader}, \citenamefont {Baugher}, \citenamefont {Bazin}, \citenamefont {Gade}, \citenamefont {Ginter}, \citenamefont {Iwasaki}, \citenamefont {Loelius}, \citenamefont {Morse}, \citenamefont {Recchia}, \citenamefont {Smalley}, \citenamefont {Stroberg}, \citenamefont {Whitmore}, \citenamefont {Weisshaar}, \citenamefont {Lemasson}, \citenamefont {Crawford}, \citenamefont {Macchiavelli},\ and\ \citenamefont {Wimmer}}]{Parker2017}%
  \BibitemOpen
  \bibfield  {author} {\bibinfo {author} {\bibfnamefont {J.~J.}\ \bibnamefont {Parker}}, \bibinfo {author} {\bibfnamefont {I.}~\bibnamefont {Wiedenh\"over}}, \bibinfo {author} {\bibfnamefont {P.~D.}\ \bibnamefont {Cottle}}, \bibinfo {author} {\bibfnamefont {J.}~\bibnamefont {Baker}}, \bibinfo {author} {\bibfnamefont {D.}~\bibnamefont {McPherson}}, \bibinfo {author} {\bibfnamefont {M.~A.}\ \bibnamefont {Riley}}, \bibinfo {author} {\bibfnamefont {D.}~\bibnamefont {Santiago-Gonzalez}}, \bibinfo {author} {\bibfnamefont {A.}~\bibnamefont {Volya}}, \bibinfo {author} {\bibfnamefont {V.~M.}\ \bibnamefont {Bader}}, \bibinfo {author} {\bibfnamefont {T.}~\bibnamefont {Baugher}}, \bibinfo {author} {\bibfnamefont {D.}~\bibnamefont {Bazin}}, \bibinfo {author} {\bibfnamefont {A.}~\bibnamefont {Gade}}, \bibinfo {author} {\bibfnamefont {T.}~\bibnamefont {Ginter}}, \bibinfo {author} {\bibfnamefont {H.}~\bibnamefont {Iwasaki}}, \bibinfo {author} {\bibfnamefont {C.}~\bibnamefont {Loelius}}, \bibinfo {author} {\bibfnamefont
  {C.}~\bibnamefont {Morse}}, \bibinfo {author} {\bibfnamefont {F.}~\bibnamefont {Recchia}}, \bibinfo {author} {\bibfnamefont {D.}~\bibnamefont {Smalley}}, \bibinfo {author} {\bibfnamefont {S.~R.}\ \bibnamefont {Stroberg}}, \bibinfo {author} {\bibfnamefont {K.}~\bibnamefont {Whitmore}}, \bibinfo {author} {\bibfnamefont {D.}~\bibnamefont {Weisshaar}}, \bibinfo {author} {\bibfnamefont {A.}~\bibnamefont {Lemasson}}, \bibinfo {author} {\bibfnamefont {H.~L.}\ \bibnamefont {Crawford}}, \bibinfo {author} {\bibfnamefont {A.~O.}\ \bibnamefont {Macchiavelli}},\ and\ \bibinfo {author} {\bibfnamefont {K.}~\bibnamefont {Wimmer}},\ }\bibfield  {title} {\bibinfo {title} {Isomeric character of the lowest observed ${4}^{+}$ state in $^{44}\mathrm{S}$},\ }\href {https://doi.org/10.1103/PhysRevLett.118.052501} {\bibfield  {journal} {\bibinfo  {journal} {Phys. Rev. Lett.}\ }\textbf {\bibinfo {volume} {118}},\ \bibinfo {pages} {052501} (\bibinfo {year} {2017})}\BibitemShut {NoStop}%
\bibitem [{\citenamefont {Scheit}\ \emph {et~al.}(1996)\citenamefont {Scheit}, \citenamefont {Glasmacher}, \citenamefont {Brown}, \citenamefont {Brown}, \citenamefont {Cottle}, \citenamefont {Hansen}, \citenamefont {Harkewicz}, \citenamefont {Hellstr\"om}, \citenamefont {Ibbotson}, \citenamefont {Jewell}, \citenamefont {Kemper}, \citenamefont {Morrissey}, \citenamefont {Steiner}, \citenamefont {Thirolf},\ and\ \citenamefont {Thoennessen}}]{Scheit1996}%
  \BibitemOpen
  \bibfield  {author} {\bibinfo {author} {\bibfnamefont {H.}~\bibnamefont {Scheit}}, \bibinfo {author} {\bibfnamefont {T.}~\bibnamefont {Glasmacher}}, \bibinfo {author} {\bibfnamefont {B.~A.}\ \bibnamefont {Brown}}, \bibinfo {author} {\bibfnamefont {J.~A.}\ \bibnamefont {Brown}}, \bibinfo {author} {\bibfnamefont {P.~D.}\ \bibnamefont {Cottle}}, \bibinfo {author} {\bibfnamefont {P.~G.}\ \bibnamefont {Hansen}}, \bibinfo {author} {\bibfnamefont {R.}~\bibnamefont {Harkewicz}}, \bibinfo {author} {\bibfnamefont {M.}~\bibnamefont {Hellstr\"om}}, \bibinfo {author} {\bibfnamefont {R.~W.}\ \bibnamefont {Ibbotson}}, \bibinfo {author} {\bibfnamefont {J.~K.}\ \bibnamefont {Jewell}}, \bibinfo {author} {\bibfnamefont {K.~W.}\ \bibnamefont {Kemper}}, \bibinfo {author} {\bibfnamefont {D.~J.}\ \bibnamefont {Morrissey}}, \bibinfo {author} {\bibfnamefont {M.}~\bibnamefont {Steiner}}, \bibinfo {author} {\bibfnamefont {P.}~\bibnamefont {Thirolf}},\ and\ \bibinfo {author} {\bibfnamefont {M.}~\bibnamefont {Thoennessen}},\ }\bibfield
   {title} {\bibinfo {title} {New region of deformation: The neutron-rich sulfur isotopes},\ }\href {https://doi.org/10.1103/PhysRevLett.77.3967} {\bibfield  {journal} {\bibinfo  {journal} {Phys. Rev. Lett.}\ }\textbf {\bibinfo {volume} {77}},\ \bibinfo {pages} {3967} (\bibinfo {year} {1996})}\BibitemShut {NoStop}%
\bibitem [{\citenamefont {Gade}\ \emph {et~al.}(2003)\citenamefont {Gade}, \citenamefont {Bazin}, \citenamefont {Campbell}, \citenamefont {Church}, \citenamefont {Dinca}, \citenamefont {Enders}, \citenamefont {Glasmacher}, \citenamefont {Hu}, \citenamefont {Kemper}, \citenamefont {Mueller}, \citenamefont {Olliver}, \citenamefont {Perry}, \citenamefont {Riley}, \citenamefont {Roeder}, \citenamefont {Sherrill},\ and\ \citenamefont {Terry}}]{Gade2003}%
  \BibitemOpen
  \bibfield  {author} {\bibinfo {author} {\bibfnamefont {A.}~\bibnamefont {Gade}}, \bibinfo {author} {\bibfnamefont {D.}~\bibnamefont {Bazin}}, \bibinfo {author} {\bibfnamefont {C.~M.}\ \bibnamefont {Campbell}}, \bibinfo {author} {\bibfnamefont {J.~A.}\ \bibnamefont {Church}}, \bibinfo {author} {\bibfnamefont {D.~C.}\ \bibnamefont {Dinca}}, \bibinfo {author} {\bibfnamefont {J.}~\bibnamefont {Enders}}, \bibinfo {author} {\bibfnamefont {T.}~\bibnamefont {Glasmacher}}, \bibinfo {author} {\bibfnamefont {Z.}~\bibnamefont {Hu}}, \bibinfo {author} {\bibfnamefont {K.~W.}\ \bibnamefont {Kemper}}, \bibinfo {author} {\bibfnamefont {W.~F.}\ \bibnamefont {Mueller}}, \bibinfo {author} {\bibfnamefont {H.}~\bibnamefont {Olliver}}, \bibinfo {author} {\bibfnamefont {B.~C.}\ \bibnamefont {Perry}}, \bibinfo {author} {\bibfnamefont {L.~A.}\ \bibnamefont {Riley}}, \bibinfo {author} {\bibfnamefont {B.~T.}\ \bibnamefont {Roeder}}, \bibinfo {author} {\bibfnamefont {B.~M.}\ \bibnamefont {Sherrill}},\ and\ \bibinfo {author}
  {\bibfnamefont {J.~R.}\ \bibnamefont {Terry}},\ }\bibfield  {title} {\bibinfo {title} {Detailed experimental study on intermediate-energy coulomb excitation of $^{46}\mathrm{Ar}$},\ }\href {https://doi.org/10.1103/PhysRevC.68.014302} {\bibfield  {journal} {\bibinfo  {journal} {Phys. Rev. C}\ }\textbf {\bibinfo {volume} {68}},\ \bibinfo {pages} {014302} (\bibinfo {year} {2003})}\BibitemShut {NoStop}%
\bibitem [{\citenamefont {Calinescu}\ \emph {et~al.}(2016)\citenamefont {Calinescu}, \citenamefont {C\'aceres}, \citenamefont {Gr\'evy}, \citenamefont {Sorlin}, \citenamefont {Dombr\'adi}, \citenamefont {Stanoiu}, \citenamefont {Astabatyan}, \citenamefont {Borcea}, \citenamefont {Borcea}, \citenamefont {Bowry}, \citenamefont {Catford}, \citenamefont {Cl\'ement}, \citenamefont {Franchoo}, \citenamefont {Garcia}, \citenamefont {Gillibert}, \citenamefont {Guerin}, \citenamefont {Kuti}, \citenamefont {Lukyanov}, \citenamefont {Lepailleur}, \citenamefont {Maslov}, \citenamefont {Morfouace}, \citenamefont {Mrazek}, \citenamefont {Negoita}, \citenamefont {Niikura}, \citenamefont {Perrot}, \citenamefont {Podoly\'ak}, \citenamefont {Petrone}, \citenamefont {Penionzhkevich}, \citenamefont {Roger}, \citenamefont {Rotaru}, \citenamefont {Sohler}, \citenamefont {Stefan}, \citenamefont {Thomas}, \citenamefont {Vajta},\ and\ \citenamefont {Wilson}}]{46Ar}%
  \BibitemOpen
  \bibfield  {author} {\bibinfo {author} {\bibfnamefont {S.}~\bibnamefont {Calinescu}}, \bibinfo {author} {\bibfnamefont {L.}~\bibnamefont {C\'aceres}}, \bibinfo {author} {\bibfnamefont {S.}~\bibnamefont {Gr\'evy}}, \bibinfo {author} {\bibfnamefont {O.}~\bibnamefont {Sorlin}}, \bibinfo {author} {\bibfnamefont {Z.}~\bibnamefont {Dombr\'adi}}, \bibinfo {author} {\bibfnamefont {M.}~\bibnamefont {Stanoiu}}, \bibinfo {author} {\bibfnamefont {R.}~\bibnamefont {Astabatyan}}, \bibinfo {author} {\bibfnamefont {C.}~\bibnamefont {Borcea}}, \bibinfo {author} {\bibfnamefont {R.}~\bibnamefont {Borcea}}, \bibinfo {author} {\bibfnamefont {M.}~\bibnamefont {Bowry}}, \bibinfo {author} {\bibfnamefont {W.}~\bibnamefont {Catford}}, \bibinfo {author} {\bibfnamefont {E.}~\bibnamefont {Cl\'ement}}, \bibinfo {author} {\bibfnamefont {S.}~\bibnamefont {Franchoo}}, \bibinfo {author} {\bibfnamefont {R.}~\bibnamefont {Garcia}}, \bibinfo {author} {\bibfnamefont {R.}~\bibnamefont {Gillibert}}, \bibinfo {author} {\bibfnamefont {I.~H.}\
  \bibnamefont {Guerin}}, \bibinfo {author} {\bibfnamefont {I.}~\bibnamefont {Kuti}}, \bibinfo {author} {\bibfnamefont {S.}~\bibnamefont {Lukyanov}}, \bibinfo {author} {\bibfnamefont {A.}~\bibnamefont {Lepailleur}}, \bibinfo {author} {\bibfnamefont {V.}~\bibnamefont {Maslov}}, \bibinfo {author} {\bibfnamefont {P.}~\bibnamefont {Morfouace}}, \bibinfo {author} {\bibfnamefont {J.}~\bibnamefont {Mrazek}}, \bibinfo {author} {\bibfnamefont {F.}~\bibnamefont {Negoita}}, \bibinfo {author} {\bibfnamefont {M.}~\bibnamefont {Niikura}}, \bibinfo {author} {\bibfnamefont {L.}~\bibnamefont {Perrot}}, \bibinfo {author} {\bibfnamefont {Z.}~\bibnamefont {Podoly\'ak}}, \bibinfo {author} {\bibfnamefont {C.}~\bibnamefont {Petrone}}, \bibinfo {author} {\bibfnamefont {Y.}~\bibnamefont {Penionzhkevich}}, \bibinfo {author} {\bibfnamefont {T.}~\bibnamefont {Roger}}, \bibinfo {author} {\bibfnamefont {F.}~\bibnamefont {Rotaru}}, \bibinfo {author} {\bibfnamefont {D.}~\bibnamefont {Sohler}}, \bibinfo {author} {\bibfnamefont
  {I.}~\bibnamefont {Stefan}}, \bibinfo {author} {\bibfnamefont {J.~C.}\ \bibnamefont {Thomas}}, \bibinfo {author} {\bibfnamefont {Z.}~\bibnamefont {Vajta}},\ and\ \bibinfo {author} {\bibfnamefont {E.}~\bibnamefont {Wilson}},\ }\bibfield  {title} {\bibinfo {title} {Coulomb excitation of $^{44}\mathrm{Ca}$ and $^{46}\mathrm{Ar}$},\ }\href {https://doi.org/10.1103/PhysRevC.93.044333} {\bibfield  {journal} {\bibinfo  {journal} {Phys. Rev. C}\ }\textbf {\bibinfo {volume} {93}},\ \bibinfo {pages} {044333} (\bibinfo {year} {2016})}\BibitemShut {NoStop}%
\bibitem [{\citenamefont {Winkler}\ \emph {et~al.}(2012)\citenamefont {Winkler}, \citenamefont {Gade}, \citenamefont {Baugher}, \citenamefont {Bazin}, \citenamefont {Brown}, \citenamefont {Glasmacher}, \citenamefont {Grinyer}, \citenamefont {Meharchand}, \citenamefont {McDaniel}, \citenamefont {Ratkiewicz},\ and\ \citenamefont {Weisshaar}}]{Winkler2012}%
  \BibitemOpen
  \bibfield  {author} {\bibinfo {author} {\bibfnamefont {R.}~\bibnamefont {Winkler}}, \bibinfo {author} {\bibfnamefont {A.}~\bibnamefont {Gade}}, \bibinfo {author} {\bibfnamefont {T.}~\bibnamefont {Baugher}}, \bibinfo {author} {\bibfnamefont {D.}~\bibnamefont {Bazin}}, \bibinfo {author} {\bibfnamefont {B.~A.}\ \bibnamefont {Brown}}, \bibinfo {author} {\bibfnamefont {T.}~\bibnamefont {Glasmacher}}, \bibinfo {author} {\bibfnamefont {G.~F.}\ \bibnamefont {Grinyer}}, \bibinfo {author} {\bibfnamefont {R.}~\bibnamefont {Meharchand}}, \bibinfo {author} {\bibfnamefont {S.}~\bibnamefont {McDaniel}}, \bibinfo {author} {\bibfnamefont {A.}~\bibnamefont {Ratkiewicz}},\ and\ \bibinfo {author} {\bibfnamefont {D.}~\bibnamefont {Weisshaar}},\ }\bibfield  {title} {\bibinfo {title} {Quadrupole collectivity beyond $n=28$: Intermediate-energy coulomb excitation of $^{47,48}\mathrm{Ar}$},\ }\href {https://doi.org/10.1103/PhysRevLett.108.182501} {\bibfield  {journal} {\bibinfo  {journal} {Phys. Rev. Lett.}\ }\textbf {\bibinfo
  {volume} {108}},\ \bibinfo {pages} {182501} (\bibinfo {year} {2012})}\BibitemShut {NoStop}%
\bibitem [{\citenamefont {Ibbotson}\ \emph {et~al.}(1999)\citenamefont {Ibbotson}, \citenamefont {Glasmacher}, \citenamefont {Mantica},\ and\ \citenamefont {Scheit}}]{Ibbotson1999}%
  \BibitemOpen
  \bibfield  {author} {\bibinfo {author} {\bibfnamefont {R.~W.}\ \bibnamefont {Ibbotson}}, \bibinfo {author} {\bibfnamefont {T.}~\bibnamefont {Glasmacher}}, \bibinfo {author} {\bibfnamefont {P.~F.}\ \bibnamefont {Mantica}},\ and\ \bibinfo {author} {\bibfnamefont {H.}~\bibnamefont {Scheit}},\ }\bibfield  {title} {\bibinfo {title} {Coulomb excitation of odd-a neutron-rich $\ensuremath{\pi}(s\ensuremath{-}d)$ and $\ensuremath{\nu}(f\ensuremath{-}p)$ shell nuclei},\ }\href {https://doi.org/10.1103/PhysRevC.59.642} {\bibfield  {journal} {\bibinfo  {journal} {Phys. Rev. C}\ }\textbf {\bibinfo {volume} {59}},\ \bibinfo {pages} {642} (\bibinfo {year} {1999})}\BibitemShut {NoStop}%
\bibitem [{\citenamefont {Longfellow}\ \emph {et~al.}(2020)\citenamefont {Longfellow}, \citenamefont {Weisshaar}, \citenamefont {Gade}, \citenamefont {Brown}, \citenamefont {Bazin}, \citenamefont {Brown}, \citenamefont {Elman}, \citenamefont {Pereira}, \citenamefont {Rhodes},\ and\ \citenamefont {Spieker}}]{43S_longfellow}%
  \BibitemOpen
  \bibfield  {author} {\bibinfo {author} {\bibfnamefont {B.}~\bibnamefont {Longfellow}}, \bibinfo {author} {\bibfnamefont {D.}~\bibnamefont {Weisshaar}}, \bibinfo {author} {\bibfnamefont {A.}~\bibnamefont {Gade}}, \bibinfo {author} {\bibfnamefont {B.~A.}\ \bibnamefont {Brown}}, \bibinfo {author} {\bibfnamefont {D.}~\bibnamefont {Bazin}}, \bibinfo {author} {\bibfnamefont {K.~W.}\ \bibnamefont {Brown}}, \bibinfo {author} {\bibfnamefont {B.}~\bibnamefont {Elman}}, \bibinfo {author} {\bibfnamefont {J.}~\bibnamefont {Pereira}}, \bibinfo {author} {\bibfnamefont {D.}~\bibnamefont {Rhodes}},\ and\ \bibinfo {author} {\bibfnamefont {M.}~\bibnamefont {Spieker}},\ }\bibfield  {title} {\bibinfo {title} {Shape changes in the $n=28$ island of inversion: Collective structures built on configuration-coexisting states in $^{43}\mathrm{S}$},\ }\href {https://doi.org/10.1103/PhysRevLett.125.232501} {\bibfield  {journal} {\bibinfo  {journal} {Phys. Rev. Lett.}\ }\textbf {\bibinfo {volume} {125}},\ \bibinfo {pages} {232501}
  (\bibinfo {year} {2020})}\BibitemShut {NoStop}%
\bibitem [{\citenamefont {Dombrádi}\ \emph {et~al.}(2003)\citenamefont {Dombrádi}, \citenamefont {Sohler}, \citenamefont {Sorlin}, \citenamefont {Azaiez}, \citenamefont {Nowacki}, \citenamefont {Stanoiu}, \citenamefont {Penionzhkevich}, \citenamefont {Timár}, \citenamefont {Amorini}, \citenamefont {Baiborodin}, \citenamefont {Bauchet}, \citenamefont {Becker}, \citenamefont {Belleguic}, \citenamefont {Borcea}, \citenamefont {Bourgeois}, \citenamefont {Dlouhy}, \citenamefont {Donzaud}, \citenamefont {Duprat}, \citenamefont {Elekes}, \citenamefont {Guillemaud-Mueller}, \citenamefont {Ibrahim}, \citenamefont {Lewitowicz}, \citenamefont {Lopez}, \citenamefont {Lucas}, \citenamefont {Lukyanov}, \citenamefont {Maslov}, \citenamefont {Moore}, \citenamefont {Mrazek}, \citenamefont {Saint-Laurent}, \citenamefont {Sarazin}, \citenamefont {Scarpaci}, \citenamefont {Sletten}, \citenamefont {Stodel}, \citenamefont {Taylor}, \citenamefont {Theisen},\ and\ \citenamefont {Voltolini}}]{Dombradi2003}%
  \BibitemOpen
  \bibfield  {author} {\bibinfo {author} {\bibfnamefont {Z.}~\bibnamefont {Dombrádi}}, \bibinfo {author} {\bibfnamefont {D.}~\bibnamefont {Sohler}}, \bibinfo {author} {\bibfnamefont {O.}~\bibnamefont {Sorlin}}, \bibinfo {author} {\bibfnamefont {F.}~\bibnamefont {Azaiez}}, \bibinfo {author} {\bibfnamefont {F.}~\bibnamefont {Nowacki}}, \bibinfo {author} {\bibfnamefont {M.}~\bibnamefont {Stanoiu}}, \bibinfo {author} {\bibfnamefont {Y.-E.}\ \bibnamefont {Penionzhkevich}}, \bibinfo {author} {\bibfnamefont {J.}~\bibnamefont {Timár}}, \bibinfo {author} {\bibfnamefont {F.}~\bibnamefont {Amorini}}, \bibinfo {author} {\bibfnamefont {D.}~\bibnamefont {Baiborodin}}, \bibinfo {author} {\bibfnamefont {A.}~\bibnamefont {Bauchet}}, \bibinfo {author} {\bibfnamefont {F.}~\bibnamefont {Becker}}, \bibinfo {author} {\bibfnamefont {M.}~\bibnamefont {Belleguic}}, \bibinfo {author} {\bibfnamefont {C.}~\bibnamefont {Borcea}}, \bibinfo {author} {\bibfnamefont {C.}~\bibnamefont {Bourgeois}}, \bibinfo {author} {\bibfnamefont
  {Z.}~\bibnamefont {Dlouhy}}, \bibinfo {author} {\bibfnamefont {C.}~\bibnamefont {Donzaud}}, \bibinfo {author} {\bibfnamefont {J.}~\bibnamefont {Duprat}}, \bibinfo {author} {\bibfnamefont {Z.}~\bibnamefont {Elekes}}, \bibinfo {author} {\bibfnamefont {D.}~\bibnamefont {Guillemaud-Mueller}}, \bibinfo {author} {\bibfnamefont {F.}~\bibnamefont {Ibrahim}}, \bibinfo {author} {\bibfnamefont {M.}~\bibnamefont {Lewitowicz}}, \bibinfo {author} {\bibfnamefont {M.}~\bibnamefont {Lopez}}, \bibinfo {author} {\bibfnamefont {R.}~\bibnamefont {Lucas}}, \bibinfo {author} {\bibfnamefont {S.}~\bibnamefont {Lukyanov}}, \bibinfo {author} {\bibfnamefont {V.}~\bibnamefont {Maslov}}, \bibinfo {author} {\bibfnamefont {C.}~\bibnamefont {Moore}}, \bibinfo {author} {\bibfnamefont {J.}~\bibnamefont {Mrazek}}, \bibinfo {author} {\bibfnamefont {M.}~\bibnamefont {Saint-Laurent}}, \bibinfo {author} {\bibfnamefont {F.}~\bibnamefont {Sarazin}}, \bibinfo {author} {\bibfnamefont {J.}~\bibnamefont {Scarpaci}}, \bibinfo {author} {\bibfnamefont
  {G.}~\bibnamefont {Sletten}}, \bibinfo {author} {\bibfnamefont {C.}~\bibnamefont {Stodel}}, \bibinfo {author} {\bibfnamefont {M.}~\bibnamefont {Taylor}}, \bibinfo {author} {\bibfnamefont {C.}~\bibnamefont {Theisen}},\ and\ \bibinfo {author} {\bibfnamefont {G.}~\bibnamefont {Voltolini}},\ }\bibfield  {title} {\bibinfo {title} {Search for particle–hole excitations across the n=28 shell gap in 45,46ar nuclei},\ }\href {https://doi.org/https://doi.org/10.1016/j.nuclphysa.2003.08.011} {\bibfield  {journal} {\bibinfo  {journal} {Nuclear Physics A}\ }\textbf {\bibinfo {volume} {727}},\ \bibinfo {pages} {195} (\bibinfo {year} {2003})}\BibitemShut {NoStop}%
\bibitem [{\citenamefont {Riley}\ \emph {et~al.}(2005)\citenamefont {Riley}, \citenamefont {Abdelqader}, \citenamefont {Bazin}, \citenamefont {Bojazi}, \citenamefont {Brown}, \citenamefont {Campbell}, \citenamefont {Church}, \citenamefont {Cottle}, \citenamefont {Dinca}, \citenamefont {Enders}, \citenamefont {Gade}, \citenamefont {Glasmacher}, \citenamefont {Honma}, \citenamefont {Horibe}, \citenamefont {Hu}, \citenamefont {Kemper}, \citenamefont {Mueller}, \citenamefont {Olliver}, \citenamefont {Otsuka}, \citenamefont {Perry}, \citenamefont {Roeder}, \citenamefont {Sherrill}, \citenamefont {Spencer},\ and\ \citenamefont {Terry}}]{Riley2005}%
  \BibitemOpen
  \bibfield  {author} {\bibinfo {author} {\bibfnamefont {L.~A.}\ \bibnamefont {Riley}}, \bibinfo {author} {\bibfnamefont {M.~A.}\ \bibnamefont {Abdelqader}}, \bibinfo {author} {\bibfnamefont {D.}~\bibnamefont {Bazin}}, \bibinfo {author} {\bibfnamefont {M.~J.}\ \bibnamefont {Bojazi}}, \bibinfo {author} {\bibfnamefont {B.~A.}\ \bibnamefont {Brown}}, \bibinfo {author} {\bibfnamefont {C.~M.}\ \bibnamefont {Campbell}}, \bibinfo {author} {\bibfnamefont {J.~A.}\ \bibnamefont {Church}}, \bibinfo {author} {\bibfnamefont {P.~D.}\ \bibnamefont {Cottle}}, \bibinfo {author} {\bibfnamefont {D.~C.}\ \bibnamefont {Dinca}}, \bibinfo {author} {\bibfnamefont {J.}~\bibnamefont {Enders}}, \bibinfo {author} {\bibfnamefont {A.}~\bibnamefont {Gade}}, \bibinfo {author} {\bibfnamefont {T.}~\bibnamefont {Glasmacher}}, \bibinfo {author} {\bibfnamefont {M.}~\bibnamefont {Honma}}, \bibinfo {author} {\bibfnamefont {S.}~\bibnamefont {Horibe}}, \bibinfo {author} {\bibfnamefont {Z.}~\bibnamefont {Hu}}, \bibinfo {author} {\bibfnamefont
  {K.~W.}\ \bibnamefont {Kemper}}, \bibinfo {author} {\bibfnamefont {W.~F.}\ \bibnamefont {Mueller}}, \bibinfo {author} {\bibfnamefont {H.}~\bibnamefont {Olliver}}, \bibinfo {author} {\bibfnamefont {T.}~\bibnamefont {Otsuka}}, \bibinfo {author} {\bibfnamefont {B.~C.}\ \bibnamefont {Perry}}, \bibinfo {author} {\bibfnamefont {B.~T.}\ \bibnamefont {Roeder}}, \bibinfo {author} {\bibfnamefont {B.~M.}\ \bibnamefont {Sherrill}}, \bibinfo {author} {\bibfnamefont {T.~P.}\ \bibnamefont {Spencer}},\ and\ \bibinfo {author} {\bibfnamefont {J.~R.}\ \bibnamefont {Terry}},\ }\bibfield  {title} {\bibinfo {title} {Thick-target inverse-kinematics proton scattering from $^{46}\mathrm{Ar}$ and the $n=28$ shell below $^{48}\mathrm{Ca}$},\ }\href {https://doi.org/10.1103/PhysRevC.72.024311} {\bibfield  {journal} {\bibinfo  {journal} {Phys. Rev. C}\ }\textbf {\bibinfo {volume} {72}},\ \bibinfo {pages} {024311} (\bibinfo {year} {2005})}\BibitemShut {NoStop}%
\bibitem [{\citenamefont {Nowak}\ \emph {et~al.}(2016)\citenamefont {Nowak}, \citenamefont {Wimmer}, \citenamefont {Hellgartner}, \citenamefont {M\"ucher}, \citenamefont {Bildstein}, \citenamefont {Diriken}, \citenamefont {Elseviers}, \citenamefont {Gaffney}, \citenamefont {Gernh\"auser}, \citenamefont {Iwanicki}, \citenamefont {Johansen}, \citenamefont {Huyse}, \citenamefont {Konki}, \citenamefont {Kr\"oll}, \citenamefont {Kr\"ucken}, \citenamefont {Lutter}, \citenamefont {Orlandi}, \citenamefont {Pakarinen}, \citenamefont {Raabe}, \citenamefont {Reiter}, \citenamefont {Roger}, \citenamefont {Schrieder}, \citenamefont {Seidlitz}, \citenamefont {Sorlin}, \citenamefont {Van~Duppen}, \citenamefont {Warr}, \citenamefont {De~Witte},\ and\ \citenamefont {Zieli\ifmmode~\acute{n}\else \'{n}\fi{}ska}}]{Nowak2016}%
  \BibitemOpen
  \bibfield  {author} {\bibinfo {author} {\bibfnamefont {K.}~\bibnamefont {Nowak}}, \bibinfo {author} {\bibfnamefont {K.}~\bibnamefont {Wimmer}}, \bibinfo {author} {\bibfnamefont {S.}~\bibnamefont {Hellgartner}}, \bibinfo {author} {\bibfnamefont {D.}~\bibnamefont {M\"ucher}}, \bibinfo {author} {\bibfnamefont {V.}~\bibnamefont {Bildstein}}, \bibinfo {author} {\bibfnamefont {J.}~\bibnamefont {Diriken}}, \bibinfo {author} {\bibfnamefont {J.}~\bibnamefont {Elseviers}}, \bibinfo {author} {\bibfnamefont {L.~P.}\ \bibnamefont {Gaffney}}, \bibinfo {author} {\bibfnamefont {R.}~\bibnamefont {Gernh\"auser}}, \bibinfo {author} {\bibfnamefont {J.}~\bibnamefont {Iwanicki}}, \bibinfo {author} {\bibfnamefont {J.~G.}\ \bibnamefont {Johansen}}, \bibinfo {author} {\bibfnamefont {M.}~\bibnamefont {Huyse}}, \bibinfo {author} {\bibfnamefont {J.}~\bibnamefont {Konki}}, \bibinfo {author} {\bibfnamefont {T.}~\bibnamefont {Kr\"oll}}, \bibinfo {author} {\bibfnamefont {R.}~\bibnamefont {Kr\"ucken}}, \bibinfo {author} {\bibfnamefont
  {R.}~\bibnamefont {Lutter}}, \bibinfo {author} {\bibfnamefont {R.}~\bibnamefont {Orlandi}}, \bibinfo {author} {\bibfnamefont {J.}~\bibnamefont {Pakarinen}}, \bibinfo {author} {\bibfnamefont {R.}~\bibnamefont {Raabe}}, \bibinfo {author} {\bibfnamefont {P.}~\bibnamefont {Reiter}}, \bibinfo {author} {\bibfnamefont {T.}~\bibnamefont {Roger}}, \bibinfo {author} {\bibfnamefont {G.}~\bibnamefont {Schrieder}}, \bibinfo {author} {\bibfnamefont {M.}~\bibnamefont {Seidlitz}}, \bibinfo {author} {\bibfnamefont {O.}~\bibnamefont {Sorlin}}, \bibinfo {author} {\bibfnamefont {P.}~\bibnamefont {Van~Duppen}}, \bibinfo {author} {\bibfnamefont {N.}~\bibnamefont {Warr}}, \bibinfo {author} {\bibfnamefont {H.}~\bibnamefont {De~Witte}},\ and\ \bibinfo {author} {\bibfnamefont {M.}~\bibnamefont {Zieli\ifmmode~\acute{n}\else \'{n}\fi{}ska}},\ }\bibfield  {title} {\bibinfo {title} {Spectroscopy of $^{46}\mathrm{Ar}$ by the $(t,\phantom{\rule{0.16em}{0ex}}p)$ two-neutron transfer reaction},\ }\href
  {https://doi.org/10.1103/PhysRevC.93.044335} {\bibfield  {journal} {\bibinfo  {journal} {Phys. Rev. C}\ }\textbf {\bibinfo {volume} {93}},\ \bibinfo {pages} {044335} (\bibinfo {year} {2016})}\BibitemShut {NoStop}%
\bibitem [{\citenamefont {Lubna}\ \emph {et~al.}(2020)\citenamefont {Lubna}, \citenamefont {Kravvaris}, \citenamefont {Tabor}, \citenamefont {Tripathi}, \citenamefont {Rubino},\ and\ \citenamefont {Volya}}]{FSU}%
  \BibitemOpen
  \bibfield  {author} {\bibinfo {author} {\bibfnamefont {R.~S.}\ \bibnamefont {Lubna}}, \bibinfo {author} {\bibfnamefont {K.}~\bibnamefont {Kravvaris}}, \bibinfo {author} {\bibfnamefont {S.~L.}\ \bibnamefont {Tabor}}, \bibinfo {author} {\bibfnamefont {V.}~\bibnamefont {Tripathi}}, \bibinfo {author} {\bibfnamefont {E.}~\bibnamefont {Rubino}},\ and\ \bibinfo {author} {\bibfnamefont {A.}~\bibnamefont {Volya}},\ }\bibfield  {title} {\bibinfo {title} {Evolution of the $n=20$ and 28 shell gaps and two-particle-two-hole states in the fsu interaction},\ }\href {https://doi.org/10.1103/PhysRevResearch.2.043342} {\bibfield  {journal} {\bibinfo  {journal} {Phys. Rev. Res.}\ }\textbf {\bibinfo {volume} {2}},\ \bibinfo {pages} {043342} (\bibinfo {year} {2020})}\BibitemShut {NoStop}%
\bibitem [{\citenamefont {Yoshida}\ \emph {et~al.}(2018)\citenamefont {Yoshida}, \citenamefont {Utsuno}, \citenamefont {Shimizu},\ and\ \citenamefont {Otsuka}}]{utsuno_new}%
  \BibitemOpen
  \bibfield  {author} {\bibinfo {author} {\bibfnamefont {S.}~\bibnamefont {Yoshida}}, \bibinfo {author} {\bibfnamefont {Y.}~\bibnamefont {Utsuno}}, \bibinfo {author} {\bibfnamefont {N.}~\bibnamefont {Shimizu}},\ and\ \bibinfo {author} {\bibfnamefont {T.}~\bibnamefont {Otsuka}},\ }\bibfield  {title} {\bibinfo {title} {Systematic shell-model study of $\ensuremath{\beta}$-decay properties and gamow-teller strength distributions in $a\ensuremath{\approx}40$ neutron-rich nuclei},\ }\href {https://doi.org/10.1103/PhysRevC.97.054321} {\bibfield  {journal} {\bibinfo  {journal} {Phys. Rev. C}\ }\textbf {\bibinfo {volume} {97}},\ \bibinfo {pages} {054321} (\bibinfo {year} {2018})}\BibitemShut {NoStop}%
\bibitem [{\citenamefont {Gade}\ and\ \citenamefont {Sherill}(2016)}]{brad}%
  \BibitemOpen
  \bibfield  {author} {\bibinfo {author} {\bibfnamefont {A.}~\bibnamefont {Gade}}\ and\ \bibinfo {author} {\bibfnamefont {B.}~\bibnamefont {Sherill}},\ }\href@noop {} {\bibfield  {journal} {\bibinfo  {journal} {Phys. Scr.}\ }\textbf {\bibinfo {volume} {91}},\ \bibinfo {pages} {053003} (\bibinfo {year} {2016})}\BibitemShut {NoStop}%
\bibitem [{\citenamefont {Morrissey}\ \emph {et~al.}(2003)\citenamefont {Morrissey}, \citenamefont {Sherrill}, \citenamefont {Steiner}, \citenamefont {Stolz},\ and\ \citenamefont {Wiedenhoever}}]{A1900}%
  \BibitemOpen
  \bibfield  {author} {\bibinfo {author} {\bibfnamefont {D.}~\bibnamefont {Morrissey}}, \bibinfo {author} {\bibfnamefont {B.}~\bibnamefont {Sherrill}}, \bibinfo {author} {\bibfnamefont {M.}~\bibnamefont {Steiner}}, \bibinfo {author} {\bibfnamefont {A.}~\bibnamefont {Stolz}},\ and\ \bibinfo {author} {\bibfnamefont {I.}~\bibnamefont {Wiedenhoever}},\ }\bibfield  {title} {\bibinfo {title} {Commissioning the a1900 projectile fragment separator},\ }\href {https://doi.org/https://doi.org/10.1016/S0168-583X(02)01895-5} {\bibfield  {journal} {\bibinfo  {journal} {Nuclear Instruments and Methods in Physics Research Section B: Beam Interactions with Materials and Atoms}\ }\textbf {\bibinfo {volume} {204}},\ \bibinfo {pages} {90} (\bibinfo {year} {2003})},\ \bibinfo {note} {14th International Conference on Electromagnetic Isotope Separators and Techniques Related to their Applications}\BibitemShut {NoStop}%
\bibitem [{\citenamefont {Tripathi}\ \emph {et~al.}(2022)\citenamefont {Tripathi}, \citenamefont {Bhattacharya}, \citenamefont {Rubino}, \citenamefont {Benetti}, \citenamefont {Perello}, \citenamefont {Tabor}, \citenamefont {Liddick}, \citenamefont {Bender}, \citenamefont {Carpenter}, \citenamefont {Carroll}, \citenamefont {Chester}, \citenamefont {Chiara}, \citenamefont {Childers}, \citenamefont {Clark}, \citenamefont {Crider}, \citenamefont {Harke}, \citenamefont {Longfellow}, \citenamefont {Lubna}, \citenamefont {Luitel}, \citenamefont {Ogunbeku}, \citenamefont {Richard}, \citenamefont {Saha}, \citenamefont {Shimizu}, \citenamefont {Shehu}, \citenamefont {Utsuno}, \citenamefont {Unz}, \citenamefont {Xiao}, \citenamefont {Yoshida},\ and\ \citenamefont {Zhu}}]{tripathi_new}%
  \BibitemOpen
  \bibfield  {author} {\bibinfo {author} {\bibfnamefont {V.}~\bibnamefont {Tripathi}}, \bibinfo {author} {\bibfnamefont {S.}~\bibnamefont {Bhattacharya}}, \bibinfo {author} {\bibfnamefont {E.}~\bibnamefont {Rubino}}, \bibinfo {author} {\bibfnamefont {C.}~\bibnamefont {Benetti}}, \bibinfo {author} {\bibfnamefont {J.~F.}\ \bibnamefont {Perello}}, \bibinfo {author} {\bibfnamefont {S.~L.}\ \bibnamefont {Tabor}}, \bibinfo {author} {\bibfnamefont {S.~N.}\ \bibnamefont {Liddick}}, \bibinfo {author} {\bibfnamefont {P.~C.}\ \bibnamefont {Bender}}, \bibinfo {author} {\bibfnamefont {M.~P.}\ \bibnamefont {Carpenter}}, \bibinfo {author} {\bibfnamefont {J.~J.}\ \bibnamefont {Carroll}}, \bibinfo {author} {\bibfnamefont {A.}~\bibnamefont {Chester}}, \bibinfo {author} {\bibfnamefont {C.~J.}\ \bibnamefont {Chiara}}, \bibinfo {author} {\bibfnamefont {K.}~\bibnamefont {Childers}}, \bibinfo {author} {\bibfnamefont {B.~R.}\ \bibnamefont {Clark}}, \bibinfo {author} {\bibfnamefont {B.~P.}\ \bibnamefont {Crider}}, \bibinfo {author}
  {\bibfnamefont {J.~T.}\ \bibnamefont {Harke}}, \bibinfo {author} {\bibfnamefont {B.}~\bibnamefont {Longfellow}}, \bibinfo {author} {\bibfnamefont {R.~S.}\ \bibnamefont {Lubna}}, \bibinfo {author} {\bibfnamefont {S.}~\bibnamefont {Luitel}}, \bibinfo {author} {\bibfnamefont {T.~H.}\ \bibnamefont {Ogunbeku}}, \bibinfo {author} {\bibfnamefont {A.~L.}\ \bibnamefont {Richard}}, \bibinfo {author} {\bibfnamefont {S.}~\bibnamefont {Saha}}, \bibinfo {author} {\bibfnamefont {N.}~\bibnamefont {Shimizu}}, \bibinfo {author} {\bibfnamefont {O.~A.}\ \bibnamefont {Shehu}}, \bibinfo {author} {\bibfnamefont {Y.}~\bibnamefont {Utsuno}}, \bibinfo {author} {\bibfnamefont {R.}~\bibnamefont {Unz}}, \bibinfo {author} {\bibfnamefont {Y.}~\bibnamefont {Xiao}}, \bibinfo {author} {\bibfnamefont {S.}~\bibnamefont {Yoshida}},\ and\ \bibinfo {author} {\bibfnamefont {Y.}~\bibnamefont {Zhu}},\ }\bibfield  {title} {\bibinfo {title} {${\ensuremath{\beta}}^{\ensuremath{-}}$ decay of exotic p and s isotopes with neutron number near 28},\ }\href
  {https://doi.org/10.1103/PhysRevC.106.064314} {\bibfield  {journal} {\bibinfo  {journal} {Phys. Rev. C}\ }\textbf {\bibinfo {volume} {106}},\ \bibinfo {pages} {064314} (\bibinfo {year} {2022})}\BibitemShut {NoStop}%
\bibitem [{\citenamefont {Bhattacharya}\ \emph {et~al.}(2023)\citenamefont {Bhattacharya}, \citenamefont {Tripathi}, \citenamefont {Tabor}, \citenamefont {Volya}, \citenamefont {Bender}, \citenamefont {Benetti}, \citenamefont {Carpenter}, \citenamefont {Carroll}, \citenamefont {Chester}, \citenamefont {Chiara}, \citenamefont {Childers}, \citenamefont {Clark}, \citenamefont {Crider}, \citenamefont {Harke}, \citenamefont {Jain}, \citenamefont {Liddick}, \citenamefont {Lubna}, \citenamefont {Luitel}, \citenamefont {Longfellow}, \citenamefont {Mogannam}, \citenamefont {Ogunbeku}, \citenamefont {Perello}, \citenamefont {Richard}, \citenamefont {Rubino}, \citenamefont {Saha}, \citenamefont {Shehu}, \citenamefont {Unz}, \citenamefont {Xiao},\ and\ \citenamefont {Zhu}}]{Soumik2023}%
  \BibitemOpen
  \bibfield  {author} {\bibinfo {author} {\bibfnamefont {S.}~\bibnamefont {Bhattacharya}}, \bibinfo {author} {\bibfnamefont {V.}~\bibnamefont {Tripathi}}, \bibinfo {author} {\bibfnamefont {S.~L.}\ \bibnamefont {Tabor}}, \bibinfo {author} {\bibfnamefont {A.}~\bibnamefont {Volya}}, \bibinfo {author} {\bibfnamefont {P.~C.}\ \bibnamefont {Bender}}, \bibinfo {author} {\bibfnamefont {C.}~\bibnamefont {Benetti}}, \bibinfo {author} {\bibfnamefont {M.~P.}\ \bibnamefont {Carpenter}}, \bibinfo {author} {\bibfnamefont {J.~J.}\ \bibnamefont {Carroll}}, \bibinfo {author} {\bibfnamefont {A.}~\bibnamefont {Chester}}, \bibinfo {author} {\bibfnamefont {C.~J.}\ \bibnamefont {Chiara}}, \bibinfo {author} {\bibfnamefont {K.}~\bibnamefont {Childers}}, \bibinfo {author} {\bibfnamefont {B.~R.}\ \bibnamefont {Clark}}, \bibinfo {author} {\bibfnamefont {B.~P.}\ \bibnamefont {Crider}}, \bibinfo {author} {\bibfnamefont {J.~T.}\ \bibnamefont {Harke}}, \bibinfo {author} {\bibfnamefont {R.}~\bibnamefont {Jain}}, \bibinfo {author}
  {\bibfnamefont {S.~N.}\ \bibnamefont {Liddick}}, \bibinfo {author} {\bibfnamefont {R.~S.}\ \bibnamefont {Lubna}}, \bibinfo {author} {\bibfnamefont {S.}~\bibnamefont {Luitel}}, \bibinfo {author} {\bibfnamefont {B.}~\bibnamefont {Longfellow}}, \bibinfo {author} {\bibfnamefont {M.~J.}\ \bibnamefont {Mogannam}}, \bibinfo {author} {\bibfnamefont {T.~H.}\ \bibnamefont {Ogunbeku}}, \bibinfo {author} {\bibfnamefont {J.}~\bibnamefont {Perello}}, \bibinfo {author} {\bibfnamefont {A.~L.}\ \bibnamefont {Richard}}, \bibinfo {author} {\bibfnamefont {E.}~\bibnamefont {Rubino}}, \bibinfo {author} {\bibfnamefont {S.}~\bibnamefont {Saha}}, \bibinfo {author} {\bibfnamefont {O.~A.}\ \bibnamefont {Shehu}}, \bibinfo {author} {\bibfnamefont {R.}~\bibnamefont {Unz}}, \bibinfo {author} {\bibfnamefont {Y.}~\bibnamefont {Xiao}},\ and\ \bibinfo {author} {\bibfnamefont {Y.}~\bibnamefont {Zhu}},\ }\bibfield  {title} {\bibinfo {title} {${\ensuremath{\beta}}^{\ensuremath{-}}$ decay of neutron-rich $^{45}\mathrm{Cl}$ located at the magic
  number $n=28$},\ }\href {https://doi.org/10.1103/PhysRevC.108.024312} {\bibfield  {journal} {\bibinfo  {journal} {Phys. Rev. C}\ }\textbf {\bibinfo {volume} {108}},\ \bibinfo {pages} {024312} (\bibinfo {year} {2023})}\BibitemShut {NoStop}%
\bibitem [{\citenamefont {Tripathi}\ \emph {et~al.}(2024)\citenamefont {Tripathi}, \citenamefont {Bhattacharya}, \citenamefont {Rubino}, \citenamefont {Benetti}, \citenamefont {Perello}, \citenamefont {Tabor}, \citenamefont {Liddick}, \citenamefont {Bender}, \citenamefont {Carpenter}, \citenamefont {Carroll}, \citenamefont {Chester}, \citenamefont {Chiara}, \citenamefont {Childers}, \citenamefont {Clark}, \citenamefont {Crider}, \citenamefont {Harke}, \citenamefont {Jain}, \citenamefont {Longfellow}, \citenamefont {Luitel}, \citenamefont {Mogannam}, \citenamefont {Ogunbeku}, \citenamefont {Richard}, \citenamefont {Saha}, \citenamefont {Shimizu}, \citenamefont {Shehu}, \citenamefont {Utsuno}, \citenamefont {Unz}, \citenamefont {Xiao}, \citenamefont {Yoshida},\ and\ \citenamefont {Zhu}}]{tripathi_submitted}%
  \BibitemOpen
  \bibfield  {author} {\bibinfo {author} {\bibfnamefont {V.}~\bibnamefont {Tripathi}}, \bibinfo {author} {\bibfnamefont {S.}~\bibnamefont {Bhattacharya}}, \bibinfo {author} {\bibfnamefont {E.}~\bibnamefont {Rubino}}, \bibinfo {author} {\bibfnamefont {C.}~\bibnamefont {Benetti}}, \bibinfo {author} {\bibfnamefont {J.~F.}\ \bibnamefont {Perello}}, \bibinfo {author} {\bibfnamefont {S.~L.}\ \bibnamefont {Tabor}}, \bibinfo {author} {\bibfnamefont {S.~N.}\ \bibnamefont {Liddick}}, \bibinfo {author} {\bibfnamefont {P.~C.}\ \bibnamefont {Bender}}, \bibinfo {author} {\bibfnamefont {M.~P.}\ \bibnamefont {Carpenter}}, \bibinfo {author} {\bibfnamefont {J.~J.}\ \bibnamefont {Carroll}}, \bibinfo {author} {\bibfnamefont {A.}~\bibnamefont {Chester}}, \bibinfo {author} {\bibfnamefont {C.~J.}\ \bibnamefont {Chiara}}, \bibinfo {author} {\bibfnamefont {K.}~\bibnamefont {Childers}}, \bibinfo {author} {\bibfnamefont {B.~R.}\ \bibnamefont {Clark}}, \bibinfo {author} {\bibfnamefont {B.~P.}\ \bibnamefont {Crider}}, \bibinfo {author}
  {\bibfnamefont {J.~T.}\ \bibnamefont {Harke}}, \bibinfo {author} {\bibfnamefont {R.}~\bibnamefont {Jain}}, \bibinfo {author} {\bibfnamefont {B.}~\bibnamefont {Longfellow}}, \bibinfo {author} {\bibfnamefont {S.}~\bibnamefont {Luitel}}, \bibinfo {author} {\bibfnamefont {M.}~\bibnamefont {Mogannam}}, \bibinfo {author} {\bibfnamefont {T.~H.}\ \bibnamefont {Ogunbeku}}, \bibinfo {author} {\bibfnamefont {A.~L.}\ \bibnamefont {Richard}}, \bibinfo {author} {\bibfnamefont {S.}~\bibnamefont {Saha}}, \bibinfo {author} {\bibfnamefont {N.}~\bibnamefont {Shimizu}}, \bibinfo {author} {\bibfnamefont {O.~A.}\ \bibnamefont {Shehu}}, \bibinfo {author} {\bibfnamefont {Y.}~\bibnamefont {Utsuno}}, \bibinfo {author} {\bibfnamefont {R.}~\bibnamefont {Unz}}, \bibinfo {author} {\bibfnamefont {Y.}~\bibnamefont {Xiao}}, \bibinfo {author} {\bibfnamefont {S.}~\bibnamefont {Yoshida}},\ and\ \bibinfo {author} {\bibfnamefont {Y.}~\bibnamefont {Zhu}},\ }\bibfield  {title} {\bibinfo {title} {Low spin spectroscopy of neutron-rich
  $^{43,44,45}\mathrm{Cl}$ via ${\ensuremath{\beta}}^{\ensuremath{-}}$ and $\ensuremath{\beta}n$ decay},\ }\href {https://doi.org/10.1103/PhysRevC.109.044320} {\bibfield  {journal} {\bibinfo  {journal} {Phys. Rev. C}\ }\textbf {\bibinfo {volume} {109}},\ \bibinfo {pages} {044320} (\bibinfo {year} {2024})}\BibitemShut {NoStop}%
\bibitem [{\citenamefont {Prisciandaro}\ \emph {et~al.}(2003)\citenamefont {Prisciandaro}, \citenamefont {Morton},\ and\ \citenamefont {Mantica}}]{BCS}%
  \BibitemOpen
  \bibfield  {author} {\bibinfo {author} {\bibfnamefont {J.}~\bibnamefont {Prisciandaro}}, \bibinfo {author} {\bibfnamefont {A.}~\bibnamefont {Morton}},\ and\ \bibinfo {author} {\bibfnamefont {P.}~\bibnamefont {Mantica}},\ }\bibfield  {title} {\bibinfo {title} {Beta counting system for fast fragmentation beams},\ }\href {https://doi.org/https://doi.org/10.1016/S0168-9002(03)01037-4} {\bibfield  {journal} {\bibinfo  {journal} {Nuclear Instruments and Methods in Physics Research Section A: Accelerators, Spectrometers, Detectors and Associated Equipment}\ }\textbf {\bibinfo {volume} {505}},\ \bibinfo {pages} {140} (\bibinfo {year} {2003})},\ \bibinfo {note} {proceedings of the tenth Symposium on Radiation Measurements and Applications}\BibitemShut {NoStop}%
\bibitem [{\citenamefont {Prokop}\ \emph {et~al.}(2014)\citenamefont {Prokop}, \citenamefont {Liddick}, \citenamefont {Abromeit}, \citenamefont {Chemey}, \citenamefont {Larson}, \citenamefont {Suchyta},\ and\ \citenamefont {Tompkins}}]{prokop}%
  \BibitemOpen
  \bibfield  {author} {\bibinfo {author} {\bibfnamefont {C.}~\bibnamefont {Prokop}}, \bibinfo {author} {\bibfnamefont {S.}~\bibnamefont {Liddick}}, \bibinfo {author} {\bibfnamefont {B.}~\bibnamefont {Abromeit}}, \bibinfo {author} {\bibfnamefont {A.}~\bibnamefont {Chemey}}, \bibinfo {author} {\bibfnamefont {N.}~\bibnamefont {Larson}}, \bibinfo {author} {\bibfnamefont {S.}~\bibnamefont {Suchyta}},\ and\ \bibinfo {author} {\bibfnamefont {J.}~\bibnamefont {Tompkins}},\ }\bibfield  {title} {\bibinfo {title} {Digital data acquisition system implementation at the national superconducting cyclotron laboratory},\ }\href {https://doi.org/https://doi.org/10.1016/j.nima.2013.12.044} {\bibfield  {journal} {\bibinfo  {journal} {Nuclear Instruments and Methods in Physics Research Section A: Accelerators, Spectrometers, Detectors and Associated Equipment}\ }\textbf {\bibinfo {volume} {741}},\ \bibinfo {pages} {163} (\bibinfo {year} {2014})}\BibitemShut {NoStop}%
\bibitem [{\citenamefont {NNDC}()}]{nndc}%
  \BibitemOpen
  \bibfield  {author} {\bibinfo {author} {\bibnamefont {NNDC}},\ }\href {https://www.nndc.bnl.gov/nudat3} {\bibinfo {title} {https://www.nndc.bnl.gov/nudat3/}}\BibitemShut {NoStop}%
\bibitem [{\citenamefont {Sorlin}\ \emph {et~al.}(1993)\citenamefont {Sorlin}, \citenamefont {Guillemaud-Mueller}, \citenamefont {Mueller}, \citenamefont {Borrel}, \citenamefont {Dogny}, \citenamefont {Pougheon}, \citenamefont {Kratz}, \citenamefont {Gabelmann}, \citenamefont {Pfeiffer}, \citenamefont {W\"ohr}, \citenamefont {Ziegert}, \citenamefont {Penionzhkevich}, \citenamefont {Lukyanov}, \citenamefont {Salamatin}, \citenamefont {Anne}, \citenamefont {Borcea}, \citenamefont {Fifield}, \citenamefont {Lewitowicz}, \citenamefont {Saint-Laurent}, \citenamefont {Bazin}, \citenamefont {D\'etraz}, \citenamefont {Thielemann},\ and\ \citenamefont {Hillebrandt}}]{Sorlin1993}%
  \BibitemOpen
  \bibfield  {author} {\bibinfo {author} {\bibfnamefont {O.}~\bibnamefont {Sorlin}}, \bibinfo {author} {\bibfnamefont {D.}~\bibnamefont {Guillemaud-Mueller}}, \bibinfo {author} {\bibfnamefont {A.~C.}\ \bibnamefont {Mueller}}, \bibinfo {author} {\bibfnamefont {V.}~\bibnamefont {Borrel}}, \bibinfo {author} {\bibfnamefont {S.}~\bibnamefont {Dogny}}, \bibinfo {author} {\bibfnamefont {F.}~\bibnamefont {Pougheon}}, \bibinfo {author} {\bibfnamefont {K.-L.}\ \bibnamefont {Kratz}}, \bibinfo {author} {\bibfnamefont {H.}~\bibnamefont {Gabelmann}}, \bibinfo {author} {\bibfnamefont {B.}~\bibnamefont {Pfeiffer}}, \bibinfo {author} {\bibfnamefont {A.}~\bibnamefont {W\"ohr}}, \bibinfo {author} {\bibfnamefont {W.}~\bibnamefont {Ziegert}}, \bibinfo {author} {\bibfnamefont {Y.~E.}\ \bibnamefont {Penionzhkevich}}, \bibinfo {author} {\bibfnamefont {S.~M.}\ \bibnamefont {Lukyanov}}, \bibinfo {author} {\bibfnamefont {V.~S.}\ \bibnamefont {Salamatin}}, \bibinfo {author} {\bibfnamefont {R.}~\bibnamefont {Anne}}, \bibinfo {author}
  {\bibfnamefont {C.}~\bibnamefont {Borcea}}, \bibinfo {author} {\bibfnamefont {L.~K.}\ \bibnamefont {Fifield}}, \bibinfo {author} {\bibfnamefont {M.}~\bibnamefont {Lewitowicz}}, \bibinfo {author} {\bibfnamefont {M.~G.}\ \bibnamefont {Saint-Laurent}}, \bibinfo {author} {\bibfnamefont {D.}~\bibnamefont {Bazin}}, \bibinfo {author} {\bibfnamefont {C.}~\bibnamefont {D\'etraz}}, \bibinfo {author} {\bibfnamefont {F.-K.}\ \bibnamefont {Thielemann}},\ and\ \bibinfo {author} {\bibfnamefont {W.}~\bibnamefont {Hillebrandt}},\ }\bibfield  {title} {\bibinfo {title} {Decay properties of exotic n\ensuremath{\simeq}28 s and cl nuclei and the $^{48}\mathrm{Ca}$${/}^{46}$ca abundance ratio},\ }\href {https://doi.org/10.1103/PhysRevC.47.2941} {\bibfield  {journal} {\bibinfo  {journal} {Phys. Rev. C}\ }\textbf {\bibinfo {volume} {47}},\ \bibinfo {pages} {2941} (\bibinfo {year} {1993})}\BibitemShut {NoStop}%
\bibitem [{\citenamefont {Sorlin}\ \emph {et~al.}(1995)\citenamefont {Sorlin}, \citenamefont {Guillemaud-Mueller}, \citenamefont {Anne}, \citenamefont {Axelsson}, \citenamefont {Bazin}, \citenamefont {Böhmer}, \citenamefont {Borrel}, \citenamefont {Jading}, \citenamefont {Keller}, \citenamefont {Kratz}, \citenamefont {Lewitowicz}, \citenamefont {Lukyanov}, \citenamefont {Mehren}, \citenamefont {Mueller}, \citenamefont {Penionzhkevich}, \citenamefont {Pougheon}, \citenamefont {Saint-Laurent}, \citenamefont {Salamatin}, \citenamefont {Shoedder},\ and\ \citenamefont {Wöhr}}]{sorlin_Pn}%
  \BibitemOpen
  \bibfield  {author} {\bibinfo {author} {\bibfnamefont {O.}~\bibnamefont {Sorlin}}, \bibinfo {author} {\bibfnamefont {D.}~\bibnamefont {Guillemaud-Mueller}}, \bibinfo {author} {\bibfnamefont {R.}~\bibnamefont {Anne}}, \bibinfo {author} {\bibfnamefont {L.}~\bibnamefont {Axelsson}}, \bibinfo {author} {\bibfnamefont {D.}~\bibnamefont {Bazin}}, \bibinfo {author} {\bibfnamefont {W.}~\bibnamefont {Böhmer}}, \bibinfo {author} {\bibfnamefont {V.}~\bibnamefont {Borrel}}, \bibinfo {author} {\bibfnamefont {Y.}~\bibnamefont {Jading}}, \bibinfo {author} {\bibfnamefont {H.}~\bibnamefont {Keller}}, \bibinfo {author} {\bibfnamefont {K.-L.}\ \bibnamefont {Kratz}}, \bibinfo {author} {\bibfnamefont {M.}~\bibnamefont {Lewitowicz}}, \bibinfo {author} {\bibfnamefont {S.}~\bibnamefont {Lukyanov}}, \bibinfo {author} {\bibfnamefont {T.}~\bibnamefont {Mehren}}, \bibinfo {author} {\bibfnamefont {A.}~\bibnamefont {Mueller}}, \bibinfo {author} {\bibfnamefont {Y.}~\bibnamefont {Penionzhkevich}}, \bibinfo {author} {\bibfnamefont
  {F.}~\bibnamefont {Pougheon}}, \bibinfo {author} {\bibfnamefont {M.}~\bibnamefont {Saint-Laurent}}, \bibinfo {author} {\bibfnamefont {V.}~\bibnamefont {Salamatin}}, \bibinfo {author} {\bibfnamefont {S.}~\bibnamefont {Shoedder}},\ and\ \bibinfo {author} {\bibfnamefont {A.}~\bibnamefont {Wöhr}},\ }\bibfield  {title} {\bibinfo {title} {Beta-decay studies of far from stability nuclei near n = 28},\ }\href {https://doi.org/https://doi.org/10.1016/0375-9474(94)00755-C} {\bibfield  {journal} {\bibinfo  {journal} {Nuclear Physics A}\ }\textbf {\bibinfo {volume} {583}},\ \bibinfo {pages} {763} (\bibinfo {year} {1995})},\ \bibinfo {note} {nucleus-Nucleus Collisions}\BibitemShut {NoStop}%
\bibitem [{\citenamefont {Grévy}\ \emph {et~al.}(2004)\citenamefont {Grévy}, \citenamefont {Angélique}, \citenamefont {Baumann}, \citenamefont {Borcea}, \citenamefont {Buta}, \citenamefont {Canchel}, \citenamefont {Catford}, \citenamefont {Courtin}, \citenamefont {Daugas}, \citenamefont {{de Oliveira}}, \citenamefont {Dessagne}, \citenamefont {Dlouhy}, \citenamefont {Knipper}, \citenamefont {Kratz}, \citenamefont {Lecolley}, \citenamefont {Lecouey}, \citenamefont {Lehrsenneau}, \citenamefont {Lewitowicz}, \citenamefont {Liénard}, \citenamefont {Lukyanov}, \citenamefont {Maréchal}, \citenamefont {Miehé}, \citenamefont {Mrazek}, \citenamefont {Negoita}, \citenamefont {Orr}, \citenamefont {Pantelica}, \citenamefont {Penionzhkevich}, \citenamefont {Péter}, \citenamefont {Pfeiffer}, \citenamefont {Pietri}, \citenamefont {Poirier}, \citenamefont {Sorlin}, \citenamefont {Stanoiu}, \citenamefont {Stefan}, \citenamefont {Stodel},\ and\ \citenamefont {Timis}}]{grevy}%
  \BibitemOpen
  \bibfield  {author} {\bibinfo {author} {\bibfnamefont {S.}~\bibnamefont {Grévy}}, \bibinfo {author} {\bibfnamefont {J.}~\bibnamefont {Angélique}}, \bibinfo {author} {\bibfnamefont {P.}~\bibnamefont {Baumann}}, \bibinfo {author} {\bibfnamefont {C.}~\bibnamefont {Borcea}}, \bibinfo {author} {\bibfnamefont {A.}~\bibnamefont {Buta}}, \bibinfo {author} {\bibfnamefont {G.}~\bibnamefont {Canchel}}, \bibinfo {author} {\bibfnamefont {W.}~\bibnamefont {Catford}}, \bibinfo {author} {\bibfnamefont {S.}~\bibnamefont {Courtin}}, \bibinfo {author} {\bibfnamefont {J.}~\bibnamefont {Daugas}}, \bibinfo {author} {\bibfnamefont {F.}~\bibnamefont {{de Oliveira}}}, \bibinfo {author} {\bibfnamefont {P.}~\bibnamefont {Dessagne}}, \bibinfo {author} {\bibfnamefont {Z.}~\bibnamefont {Dlouhy}}, \bibinfo {author} {\bibfnamefont {A.}~\bibnamefont {Knipper}}, \bibinfo {author} {\bibfnamefont {K.}~\bibnamefont {Kratz}}, \bibinfo {author} {\bibfnamefont {F.}~\bibnamefont {Lecolley}}, \bibinfo {author} {\bibfnamefont {J.}~\bibnamefont
  {Lecouey}}, \bibinfo {author} {\bibfnamefont {G.}~\bibnamefont {Lehrsenneau}}, \bibinfo {author} {\bibfnamefont {M.}~\bibnamefont {Lewitowicz}}, \bibinfo {author} {\bibfnamefont {E.}~\bibnamefont {Liénard}}, \bibinfo {author} {\bibfnamefont {S.}~\bibnamefont {Lukyanov}}, \bibinfo {author} {\bibfnamefont {F.}~\bibnamefont {Maréchal}}, \bibinfo {author} {\bibfnamefont {C.}~\bibnamefont {Miehé}}, \bibinfo {author} {\bibfnamefont {J.}~\bibnamefont {Mrazek}}, \bibinfo {author} {\bibfnamefont {F.}~\bibnamefont {Negoita}}, \bibinfo {author} {\bibfnamefont {N.}~\bibnamefont {Orr}}, \bibinfo {author} {\bibfnamefont {D.}~\bibnamefont {Pantelica}}, \bibinfo {author} {\bibfnamefont {Y.}~\bibnamefont {Penionzhkevich}}, \bibinfo {author} {\bibfnamefont {J.}~\bibnamefont {Péter}}, \bibinfo {author} {\bibfnamefont {B.}~\bibnamefont {Pfeiffer}}, \bibinfo {author} {\bibfnamefont {S.}~\bibnamefont {Pietri}}, \bibinfo {author} {\bibfnamefont {E.}~\bibnamefont {Poirier}}, \bibinfo {author} {\bibfnamefont {O.}~\bibnamefont
  {Sorlin}}, \bibinfo {author} {\bibfnamefont {M.}~\bibnamefont {Stanoiu}}, \bibinfo {author} {\bibfnamefont {I.}~\bibnamefont {Stefan}}, \bibinfo {author} {\bibfnamefont {C.}~\bibnamefont {Stodel}},\ and\ \bibinfo {author} {\bibfnamefont {C.}~\bibnamefont {Timis}},\ }\bibfield  {title} {\bibinfo {title} {Beta-decay half-lives at the n=28 shell closure},\ }\href {https://doi.org/https://doi.org/10.1016/j.physletb.2004.06.005} {\bibfield  {journal} {\bibinfo  {journal} {Physics Letters B}\ }\textbf {\bibinfo {volume} {594}},\ \bibinfo {pages} {252} (\bibinfo {year} {2004})}\BibitemShut {NoStop}%
\bibitem [{\citenamefont {Weissman}\ \emph {et~al.}(2004)\citenamefont {Weissman}, \citenamefont {Arnd}, \citenamefont {Bergmann}, \citenamefont {Brown}, \citenamefont {Catherall}, \citenamefont {Cederkall}, \citenamefont {Dillmann}, \citenamefont {Hallmann}, \citenamefont {Fraile}, \citenamefont {Franchoo}, \citenamefont {Gaudefroy}, \citenamefont {K\"oster}, \citenamefont {Kratz}, \citenamefont {Pfeiffer},\ and\ \citenamefont {Sorlin}}]{Weissman2004}%
  \BibitemOpen
  \bibfield  {author} {\bibinfo {author} {\bibfnamefont {L.}~\bibnamefont {Weissman}}, \bibinfo {author} {\bibfnamefont {O.}~\bibnamefont {Arnd}}, \bibinfo {author} {\bibfnamefont {U.}~\bibnamefont {Bergmann}}, \bibinfo {author} {\bibfnamefont {A.}~\bibnamefont {Brown}}, \bibinfo {author} {\bibfnamefont {R.}~\bibnamefont {Catherall}}, \bibinfo {author} {\bibfnamefont {J.}~\bibnamefont {Cederkall}}, \bibinfo {author} {\bibfnamefont {I.}~\bibnamefont {Dillmann}}, \bibinfo {author} {\bibfnamefont {O.}~\bibnamefont {Hallmann}}, \bibinfo {author} {\bibfnamefont {L.}~\bibnamefont {Fraile}}, \bibinfo {author} {\bibfnamefont {S.}~\bibnamefont {Franchoo}}, \bibinfo {author} {\bibfnamefont {L.}~\bibnamefont {Gaudefroy}}, \bibinfo {author} {\bibfnamefont {U.}~\bibnamefont {K\"oster}}, \bibinfo {author} {\bibfnamefont {K.-L.}\ \bibnamefont {Kratz}}, \bibinfo {author} {\bibfnamefont {B.}~\bibnamefont {Pfeiffer}},\ and\ \bibinfo {author} {\bibfnamefont {O.}~\bibnamefont {Sorlin}},\ }\bibfield  {title} {\bibinfo {title}
  {$\ensuremath{\beta}$ decay of $^{47}\mathrm{Ar}$},\ }\href {https://doi.org/10.1103/PhysRevC.70.024304} {\bibfield  {journal} {\bibinfo  {journal} {Phys. Rev. C}\ }\textbf {\bibinfo {volume} {70}},\ \bibinfo {pages} {024304} (\bibinfo {year} {2004})}\BibitemShut {NoStop}%
\bibitem [{\citenamefont {Mrázek}\ \emph {et~al.}(2004)\citenamefont {Mrázek}, \citenamefont {Grévy}, \citenamefont {Iulian}, \citenamefont {Buta}, \citenamefont {Negoita}, \citenamefont {Angélique}, \citenamefont {Baumann}, \citenamefont {Borcea}, \citenamefont {Canchel}, \citenamefont {Catford}, \citenamefont {Courtin}, \citenamefont {Daugas}, \citenamefont {Dlouhý}, \citenamefont {Dessagne}, \citenamefont {Knipper}, \citenamefont {Lehrsenneau}, \citenamefont {Lecolley}, \citenamefont {Lecouey}, \citenamefont {Lewitowicz}, \citenamefont {Liénard}, \citenamefont {Lukyanov}, \citenamefont {Maréchal}, \citenamefont {Miehe}, \citenamefont {{de Oliveira}}, \citenamefont {Orr}, \citenamefont {Pantelica}, \citenamefont {Penionzhkevich}, \citenamefont {Peter}, \citenamefont {Pietri}, \citenamefont {Poirier}, \citenamefont {Sorlin}, \citenamefont {Stanoiu}, \citenamefont {Stodel}, \citenamefont {Tarasov},\ and\ \citenamefont {Timis}}]{Mrazek2004}%
  \BibitemOpen
  \bibfield  {author} {\bibinfo {author} {\bibfnamefont {J.}~\bibnamefont {Mrázek}}, \bibinfo {author} {\bibfnamefont {S.}~\bibnamefont {Grévy}}, \bibinfo {author} {\bibfnamefont {S.}~\bibnamefont {Iulian}}, \bibinfo {author} {\bibfnamefont {A.}~\bibnamefont {Buta}}, \bibinfo {author} {\bibfnamefont {F.}~\bibnamefont {Negoita}}, \bibinfo {author} {\bibfnamefont {J.}~\bibnamefont {Angélique}}, \bibinfo {author} {\bibfnamefont {P.}~\bibnamefont {Baumann}}, \bibinfo {author} {\bibfnamefont {C.}~\bibnamefont {Borcea}}, \bibinfo {author} {\bibfnamefont {G.}~\bibnamefont {Canchel}}, \bibinfo {author} {\bibfnamefont {W.}~\bibnamefont {Catford}}, \bibinfo {author} {\bibfnamefont {S.}~\bibnamefont {Courtin}}, \bibinfo {author} {\bibfnamefont {J.}~\bibnamefont {Daugas}}, \bibinfo {author} {\bibfnamefont {Z.}~\bibnamefont {Dlouhý}}, \bibinfo {author} {\bibfnamefont {P.}~\bibnamefont {Dessagne}}, \bibinfo {author} {\bibfnamefont {A.}~\bibnamefont {Knipper}}, \bibinfo {author} {\bibfnamefont {G.}~\bibnamefont
  {Lehrsenneau}}, \bibinfo {author} {\bibfnamefont {F.}~\bibnamefont {Lecolley}}, \bibinfo {author} {\bibfnamefont {J.}~\bibnamefont {Lecouey}}, \bibinfo {author} {\bibfnamefont {M.}~\bibnamefont {Lewitowicz}}, \bibinfo {author} {\bibfnamefont {E.}~\bibnamefont {Liénard}}, \bibinfo {author} {\bibfnamefont {S.}~\bibnamefont {Lukyanov}}, \bibinfo {author} {\bibfnamefont {F.}~\bibnamefont {Maréchal}}, \bibinfo {author} {\bibfnamefont {C.}~\bibnamefont {Miehe}}, \bibinfo {author} {\bibfnamefont {F.}~\bibnamefont {{de Oliveira}}}, \bibinfo {author} {\bibfnamefont {N.}~\bibnamefont {Orr}}, \bibinfo {author} {\bibfnamefont {D.}~\bibnamefont {Pantelica}}, \bibinfo {author} {\bibfnamefont {Y.}~\bibnamefont {Penionzhkevich}}, \bibinfo {author} {\bibfnamefont {J.}~\bibnamefont {Peter}}, \bibinfo {author} {\bibfnamefont {S.}~\bibnamefont {Pietri}}, \bibinfo {author} {\bibfnamefont {E.}~\bibnamefont {Poirier}}, \bibinfo {author} {\bibfnamefont {O.}~\bibnamefont {Sorlin}}, \bibinfo {author} {\bibfnamefont
  {M.}~\bibnamefont {Stanoiu}}, \bibinfo {author} {\bibfnamefont {O.}~\bibnamefont {Stodel}}, \bibinfo {author} {\bibfnamefont {O.}~\bibnamefont {Tarasov}},\ and\ \bibinfo {author} {\bibfnamefont {C.}~\bibnamefont {Timis}},\ }\bibfield  {title} {\bibinfo {title} {Study of neutron-rich argon isotopes in $\beta$-decay},\ }\href {https://doi.org/https://doi.org/10.1016/j.nuclphysa.2004.03.021} {\bibfield  {journal} {\bibinfo  {journal} {Nuclear Physics A}\ }\textbf {\bibinfo {volume} {734}},\ \bibinfo {pages} {E65} (\bibinfo {year} {2004})},\ \bibinfo {note} {proceedings of the Eighth International Conference on Nucleus-Nucleus Collisions (NN2003)}\BibitemShut {NoStop}%
\bibitem [{\citenamefont {Wang}\ \emph {et~al.}(2021)\citenamefont {Wang}, \citenamefont {Huang}, \citenamefont {Kondev}, \citenamefont {Audi},\ and\ \citenamefont {Naimi}}]{AME2020}%
  \BibitemOpen
  \bibfield  {author} {\bibinfo {author} {\bibfnamefont {M.}~\bibnamefont {Wang}}, \bibinfo {author} {\bibfnamefont {W.}~\bibnamefont {Huang}}, \bibinfo {author} {\bibfnamefont {F.}~\bibnamefont {Kondev}}, \bibinfo {author} {\bibfnamefont {G.}~\bibnamefont {Audi}},\ and\ \bibinfo {author} {\bibfnamefont {S.}~\bibnamefont {Naimi}},\ }\bibfield  {title} {\bibinfo {title} {The ame 2020 atomic mass evaluation (ii)},\ }\href {https://doi.org/10.1088/1674-1137/abddaf} {\bibfield  {journal} {\bibinfo  {journal} {Chinese Physics C}\ }\textbf {\bibinfo {volume} {45}},\ \bibinfo {pages} {030003} (\bibinfo {year} {2021})}\BibitemShut {NoStop}%
\bibitem [{\citenamefont {Turkat}\ \emph {et~al.}(2023)\citenamefont {Turkat}, \citenamefont {Mougeot}, \citenamefont {Singh},\ and\ \citenamefont {Zuber}}]{forbidden_TURKAT}%
  \BibitemOpen
  \bibfield  {author} {\bibinfo {author} {\bibfnamefont {S.}~\bibnamefont {Turkat}}, \bibinfo {author} {\bibfnamefont {X.}~\bibnamefont {Mougeot}}, \bibinfo {author} {\bibfnamefont {B.}~\bibnamefont {Singh}},\ and\ \bibinfo {author} {\bibfnamefont {K.}~\bibnamefont {Zuber}},\ }\bibfield  {title} {\bibinfo {title} {Systematics of logft values for ${\ensuremath{\beta}}^{\ensuremath{-}}$, and ec/${\ensuremath{\beta}}^{\ensuremath{+}}$ transitions},\ }\href {https://doi.org/https://doi.org/10.1016/j.adt.2023.101584} {\bibfield  {journal} {\bibinfo  {journal} {Atomic Data and Nuclear Data Tables}\ }\textbf {\bibinfo {volume} {152}},\ \bibinfo {pages} {101584} (\bibinfo {year} {2023})}\BibitemShut {NoStop}%
\bibitem [{\citenamefont {Gade}\ \emph {et~al.}(2016)\citenamefont {Gade}, \citenamefont {Tostevin}, \citenamefont {Bader}, \citenamefont {Baugher}, \citenamefont {Bazin}, \citenamefont {Berryman}, \citenamefont {Brown}, \citenamefont {Diget}, \citenamefont {Glasmacher}, \citenamefont {Hartley}, \citenamefont {Lunderberg}, \citenamefont {Stroberg}, \citenamefont {Recchia}, \citenamefont {Ratkiewicz}, \citenamefont {Weisshaar},\ and\ \citenamefont {Wimmer}}]{Gade2016}%
  \BibitemOpen
  \bibfield  {author} {\bibinfo {author} {\bibfnamefont {A.}~\bibnamefont {Gade}}, \bibinfo {author} {\bibfnamefont {J.~A.}\ \bibnamefont {Tostevin}}, \bibinfo {author} {\bibfnamefont {V.}~\bibnamefont {Bader}}, \bibinfo {author} {\bibfnamefont {T.}~\bibnamefont {Baugher}}, \bibinfo {author} {\bibfnamefont {D.}~\bibnamefont {Bazin}}, \bibinfo {author} {\bibfnamefont {J.~S.}\ \bibnamefont {Berryman}}, \bibinfo {author} {\bibfnamefont {B.~A.}\ \bibnamefont {Brown}}, \bibinfo {author} {\bibfnamefont {C.~A.}\ \bibnamefont {Diget}}, \bibinfo {author} {\bibfnamefont {T.}~\bibnamefont {Glasmacher}}, \bibinfo {author} {\bibfnamefont {D.~J.}\ \bibnamefont {Hartley}}, \bibinfo {author} {\bibfnamefont {E.}~\bibnamefont {Lunderberg}}, \bibinfo {author} {\bibfnamefont {S.~R.}\ \bibnamefont {Stroberg}}, \bibinfo {author} {\bibfnamefont {F.}~\bibnamefont {Recchia}}, \bibinfo {author} {\bibfnamefont {A.}~\bibnamefont {Ratkiewicz}}, \bibinfo {author} {\bibfnamefont {D.}~\bibnamefont {Weisshaar}},\ and\ \bibinfo {author}
  {\bibfnamefont {K.}~\bibnamefont {Wimmer}},\ }\bibfield  {title} {\bibinfo {title} {Single-particle structure at $n=29$: The structure of $^{47}\mathrm{Ar}$ and first spectroscopy of $^{45}\mathrm{S}$},\ }\href {https://doi.org/10.1103/PhysRevC.93.054315} {\bibfield  {journal} {\bibinfo  {journal} {Phys. Rev. C}\ }\textbf {\bibinfo {volume} {93}},\ \bibinfo {pages} {054315} (\bibinfo {year} {2016})}\BibitemShut {NoStop}%
\bibitem [{\citenamefont {Maheshwari}\ and\ \citenamefont {Nomura}(2024)}]{IPM_Maheshwari_2024}%
  \BibitemOpen
  \bibfield  {author} {\bibinfo {author} {\bibfnamefont {B.}~\bibnamefont {Maheshwari}}\ and\ \bibinfo {author} {\bibfnamefont {K.}~\bibnamefont {Nomura}},\ }\bibfield  {title} {\bibinfo {title} {Corrigendum: ‘weakening of n = 28 shell gap and the nature of 02+ states’ (2024 j. phys. g: Nucl. part. phys. 51 095101)},\ }\href {https://doi.org/10.1088/1361-6471/ad6a2c} {\bibfield  {journal} {\bibinfo  {journal} {Journal of Physics G: Nuclear and Particle Physics}\ }\textbf {\bibinfo {volume} {51}},\ \bibinfo {pages} {109501} (\bibinfo {year} {2024})}\BibitemShut {NoStop}%
\bibitem [{\citenamefont {Cox}\ \emph {et~al.}(2024)\citenamefont {Cox}, \citenamefont {Xu}, \citenamefont {Grzywacz}, \citenamefont {Ong}, \citenamefont {Rasco}, \citenamefont {Kitamura}, \citenamefont {Hoskins}, \citenamefont {Neupane}, \citenamefont {Ruland}, \citenamefont {Allmond}, \citenamefont {King}, \citenamefont {Lubna}, \citenamefont {Rykaczewski}, \citenamefont {Schatz}, \citenamefont {Sherrill}, \citenamefont {Tarasov}, \citenamefont {Ayangeakaa}, \citenamefont {Berg}, \citenamefont {Bleuel}, \citenamefont {Cerizza}, \citenamefont {Christie}, \citenamefont {Chester}, \citenamefont {Davis}, \citenamefont {Dembski}, \citenamefont {Doetsch}, \citenamefont {Duarte}, \citenamefont {Estrade}, \citenamefont {Fija\l{}kowska}, \citenamefont {Gray}, \citenamefont {Good}, \citenamefont {Haak}, \citenamefont {Hanai}, \citenamefont {Harke}, \citenamefont {Harris}, \citenamefont {Hermansen}, \citenamefont {Hoff}, \citenamefont {Jain}, \citenamefont {Karny}, \citenamefont {Kolos}, \citenamefont {Laminack},
  \citenamefont {Liddick}, \citenamefont {Longfellow}, \citenamefont {Lyons}, \citenamefont {Madurga}, \citenamefont {Mogannam}, \citenamefont {Nowicki}, \citenamefont {Ogunbeku}, \citenamefont {Owens-Fryar}, \citenamefont {Rajabali}, \citenamefont {Richard}, \citenamefont {Ronning}, \citenamefont {Rose}, \citenamefont {Siegl}, \citenamefont {Singh}, \citenamefont {Spyrou}, \citenamefont {Sweet}, \citenamefont {Tsantiri}, \citenamefont {Walters},\ and\ \citenamefont {Yokoyama}}]{Cox2024}%
  \BibitemOpen
  \bibfield  {author} {\bibinfo {author} {\bibfnamefont {I.}~\bibnamefont {Cox}}, \bibinfo {author} {\bibfnamefont {Z.~Y.}\ \bibnamefont {Xu}}, \bibinfo {author} {\bibfnamefont {R.}~\bibnamefont {Grzywacz}}, \bibinfo {author} {\bibfnamefont {W.-J.}\ \bibnamefont {Ong}}, \bibinfo {author} {\bibfnamefont {B.~C.}\ \bibnamefont {Rasco}}, \bibinfo {author} {\bibfnamefont {N.}~\bibnamefont {Kitamura}}, \bibinfo {author} {\bibfnamefont {D.}~\bibnamefont {Hoskins}}, \bibinfo {author} {\bibfnamefont {S.}~\bibnamefont {Neupane}}, \bibinfo {author} {\bibfnamefont {T.~J.}\ \bibnamefont {Ruland}}, \bibinfo {author} {\bibfnamefont {J.~M.}\ \bibnamefont {Allmond}}, \bibinfo {author} {\bibfnamefont {T.~T.}\ \bibnamefont {King}}, \bibinfo {author} {\bibfnamefont {R.~S.}\ \bibnamefont {Lubna}}, \bibinfo {author} {\bibfnamefont {K.~P.}\ \bibnamefont {Rykaczewski}}, \bibinfo {author} {\bibfnamefont {H.}~\bibnamefont {Schatz}}, \bibinfo {author} {\bibfnamefont {B.~M.}\ \bibnamefont {Sherrill}}, \bibinfo {author} {\bibfnamefont
  {O.~B.}\ \bibnamefont {Tarasov}}, \bibinfo {author} {\bibfnamefont {A.~D.}\ \bibnamefont {Ayangeakaa}}, \bibinfo {author} {\bibfnamefont {H.~C.}\ \bibnamefont {Berg}}, \bibinfo {author} {\bibfnamefont {D.~L.}\ \bibnamefont {Bleuel}}, \bibinfo {author} {\bibfnamefont {G.}~\bibnamefont {Cerizza}}, \bibinfo {author} {\bibfnamefont {J.}~\bibnamefont {Christie}}, \bibinfo {author} {\bibfnamefont {A.}~\bibnamefont {Chester}}, \bibinfo {author} {\bibfnamefont {J.}~\bibnamefont {Davis}}, \bibinfo {author} {\bibfnamefont {C.}~\bibnamefont {Dembski}}, \bibinfo {author} {\bibfnamefont {A.~A.}\ \bibnamefont {Doetsch}}, \bibinfo {author} {\bibfnamefont {J.~G.}\ \bibnamefont {Duarte}}, \bibinfo {author} {\bibfnamefont {A.}~\bibnamefont {Estrade}}, \bibinfo {author} {\bibfnamefont {A.}~\bibnamefont {Fija\l{}kowska}}, \bibinfo {author} {\bibfnamefont {T.~J.}\ \bibnamefont {Gray}}, \bibinfo {author} {\bibfnamefont {E.~C.}\ \bibnamefont {Good}}, \bibinfo {author} {\bibfnamefont {K.}~\bibnamefont {Haak}}, \bibinfo {author}
  {\bibfnamefont {S.}~\bibnamefont {Hanai}}, \bibinfo {author} {\bibfnamefont {J.~T.}\ \bibnamefont {Harke}}, \bibinfo {author} {\bibfnamefont {C.}~\bibnamefont {Harris}}, \bibinfo {author} {\bibfnamefont {K.}~\bibnamefont {Hermansen}}, \bibinfo {author} {\bibfnamefont {D.~E.~M.}\ \bibnamefont {Hoff}}, \bibinfo {author} {\bibfnamefont {R.}~\bibnamefont {Jain}}, \bibinfo {author} {\bibfnamefont {M.}~\bibnamefont {Karny}}, \bibinfo {author} {\bibfnamefont {K.}~\bibnamefont {Kolos}}, \bibinfo {author} {\bibfnamefont {A.}~\bibnamefont {Laminack}}, \bibinfo {author} {\bibfnamefont {S.~N.}\ \bibnamefont {Liddick}}, \bibinfo {author} {\bibfnamefont {B.}~\bibnamefont {Longfellow}}, \bibinfo {author} {\bibfnamefont {S.}~\bibnamefont {Lyons}}, \bibinfo {author} {\bibfnamefont {M.}~\bibnamefont {Madurga}}, \bibinfo {author} {\bibfnamefont {M.~J.}\ \bibnamefont {Mogannam}}, \bibinfo {author} {\bibfnamefont {A.}~\bibnamefont {Nowicki}}, \bibinfo {author} {\bibfnamefont {T.~H.}\ \bibnamefont {Ogunbeku}}, \bibinfo {author}
  {\bibfnamefont {G.}~\bibnamefont {Owens-Fryar}}, \bibinfo {author} {\bibfnamefont {M.~M.}\ \bibnamefont {Rajabali}}, \bibinfo {author} {\bibfnamefont {A.~L.}\ \bibnamefont {Richard}}, \bibinfo {author} {\bibfnamefont {E.~K.}\ \bibnamefont {Ronning}}, \bibinfo {author} {\bibfnamefont {G.~E.}\ \bibnamefont {Rose}}, \bibinfo {author} {\bibfnamefont {K.}~\bibnamefont {Siegl}}, \bibinfo {author} {\bibfnamefont {M.}~\bibnamefont {Singh}}, \bibinfo {author} {\bibfnamefont {A.}~\bibnamefont {Spyrou}}, \bibinfo {author} {\bibfnamefont {A.}~\bibnamefont {Sweet}}, \bibinfo {author} {\bibfnamefont {A.}~\bibnamefont {Tsantiri}}, \bibinfo {author} {\bibfnamefont {W.~B.}\ \bibnamefont {Walters}},\ and\ \bibinfo {author} {\bibfnamefont {R.}~\bibnamefont {Yokoyama}},\ }\bibfield  {title} {\bibinfo {title} {Proton shell gaps in $n=28$ nuclei from the first complete spectroscopy study with frib decay station initiator},\ }\href {https://doi.org/10.1103/PhysRevLett.132.152503} {\bibfield  {journal} {\bibinfo  {journal} {Phys.
  Rev. Lett.}\ }\textbf {\bibinfo {volume} {132}},\ \bibinfo {pages} {152503} (\bibinfo {year} {2024})}\BibitemShut {NoStop}%
\end{thebibliography}%

\end{document}